\documentclass[11pt,a4paper]{article}
\usepackage{color}
\usepackage{wrapfig}
\usepackage{subfigure}
\usepackage{graphicx}% Include figure files
\usepackage{dcolumn}% Align table columns on decimal point
\usepackage{bm}% bold math
\usepackage{epsf}
\usepackage{amssymb,amsmath,amsfonts,amsthm, amscd, mathrsfs,helvet}

\usepackage{verbatim}
\usepackage{psfrag}
\usepackage[all]{xy}
\usepackage{cite}
\usepackage{epic}
\makeatletter
\renewcommand\section{\@startsection{section}{2}{\z@}%
                         {-3.25ex\@plus -1ex \@minus -.2ex}%
                         {1.5ex \@plus .2ex}%
                         {\normalfont\large\bfseries\mathversion{bold}}}
\renewcommand\subsection{\@startsection{subsection}{3}{\z@}%
                         {-3.25ex\@plus -1ex \@minus -.2ex}%
                         {1.5ex \@plus .2ex}%
                         {\normalfont\normalsize\bfseries\mathversion{bold}}}
\makeatother

\def\appendix#1{
  \addtocounter{section}{1}
 \setcounter{equation}{0}
  \renewcommand{\thesection}{\Alph{section}}
 \section*{Appendix \thesection\protect\indent \parbox[t]{11.715cm} {#1}}
  \addcontentsline{toc}{section}{Appendix \thesection \ \ \ #1}
   }

\newcommand{\newsection}{
\setcounter{equation}{0}
\section}

\newcommand{\captionfonts}{\footnotesize}

\makeatletter  % Allow the use of @ in command names
\long\def\@makecaption#1#2{%
  \vskip\abovecaptionskip
  \sbox\@tempboxa{{\captionfonts #1: #2}}%
  \ifdim \wd\@tempboxa >\hsize
    {\captionfonts #1: #2\par}
  \else
    \hbox to\hsize{\hfil\box\@tempboxa\hfil}%
  \fi
  \vskip\belowcaptionskip}
\makeatother
\newcommand{\Eqn}[1]{&\hspace{-0.5em}#1\hspace{-0.5em}&}

\newcommand{\hI}{{\hat{I}}}
\newcommand{\hK}{{\hat{K}}}
\newcommand{\hP}{{\hat{P}}}

\newcommand{\hS}{{\hat{S}}}
\newcommand{\hT}{{\hat{T}}}
\newcommand{\hOmega}{{\hat{\Omega}}}

\newcommand{\bbZ}{{\mathbb Z}}
\newcommand{\alg}[1]{\mathfrak{#1}}

\newcommand{\beq}{\begin{equation}}
\newcommand{\eeq}{\end{equation}}
\newcommand\beqa{\begin{eqnarray}}
\newcommand\eeqa{\end{eqnarray}}
\newcommand\bea{\begin{array}}
\newcommand\eea{\end{array}}
\newcommand\ba{\begin{array}}
\newcommand\ea{\end{array}}
\newcommand{\nn}{\nonumber}
\newcommand{\shq}{\hat q\hspace{-.55em}/}

\newcommand{\neqa}{\nonumber\end{eqnarray}}
\newcommand{\la}{\label}

\newcommand{\p}{\partial}

\newcommand{\eq}[1]{eq.(\ref{#1})}
\newcommand{\eqs}[2]{eqs.(\ref{#1},\ref{#2})}

\newcommand{\Tr}{{\rm Tr}}

\newcommand{\hf}{\frac{1}{2}}

\renewcommand{\d}{\partial}
\renewcommand{\O}{{\cal O}}

\newcommand{\<}{{\langle}}
\renewcommand{\>}{{\rangle}}

\def\s{\sigma}
\def\a{\alpha}
\def\b{\beta}
\def\th{\theta}

\def\({\left(}
\def\){\right)}
\def\[{\left[}
\def\]{\right]}

\def\<{\langle}
\def\>{\rangle}

\newcommand{\re}{\relax{\rm I\kern-.18em R}}

\renewcommand{\sp}{p\hspace{-.40em}/}
\newcommand{\stp}{\tilde p\hspace{-.40em}/}
\newcommand{\shp}{\hat p\hspace{-.40em}/}
\newcommand{\sq}{q\hspace{-.55em}/}
\renewcommand{\u}{{\rm u}}

\def\a{{\alpha}}

\def\hf{\frac{1}{2}}

\def\({\left(}
\def\){\right)}
\def\[{\left[}
\def\]{\right]}

\def\su2{{SU(2)}}

\def\tr{{\rm tr}}

\def\sG{\,\slash\!\!\!\! G}
\def\sH{\slash\!\!\!\! H}
\def\pint{-\hskip-0.41cm \int}

\begin{document}

\renewcommand{\thefootnote}{\fnsymbol{footnote}}
\setcounter{footnote}{0}

\thispagestyle{empty}
\begin{flushright}
LPTENS-06/05\\
hep-th/0603043
\end{flushright}
\begin{flushleft}
\ \\[-15mm]\hspace{22mm}\includegraphics[scale=0.2]{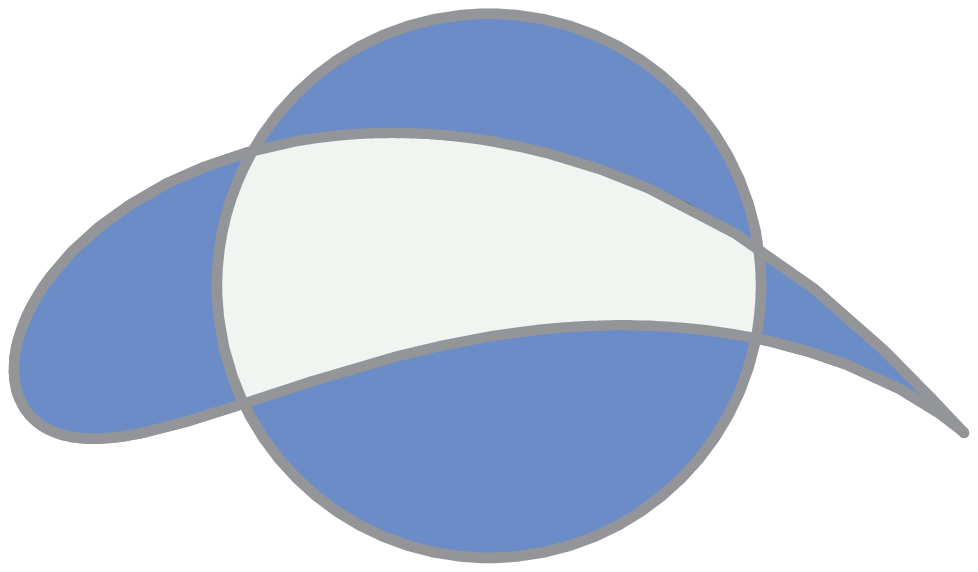}
\end{flushleft}
\vspace{.5cm} \setcounter{footnote}{0}
\begin{center}
%{ \today\\}
{\Large{\bf Strings as Multi-Particle States of Quantum Sigma-Models
\par}
   }\vspace{4mm}
{\large\rm Nikolay~Gromov$^{a,b,c}$,
Vladimir~Kazakov$^{a,b,\!}$\footnote{Membre de l'Institut
Universitaire de France} ,
Kazuhiro~Sakai$^{a,\!}$\footnote{Chercheur Associ\'e du C.N.R.S.} ,
Pedro~Vieira$^{a,b,d}$
\\[7mm]
\large\it\small ${}^a$ Laboratoire de Physique Th\'eorique\\
de l'Ecole Normale Sup\'erieure\footnote{Unit\'e mixte du C.N.R.S.
et de l' Ecole Normale Sup\'erieure, UMR 8549.} et
${}^b\,$l'Universit\'e Paris-VI,\\
Paris, 75231, France\footnote{
\tt\noindent gromov@thd.pnpi.spb.ru,\ kazakov@physique.ens.fr,\ sakai@lpt.ens.fr,\ pedro@lpt.ens.fr}\vspace{3mm}\\
\large\it\small ${}^c$ St.Petersburg INP,\\
 Gatchina, 188 300, St.Petersburg, Russia \vspace{3mm}\\
\large\it\small ${}^d$
 Departamento de F\'\i sica e Centro de F\'\i sica do Porto\\
Faculdade de Ci\^encias da Universidade do Porto\\
Rua do Campo Alegre, 687, \,4169-007 Porto, Portugal }

\noindent\\[20mm]

{\sc Abstract}\\[2mm]
\end{center}

   We study the quantum Bethe ansatz equations in
  the $O(2n)$ sigma-model for physical particles on
  a circle, with the interaction  given by the
  Zamolodchikovs' S-matrix, in view of its
  application to quantization of the string  on the
  $S^{\,2n-1}\times R_t$ space. For a finite number
  of particles, the system  looks like an
  inhomogeneous integrable $O(2n)$ spin chain.
  Similarly to  $OSp\,(2m+n|2m)$ conformal sigma-model
  considered by Mann and Polchinski, we
  reproduce in the limit of large density of
  particles  the  finite gap
  Kazakov--Marshakov--Minahan--Zarembo
  solution for the
  classical string and its generalization to the
  $S^5\times R_t$ sector of the
  Green--Schwarz--Metsaev--Tseytlin superstring. We
  also reproduce some quantum effects: the BMN limit and the
  quantum homogeneous spin chain similar to the one
  describing the bosonic sector of the one-loop
  ${\cal N}=4$ super Yang--Mills theory. We discuss the
  prospects of generalization of these Bethe
  equations to the full superstring sigma-model.

%\newpage
%\tableofcontents
\newpage

\setcounter{page}{1}
\renewcommand{\thefootnote}{\arabic{footnote}}
\setcounter{footnote}{0}

\vskip1cm

\tableofcontents

\newpage

\newsection{Introduction}

Superstring theory on non-trivial gravitational backgrounds is
extremely important for the progress of the theory of fundamental
interactions and cosmology: not only it might describe some rather
realistic  models in cosmology and black hole physics but it also
provides a dual description of some strongly interacting planar
Yang--Mills gauge theories. A set of these dualities are often named
as AdS/CFT correspondence
\cite{Maldacena:1997re,Witten:1998qj,Gubser:1998bc} (see
\cite{Aharony:1999ti} for the review). Apart from its physical
significance for the strongly interacting gauge theories a great
technical advantage of the string side of duality is that the string
theory in the tree approximation is a two dimensional
$\sigma$-model, at least in the world-sheet perturbation theory, and
the  string interactions are not relevant in the planar 't~Hooft
limit of the dual gauge theory. And since the two-dimensional
$\sigma$-models are sometimes solvable it gives us hopes that so are
some realistic string sigma-models, together with the dual
supersymmetric  Yang--Mills (SYM) theories in the large $N$ limit.

The master-example of the AdS/CFT correspondence, the 4-dimensional
conformal ${\cal N}=4$ SYM theory, and its dual, the
Green--Schwarz--Metsaev--Tseytlin superstring \cite{Metsaev:1998it}
on the $AdS_5\times S^5$ background looks particularly promising in
this respect. The 1-loop integrability was discovered in ${\cal
N}=4$ SYM in \cite{Minahan:2002ve} for the bosonic
sector\footnote{Integrable spin chains have been discovered in
(non-supersymmetric) gauge theories before
\cite{Lipatov:1994yb,Faddeev:1995zg}.} where the dilatation operator
was identified with the Hamiltonian of an integrable 1-dimensional
spin chain. It was soon followed by the demonstration of the
classical integrability of the full superstring $\sigma$-model on
$AdS_5\times S^5$\cite{Bena:2003wd}, where already some exact
particular classical string solutions were known
\cite{Gubser:2002tv,Frolov:2002av,Frolov:2003qc}. It was then shown in
\cite{Beisert:2003tq} that the integrability might persist in ${\cal
N}=4$ SYM even in higher loops, and is certainly a property of the
full theory at one loop \cite{Beisert:2003ys,Beisert:2003yb}. The
underlying (super) spin chains are solvable by the standard Bethe
ansatz technique and possess an interesting ferromagnetic continuous
limit \cite{Sutherland,Shastry,Beisert:2003xu} corresponding to very
long local single trace operators in SYM. The resulting anomalous
dimensions were directly compared for some states with the energies
of classical string solutions in the so called Frolov--Tseytlin
limit.

As a next step in this progress, the general classical  solution for
the superstring rotating on $AdS_5\times S^5$ was proposed, first
for the $S^3\times R_t$ sector \cite{Kazakov:2004qf}, then for
$AdS_3\times S^1$ \cite{Kazakov:2004nh}, $S^5\times R_t$
\cite{Beisert:2004ag}, and finally for the full $AdS_5\times S^5$
superstring \cite{Beisert:2005bm}. The full solution even bears  an
imprint of fermions in the corresponding algebraic curve describing
a finite gap solution. It was shown in
\cite{Kazakov:2004qf,Beisert:2004ag,Schafer-Nameki:2004ik,Beisert:2005di} that {\it all} states and
anomalous dimensions of the SYM theory can be mapped onto {\it all}
classical states of the string in this particular limit, giving
another impressive demonstration of the AdS/CFT correspondence.

This integrability is recently generalized to three-loops
\cite{Beisert:2003ys,Staudacher:2004tk} and even a plausible
perturbative all-loop ansatz was proposed
\cite{Beisert:2004hm,Beisert:2005fw}. The perturbative comparison of
the AdS/CFT correspondence works well for this ansatz up to two
loops but breaks down at three loops \cite{Serban:2004jf}, a
mismatch usually explained by certain non-perturbative effects which
are difficult to take into account in the comparison of the weak and
strong coupling regimes.\footnote{Nevertheless, the classical string
KMMZ solution \cite{Kazakov:2004qf} inspired a beautiful result of
\cite{Staudacher:2004tk}, where the three-loop dilatation operator
was constructed for some subsectors of N=4 SYM.} One-loop
world-sheet calculations for particular string solutions
\cite{Frolov:2003tu}, as well as the direct one-loop world-sheet
calculation of the BMN spectra \cite{Parnachev:2002kk,Callan:2003xr,Fuji:2005ry} also show a perfect AdS/CFT correspondence with respect
to similar quantum corrections to the solutions in SYM spin chains
in the scaling limit
\cite{Beisert:2005mq,Hernandez:2005nf,Beisert:2005bv,Gromov:2005gp}.

In spite of the multitude of signs of  integrability of the ${\cal
N}=4$ SYM, the ultimate progress is hardly possible without the
establishment of integrability of the full quantum string
$\sigma$-model. It is not an easy task, though far from being
hopeless (see a discussion in \cite{Polyakov:2005ss}). Similar
$\sigma$-models were already proven to be integrable starting from
 the Zamolodchikovs' discovery of the factorized unitary and
crossing-invariant S-matrix for physical particles in the $O(n)$
sigma-model \cite{Zamolodchikov:1977nu}. Then the $SU(n)_L\times
SU(n)_R$ principal chiral field was solved
\cite{Polyakov:1983tt,Wiegmann:1984ec,Wiegmann:1984pw,Faddeev:1985qu,Ogievetsky:1984pv,Wiegmann:1984mk} and the
corresponding physical S-matrix was found \cite{Wiegmann:1984ec}.
This model represents the first non-trivial example of the field
theory completely solvable in the 't Hooft limit
\cite{Fateev:1994ai}. Some supersymmetric $\sigma$-models, including
conformal ones, appeared to be integrable as well
\cite{Shankar:1977cm,Saleur:2001cw,Mann:2004jr}.

The main difficulties for establishing the integrability of the full
superstring are the following:

1. It is   a non-compact $\sigma$-model as the $AdS_5$ subspace is.
The S-matrix approach is complicated by the existence of a
continuous spectrum of asymptotic states, although we know some
integrable $\sigma$-models of this kind: quantum Liouville theory
\cite{Dorn:1994xn,Zamolodchikov:1995aa,Teschner:1997ft,FZZ} bearing
many similarities with a coset of  $AdS_3$ sigma-model
\cite{Polyakov:1998ju}, the Sine-Liouville theory and its
strong-coupling dual, the 2D dilatonic black hole \footnote{
Similarly to the AdS/CFT correspondence of superstring to the SYM,
they also have their dual, matrix models and matrix quantum
mechanics description for the $c=1$ matrix model
\cite{Kazakov:1988ch}, or the matrix black hole model of
\cite{Kazakov:2000pm} (see \cite{Maldacena:2002eg} for the
discussion of this issue).}.

 2. The Virasoro conditions  impose a complicated
 selection rule on the possible quantum states: they
eliminate the longitudinal motions
 of the string  forbidden by the world-sheet reparametrization symmetry,
 together with the corresponding quantum states.
We could start from a gauge which eliminates the conformal
redundancy from the beginning, but it hides the world-sheet Lorentz
covariance as well.

 3. It seems hopeless to find a completely closed subsector of the superstring
 on $AdS_5\times S^5$
 at the quantum mechanical and nonperturbative level, and to write
 down a selfconsistent system of Bethe equations for it
 (though see some instructive attempts in
 \cite{Arutyunov:2004vx,Schafer-Nameki:2005tn,Frolov:2006cc}).
 To obtain the right divergency free
 string sigma model we have to consider the full theory, including
 fermions.

The full Metsaev--Tseytlin superstring looks like two sigma models,
one on $AdS_5$, another on $S^5$ space, each containing 4 on-shell
bosonic fields ``glued'' together by 16 fermionic fields (8 on-shell
fields). Classically the fermions do not influence the dynamics and
the system looks like two independent bosonic $\sigma$-models. If we
 quantize the $S^5$ $\sigma$-model (with a trivial inclusion of the
 $AdS$-time coordinate, which makes it $S^5\times R_t$) it will show
 the asymptotically free behavior, even if we impose the
 analogue of Virasoro constraints (no longitudinal modes).
 Nevertheless, this  compact $O(6)$ $\sigma$-model may provide a
 valuable information on  the full superstring: its classical
 limit describes precisely the compact sector of the superstring
 \cite{Beisert:2004ag} and its quantization should
bear some important features of
 the full quantum superstring $\sigma$-model. And if we manage  to
 impose the Virasoro constraints in a natural way it can serve as a
 rather sophisticated toy model for the quantum superstring.

Our principal motivation in this paper is to study the full quantum
version of  the $O(2n)$ sigma model in the context of its
application to the string theory. Our starting point is the
Zamolodchikovs' S-matrix for physical particles. We put $L$ such
particles on a circle and study possible states by means of the
nested Bethe ansatz equations diagonalizing the underlying monodromy
matrix by the use of the quantum inverse scattering method
\cite{Faddeev:1979gh,Kulish:1981gi}. The main result may be
interesting not only to the string theorists but also to the
specialists in integrable 2D field theories: we reconstruct all
finite gap solutions of \cite{Kazakov:2004qf,Beisert:2004ag} for the
classical string motions on $S^5\times R_t$, directly from these
Bethe equations in the limit of large density of the particles with
sufficiently big momenta. In this limit, the theory is in the
extreme asymptotically free regime. We can neglect the conformal
anomaly since the particles are almost massless, and the states are
classical solitons. These solitons  are nothing  but the classical
string states. The string looks like a collective state of the
particles. The string mode numbers can be identified with certain
``magnon'' mode numbers of the Bethe equations, the number of
excited modes corresponds to the number of gaps in finite gap
solution (this number is equal to the genus of the algebraic curve
describing the solution), and the amplitude of a mode is fixed by
the number of Bethe roots forming the corresponding gap.

The system  looks like  an inhomogeneous  $SO(6)$ spin chain with
the nearest neighbor interactions defined by the rapidities of the
neighboring particles. The rapidities, in their turn, are fixed from
the periodicity condition, thus make the spin chain
dynamical\footnote{Not to confuse it with the dynamical S-matrix of
\cite{Beisert:2005tm} for the $\alg{su}(3|2)$ sector of the ${\cal
N}=4$ SYM where even the length of the chain can change, a feature
certainly to be taken into account for the full string
$\sigma$-model.}.

Qualitatively it is quite similar to the $OSp(2m+n|2m)$
supersymmetric $\sigma$-model of \cite{Mann:2005ab}, conformal in
the limit $n\to 2$. Our way to restore the classical limit is also
inspired by \cite{Mann:2005ab}, but the details are very different,
since our model is conformal only classically. However, some
features of our model look closer to compact sectors of the string
theory on $AdS_5\times S^5$. In particular, in a certain limit we
restore the homogeneous spin chain, the same (up to some
normalizations) as the one describing the 1-loop SYM theory in the
corresponding sector. To make it working for higher SYM loops it
might be necessary to incorporate into the future string
$\sigma$-model certain features of the Hubbard model which is a good
candidate for reproducing all loops of the perturbative SYM
\cite{Rej:2005qt}.

The classical string solutions  are encoded into an algebraic curve.
To each curve with the fixed moduli one can associate the unique (up
to some trivial choice of initial conditions, as proven in
\cite{Dorey:2006zj}) classical string solution with Virasoro
conditions imposed and with the number of excited oscillatory modes
corresponding to the genus of the curve.

The algebraic curve obtained in \cite{Kazakov:2004qf,Beisert:2004ag}
from the purely classical finite gap method \cite{Novikov:1984id}
turns out to be the same  as the one extracted here from the Bethe
equations, and the two different projections (Riemann surfaces) of
the curve are related by Zhukovsky map
\begin{eqnarray*}
z=x+1/x\,,
\end{eqnarray*}
which is an important part of the construction\footnote{The
Zhukovsky map  was proposed in 1906 in \cite{ZHUK} for  description
of aerodynamics of the airplane wing. It maps a circle passing
through the point $x=1$ into a wing-shaped figure in $z$-plane (see
the front page).}. It was used already in \cite{Kazakov:2004qf} for
the successful 2-loop comparison of the string and SYM results  and
its significance is better understood in
\cite{Beisert:2004hm,Arutyunov:2004vx,Dorey:2006zj}. This map might
identify the right variables to quantize in the full superstring
theory.

The Virasoro conditions deserve a special mentioning here. As it
will be seen from our classical limit procedure, the correct
classical string solutions are related  only to  Bethe states where
all rapidities of the physical particles have the same mode number.
In the classical limit they condense into a single cut, which
reproduces, after Zhukovsky transformation, the two pole structure
of the algebraic curve of the finite gap solution. We conjecture
that this is the way to impose the Virasoro-like constraints for any
general quantum state as well: the rapidities should have the same
mode number.
 We confirm it by some examples of particular solutions
with small longitudinal amplitudes.\footnote{ At the classical level
Virasoro constraints have a clear interpretation as a restriction on
the allowed solutions forbidding longitudinal motions of the string,
even if we truncate the full $AdS_5\times S^5$ theory to a
subsector, like the $S^5\times R$ subsector considered in this
paper. On the quantum level, the subsectors are not even described
by a conformal field theory and the Virasoro constraints do not have
a clear meaning and are not imposed by the reparametrization
symmetry. Therefore we rather have to speak in this case about a
natural selection rule for the quantum states which leads to the
standard Virasoro conditions in the classical limit.}

Finally, we would like to mention that the Bethe ansatz description
provides an unusual example of the holography in the string theory:
the Bethe roots look like the positions of some D-branes reducing
the quantization procedure to a certain ``discretization'' of the
underlying algebraic curves. But unlike the matrix model description
of D-branes where the eigenvalues serve as quantum coordinates of
the D-branes and should be dynamical variables, the Bethe roots of
the stringy $\sigma$-models are fixed by the periodicity condition.

The paper is organized as follows. In section 2, we define the
$O(2n)$ $\sigma$-model and remind its classical integrability
properties allowing the finite gap solution. In section 3, the
quantum Bethe ansatz solution of the  $SU(2)$ principal chiral field
model, equivalent to the $O(4)$ $\sigma$-model, is presented and its
classical limit  is considered and the corresponding algebraic curve
is constructed. In section 4, we map this algebraic curve to the
curve of the finite gap solution obtained directly from the
classical action, fixing appropriately the relations between all the
parameters of two approaches. In section 5, the construction is
generalized to the $O(6)$ sector. In conclusions we make some
remarks concerning possible generalizations of this construction to
the full superstring on $AdS_5\times S^5$ and problems related to
it. In Appendix A the solution for the $U(1)$ sector in the scaling
limit is presented in detail. In Appendix B we provide an
alternative proof of our correspondence for the $O(4)$ sector, namely through the
transformation of resolvents. In Appendix C both the bootstrap
method as well as the algebraic Bethe ansatz program are reviewed
for the $O(4)$ $\sigma$-model. In Appendix D this is generalized to
the $O(2n)$ $\sigma$-model.

%%%%%%%%%%%%%%%%%%%%%%%%%%%%%%%%%%%%%%%%%%%%%%%%%%%%%%%%%%%%%%%%%%%%%%%%%%%%%%%
\section{$O(2n)$ Sigma-Model and Classically Integrable String Theory \label{sec2}}

The main system of our interest is the superstring on the
$AdS_5\times S^5$ background. A compact bosonic subsector of it is
described by the sigma model on the subspace $S^5\times R_t$, where
$R_t$ is the coordinate corresponding to the AdS time. The time
direction will be almost completely  decoupled from the dynamics of
the rest of the string coordinates, appearing only through the
Virasoro conditions. These conditions represent only a selection
rule upon the states of the $O(6)$ sigma model allowing to choose a
particular set of states, or a particular set of solutions in the
classical limit, which do not contain any dynamics along the string.

Of course in the absence of fermions and the AdS part of the full
10D superstring theory, this model will be asymptotically free and
will not be suitable as a viable quantum string theory. In addition,
on the dual ${\cal N}=4$ Yang--Mills side, the corresponding
$\alg{so}(6)$ sector of bosons is not closed under the action of
dilatation operator (except the limit of long operators which can be
compared with the classical string \cite{Minahan:2005jq}). However,
in the classical limit we will encounter the full classical finite
gap solution of the string in $SO(6)$ sector found in
\cite{Beisert:2004ag}. Furthermore we still expect to capture some
features of the full quantum string $\sigma$-model.

In the rest of this section we will first take instead of $O(6)$ a
more general $O(2n)$ case. We will later reproduce the full
classical finite gap solution \cite{Beisert:2004ag} of this model by
means of the Zamolodchikovs' physical S-matrix which may be
considered as the most convincing demonstration of the correctness
of the quantization of the model by the bootstrap method.

%%%%%%%%%%%%%%%%%%%%%%%%%%%%%%%%%%%%%%%%%%%%%%%%%%%%%%%
\subsection{Definition of the
Sigma-Model on $S^{2n-1}$ and Virasoro Conditions}

The $S^5\times R_t$ reduction of Green-Schwarz-Metsaev-Tseytlin
superstring  on $AdS_5\times S^5$ background
\cite{Green:1981yb,Metsaev:1998it} in the orthogonal gauge has a
simple action \cite{Frolov:2003qc} in terms of homogeneous target
space coordinates $X_i(\tau,\s),\ i=1,\dots, 2n$ and a scalar
$Y(\tau,\s)$ representing the AdS time:
\beq S=\frac{\sqrt{\lambda}}{4\pi}\int_0^{2\pi}d\sigma\int
d\tau\[(\d_a X_i)^2-(\d_a Y)^2\],\;\;\;\;\;   X_iX_i=1. \eeq
The coupling constant in front of the action is identified by the
AdS/CFT correspondence  with the 't Hooft coupling $\lambda=g^2N_c$
of the ${\cal N}=4$ supersymmetric Yang--Mills (SYM) theory.

We work in the static gauge $ Y(\tau,\s)=\kappa \tau$. The Virasoro
condition reads
\beq\label{VIRASORO}  \tr (j_\pm)^2=-2\,(\p_\pm Y)^2=-2\,\kappa^2
\,\,\, , \,\,\, \d_\pm=\p_\tau\pm\p_\s\, \eeq
where the current is  a $6\times 6$ matrix defined as
\beq  \(j_\pm\)_{ab}=\(h^{-1}\partial _\pm h\)_{ab}=2\( X_a\,\d_\pm
X_b-\d_\pm X_a\,X_b\) \,\,\, , \,\,\, h_{ab}=\delta_{ab}-2X_aX_b.
\label{js} \eeq
$\Delta=\sqrt{\lambda}\,\kappa$ is identified with the dimension of
the corresponding SYM operator according to the AdS/CFT
correspondence.

To illuminate the difference between the string sigma-model on
$S^5\times R_t$ and the ordinary sigma-model on $S^5$, it is
instructive to consider slightly more general constraints. These
constraints can be viewed as the Virasoro constraints for another
gauge
\beq
 Y=\frac{1}{2}\kappa_+(\tau+\sigma)+\frac{1}{2}\kappa_-(\tau-\sigma).
\la{PERIODT}\eeq
% Non-trivial winding $\kappa_+\ne\kappa_-$ may happen when the
AdS time $Y$ is periodically compactified in this case.
\beq\la{SKEWV} \tr (j_\pm)^2=-2\,\kappa_\pm^2\,. \eeq
The complete Virasoro conditions forbidding the time
compactification will be imposed by the identification
$\kappa_+=\kappa_-=\kappa$. Note that the condition \eq{SKEWV} still
allows a nonzero longitudinal momentum, although its density, as
well as the energy density are constant along the string.

In what follows we will deal with the action
\beq\label{SIGMAM}
S=\frac{\sqrt{\lambda}}{4\pi}\int_0^{2\pi}d\sigma\int d\tau\(\d_a
X_i\)^2=-\frac{\sqrt{\lambda}}{32\pi}\int d^2\sigma\;\tr\(j_a j_a\),
\eeq
subject to the condition \eq{SKEWV}, with an arbitrary number of
fields $X_i,\quad i=1,\ldots,2n$.

The energy and the momentum are given by
\beq \la{ESO6}E^{\rm cl}\pm P^{\rm
cl}=-\frac{\sqrt{\lambda}}{32\pi}\int d^2\sigma\;\tr[j_0\pm j_1]^2,
\eeq
or, using (\ref{SKEWV}),
\beq\la{ENMOMCL}
E^{\rm
cl}=\frac{\sqrt{\lambda}}{4}\(\kappa_+^2+\kappa_-^2\),\;\;\;\;\;P^{\rm
cl}=\frac{\sqrt{\lambda}}{4}\(\kappa_+^2-\kappa_-^2\)\,.
\eeq

%%%%%%%%%%%%%%%%%%%%%%%%%%%%%%%%%%%%%%%%%%%%%%%%%%%%%%%%%%%%%%%%
\subsection{\la{AnPrSO6}Classical Integrability and Finite Gap Solution}

The equations of motion and the form of the current, $j=h^{-1}dh$,
can be encoded into a single flatness condition
\cite{Novikov:1984id}
\beq \[{\cal L}_+(x),{\cal L}_-(x)\]=0, \label{LAX} \eeq
where the Lax pair of operators, the currents deformed by  spectral
parameter $x$, are given by
\beq\la{LAXC} {\cal L}_\pm(x) =
\partial_\pm -
\frac{j_\pm}{x\mp 1}.   \eeq
The Lax operators describe a connection over the world-sheet and
define the monodromy matrix
\beq\label{MONO}
\Omega(x)=\stackrel{\leftarrow}{P}\exp\int_0^{2\pi}d\sigma\,
\frac{1}{2}\left(\frac{j_+}{x-1}+\frac{j_-}{x+1}\right). \eeq
By construction $\Omega(x)$ is a complex orthogonal matrix and thus
has the eigenvalues
\beq  \left(e^{i\hat q_1(x)},e^{-i\hat q_1(x)},e^{i\hat
q_2(x)},e^{-i\hat q_2(x)}, \ldots,e^{i\hat q_{n}(x)},e^{-i\hat
q_{n}(x)}\right)  \eeq
where $q_k(x)$ are called quasi-momenta. They do not depend on time
$\tau$ due to \eq{LAX} and provide an infinite set of classical
integrals of motion of the model. As for the global conserved
charges of the sigma-model one has
\beq
J=\frac{\sqrt{\lambda}}{4\pi\,i}\int\limits_0^{2\pi}d\sigma\,j_\tau={\rm
diag}\left(J_1,-J_1,J_2,-J_2,\dots,J_n,-J_n\right) \,. \eeq
This current is normalized so that after quantization $J_i$ are
integers for any quantum state. This is the $O(2n)$ angular momenta
quantization condition.

Let us now recall some analytic properties of these quasi-momenta
\cite{Beisert:2004ag}. Expanding the monodromy matrix in powers of
$1/x$ around $x=\infty$, we obtain
\begin{eqnarray}
\hat q_k(x)=\frac{1}{x}\,\frac{4\pi
J_k}{\sqrt{\lambda}}+\mathcal{O}(1/x^2)\,. \la{singinf}
\end{eqnarray}
 We see from (\ref{MONO}) that simple poles will appear at  $x=\pm 1$
for the quasi-momenta. Furthermore, from (\ref{js}) we see that $j$
has only 2 eigenvectors corresponding to nonzero eigenvalues. Thus
we can diagonalize $j_\pm$ in such a way that the poles at $\pm 1$
are only present in $q_1$,
\begin{eqnarray}
\hat q_k(x)=\delta_{k,1}\frac{2\pi\kappa_\pm}{x\mp
1}+\mathcal{O}\left((x\mp 1)^0\right)\,\,\, {\rm for} \,\,\,
x\rightarrow \pm 1\,. \label{singpm1}
\end{eqnarray}
Furthermore the monodromy matrix (\ref{MONO}) with inverse spectral
parameter, $\Omega(1/x)$,
can be written, using the explicit form (\ref{js}) of $j$ and $h$,
as \beq \nn
\Omega(1/x)=\stackrel{\leftarrow}{P}\exp\int_0^{2\pi}d\sigma\,\(
h^{-1}\frac{1}{2}\left(\frac{j_+}{x-1}+\frac{j_-}{x+1}\right)h-
h^{-1}\partial_\sigma h\)=h(2\pi)\Omega(x) h^{-1}(0)
 \,.\eeq
which is nothing but a similarity transformation due to the
periodicity $h(2\pi)= h(0)$. Therefore $\Omega(x)$ and $\Omega(1/x)$
have the same set of eigenvalues. For the quasi-momenta this gives
the transformation law
\begin{eqnarray}
\hat q_k(1/x)=\delta_{k,1}\(4\pi m-\hat q_1(x) \)+(1-\delta_{k,1})\,
\hat q_k(x) \label{x1/x}
\end{eqnarray}
which respects (\ref{singpm1}). This symmetry will be of upmost
importance in our further discussions. From this relation one
immediately reads off the analytic behavior of $\hat q_k$ around
$x=0$ giving  the behavior around $x=\infty$.

Finally we obtain the quasi-momentum by solving the characteristic
polynomial of $\Omega$ which is well defined in ${\mathbb C}$. This,
on the other hand, implies that the quasi-momenta will be the $2n$
branches of a single function with branch cuts $\mathcal{C}_{a}$.
Along these cuts the quasi-momenta in general get permutated and
shifted by multiples of $2\pi$ as
\begin{eqnarray}
\la{qqh}\shq_k\mp\shq_{l}&=&2\pi\,n_{a} \,\,\, , \,\,\, x\in
\mathcal{C}_{a},
\end{eqnarray}
so that we obtain some sheets connected by the cuts as in the
example in figure \ref{fig:BKS}. For each cut outside the unit
circle there will be a mirror cut inside the unit circle due to
(\ref{x1/x}).

%%%%%%%%%%%%%%%%%%%%%%%%%%%%%%%%%%%%%%%%%%%%%%%%%%%%%%%%%%%%%%
\subsection{Classical  $SU(2)$ Principal Chiral Field}

In this subsection we will concentrate on the classical finite gap
solution of the $O(4)$ sigma model formulated in terms of the
$SU(2)$ principal chiral field. We will essentially briefly repeat
the construction of \cite{Kazakov:2004qf} (with a small
generalization to the excitations of both left and right sectors) to
fix the notations for the easy comparison with the quantum Bethe
ansatz solution of the model.

%%%%%%%%%%%%%%%%%%%%%%%%%%%%%%%%%%%%%%%%%%%%
\subsection{Definition of the Model}

The $O(4)$ sigma model  represents the reduction of the sigma model
in the $AdS_5\times S^5$ background to the subsector of string
moving on $S^3\times R_t$. Classically this is a perfectly
consistent reduction while at the quantum level one might still
expect to capture some features of the full theory. The action of
the theory can be represented in terms of the $SU(2)$ group valued
field $g=X_1+i\tau_3 X_2+i\tau_2 X_3+i\tau_1 X_4$ ($\tau_i$ are the
Pauli matrices). Then the action \eq{SIGMAM} takes the form of the
$SU(2)$ principal chiral field
\beq\la{ACTIONSU2}
S=\frac{\sqrt{\lambda}}{8\pi}\int_0^{2\pi}d\sigma\int \tr(\d_a
g^\dag\d_a g) \,.\eeq
The obvious global symmetry of the action is the left and right
multiplication by $SU(2)$ group element. The currents of this
symmetry are
\beq  j^R_\pm= g^{-1}\p_\pm g\,,\qquad j^L_\pm= \d_\pm g g^{-1}\,,
\eeq
with corresponding Noether charges
\beq Q_L=\frac{\sqrt{\lambda}}{4\pi}\int d\sigma\ \tr\( i\d_0 g\,
g^\dag \tau^3\),\;\;\;\;\; Q_R=\frac{\sqrt{\lambda}}{4\pi}\int
d\sigma\ \tr\( i g^\dag \d_0 g\, \tau^3\)\,.\;\;\;\;\; \eeq
In the quantum theory these charges are positive integers. It will
be important for future comparisons to notice that the normalization
of the generators is such that the smallest possible charge is $1$.

Virasoro conditions read
\beq \tr (j_\pm^2)=-2\,\kappa_\pm^2\,. \la{VIRASORO1/2} \eeq
From the action we read off the energy and momentum as
\beq \la{EPclassic} E^{\,\rm cl}\pm P^{\,\rm
cl}=-\frac{\sqrt{\lambda}}{8\pi}\int\tr[j_0\pm
j_1]^2d\sigma=\frac{\sqrt{\lambda}}{2}\kappa^2_\pm\,. \eeq

The Lax construction (\ref{LAX}-\ref{MONO}) can be repeated for this
formulation using either the left or the right current, the two
choices being related by a simple relation
\beq\la{LRCUR}  \Omega_L(1/x)=g(2\pi)\Omega_R(x) g^{-1}(0)  \,. \eeq
Let us use $\Omega_R$, for definiteness. Now there will be only two
quasi-momenta $p_1(x),p_2(x)$ since the $2\times 2$ unitary
monodromy matrix $\Omega(x)$ will have eigenvalues
$e^{ip_1(x)},e^{ip_2(x)}$ . Unimodularity imposes
\beq\la{UNIM}   \tilde p_1(x)=-\tilde p_2(x) \mod \ \ 2\pi  \eeq
so that we can define
\beq\la{OMEGAP}   T(x)\equiv\frac{1}{2}\Tr\, \Omega_R(x)\equiv\cos
\tilde p(x) \,.    \eeq
Analyzing the singularities of \eq{MONO} we find the behavior of the
quasi-momentum at $x\to\infty$
\beq\la{GLCH} \tilde p(x)=-\frac{2\pi Q_R}{\sqrt{\lambda}\,
x}+\cdots\,,   \eeq
and at $x\to 0$ (from \eq{LRCUR})
\beq \la{PXZERO}   \tilde p(x)\sim 2\pi m+{2\pi Q_L\over
\sqrt{\lambda}}x+\cdots\,, \eeq
where we also used the fact that due to the periodicity of
$g(\sigma)$, $\Omega(0)=1$ and hence $\tilde p(0)=2\pi m$. Finally,
at $x\to \pm 1$
\beq \la{LCH}    \tilde p(x)=-\frac{\pi\kappa_\pm}{x\mp 1}+\cdots\,,
 \eeq
By construction $\Omega_R(x)$ is analytical in the whole plane
except $x=\pm 1$ where one has essential singularities. Then, from
\eq{OMEGAP}, one concludes that for $x\neq \pm1$ the only
singularities of
\beq\la{PPRIM}  \tilde  p\,'(x)= -\frac{T'(x)}{\sqrt{1-T^2(x)}}
\eeq
can be of the type $ \tilde p\,'\,(x\to x_k)\sim
\frac{1}{\sqrt{x-x_k}}$. If we look for the finite gap solution the
number of these branch points is finite and even, and we conclude
from (\ref{GLCH}-\ref{PPRIM}) that the $\tilde  p_1'(x),\tilde
p_2'(x)$ are two branches  of an analytical function  defined by a
hyperelliptic curve of  genus $g$:
\beq\la{FINGAPS} [\tilde
p'(x)]^2=\(\frac{b_+}{(x-1)^2}+\frac{b_-}{(x+1)^2}+\frac{c_+}{x-1}
+\frac{c_-}{x+1}+
P_{g-1}(x)\)^2\frac{1}{\prod_{k=1}^{2(g+1)}(x-x_k)} \eeq
The coefficients of the polynomial $P_{g-1}(x)$ of degree $g-1$ and
$b_\pm,c_\pm$, and the positions of branch points $x_k$ make a total
of $3g+6$ constants. They are fixed by
\begin{enumerate}
\item{The asymptotics (\ref{GLCH}-\ref{LCH}) and the absence of simple
poles at $x=\pm 1$ ($5$ conditions).}
\item{The singlevaluedness conditions
\beq\la{SINGLEV} \oint_{A_j}d\tilde p = 0\qquad  j=1,\ldots,g
\eeq
where the homomorphic integrals around the cycles $A_i$ are
essentially the contour integrals around the cuts $C_k$ ($g$
conditions).}
\item{The integer $B$-period conditions, analogous to \eq{qqh}
\beq\la{BCYCLE} \oint_{B_{j}}d\tilde p = 2\pi n_j\qquad
j=1,\ldots,g+1 \,.\eeq following from the ambiguity of the
definition of the  branches of the quasi-momentum  \eq{UNIM} ($g+1$
conditions). }
\end{enumerate}

The cycle $A_i$ is the contour which encircles the cut $C_i$ -- see figure \ref{fig:uvKMMZ} -- in the
counterclockwise direction in the upper sheet if the cut is outside the unit circle or in the clockwise
direction in the lower sheet if it is inside the unit circle.

The $B_k$ cycle consists of the contour starting at $\infty_+$ in
the upper sheet, changing sheet in at the cut $C_k$ and going from
here to $\infty_-$ in this lower sheet.

We can also write (\ref{PXZERO}) as \beq\la{BCYCLE}
\oint_{B_{\th}}d\tilde p = 2\pi m \,.\eeq where this new cycle goes
from $\infty$ to $0$ in the upper sheet.

Let us distinguish the cuts  inside and outside the
unit circle by superscripts $u$ and $v$. The remaining $g$ constants
can be parametrized by the filling fraction numbers which we define
as
\beqa\la{ACYCLE} S_i^v=-\frac{\sqrt{\lambda}}{8\pi^2
i}\oint_{A^v_i}\tilde p(x)\(1-\frac{1}{x^2}\)dx,\;\;\;\;\;
S_i^u=\frac{\sqrt{\lambda}}{8\pi^2 i}\oint_{A^u_i}\tilde
p(x)\(1-\frac{1}{x^2}\)dx \eeqa
From the AdS/CFT correspondence these filling fractions are expected
to be integers since this is obvious on the SYM side
\cite{Kazakov:2004qf,Beisert:2005di}. In fact, it was pointed out in
\cite{Beisert:2004ag,Beisert:2005di} and shown in
\cite{Dorey:2006zj} that $S_i^{u,v}$ are action variables so that
quasi-classically they indeed become integers. We also find a
striking evidence for this quantization in the string side when
finding the classics from the quantum Bethe ansatz where these
quantities are naturally quantized.

The conditions \eq{BCYCLE} can also be represented as
\beq  \stp(x)= 2\pi n_k,\qquad   x\in C_k    \eeq
where we defined
\beq\la{SLASHP} \stp(x)=\hf \(\tilde p(x+i0)+\tilde p(x-i0)\).
\eeq

%%%%%%%%%%%%%%%%%%%%%%%%%%%%%%%%%%%%%%%%%%%%%%%%%%%%%%%%%%%%%%%%%%%
\newsection{Quantum Bethe Ansatz and Classical Limit: $O(4)$ Sigma-Model \label{sec3}}

We will describe a  quantum state of the $O(n)$ sigma model (the
analogue of the "closed string" state for the $S^3\times R_1$ sector
of the superstring theory)  by a system of $L$ physical particles of
the mass $m_0$ put on a circle of the length ${\cal L}$.  These
particles transforms in the vector representation under $O(4)$
symmetry group or in the bi-fundamental representations of
$SU(2)_R\times SU(2)_L$. The scattering of the particles in this
theory is known to be elastic and factorizable: the relativistic
S-matrix $\hat S\(\th_1-\th_2\)$  depends only on the difference of
 rapidities of scattering particles $\th_1$ and $\th_2$ and obeys the
Yang--Baxter equations. As was shown in \cite{Zamolodchikov:1977nu}
(and in \cite{Wiegmann:1984mk,Wiegmann:1984ec,Wiegmann:1984pw,Ogievetsky:1984pv} for the general principle chiral field) these
properties, together with the unitarity and crossing-invariance,
define essentially unambiguously the S-matrix. In Appendix C we
review the Bootstrap program of \cite{Zamolodchikov:1977nu} as well
and the algebraic Bethe ansatz construction for this system
\cite{Zamolodchikov:1992zr}.

There should be no confusion with the employed notation. Having in
mind the comparison in section \ref{sec4} we will use the same
letters in sections \ref{sec2} and \ref{sec3} for quantities that we
be later proven to be the same. For what concerns this section they
are just definitions.

%%%%%%%%%%%%%%%%%%%%%%%%%%%%%%%%%%%%%%%%%%%%%%%%%%%%%%%%
\subsection{Bethe Equations for Particles on a Circle}

 When this system of particles is put into a
finite 1-dimensional periodic box of the length $\cal L$ the set of
rapidities of the particles $\{\theta_\a\}$ is constrained by the
condition of  periodicity of the wave function $|\psi\rangle$ of the
system,
\begin{eqnarray}
|\psi\rangle=e^{i\mu
\sinh\pi\theta_\a}\overleftarrow{\prod_{1}^{\a-1}}\ \hat
S\(\th_\a-\th_\b\) \overrightarrow{\prod_{N}^{\a+1}}\ \hat
S\(\th_\a-\th_\b\) |\psi\rangle \label{MBAE}
\end{eqnarray}
where the first term
is due to the free phase of the particle and the second is the
product of the scattering phases with the other particles. Arrows
stand for ordering of the terms in the product.  $\mu=m_0{\cal L}$
is a  dimensionless parameter. The physical $SU(2)$ PCF S-matrix is
a product of left and right S-matrices, $\hat S=\hat S_L\times\hat
S_R$ \cite{Zamolodchikov:1977nu,Wiegmann:1984ec}
\begin{eqnarray*}
\hat S_{L,R}(\theta)=S_0(\theta)\( P_{L,R}^+ +
\frac{\theta+i}{\theta-i}P_{L,R}^-\)\;\;\;\;\; {\rm or} \;\;\;\;\;
\(\hat
S_{R,L}(\theta)\)^{a'b'}_{a\,b}=\frac{S_0(\theta)}{\theta-i}\left(\theta\,
\delta^{a'}_{a}\delta^{b'}_{b}-i\,\delta^{b'}_{a}\delta^{a'}_{b}\right)
\end{eqnarray*}
and
\begin{eqnarray}
S_0(\theta)=i\frac{\Gamma\(-\frac{\theta}{2i}\)\Gamma\(\frac{1}{2}
+\frac{\theta}{2i}\)}{\Gamma\(\frac{\theta}{2i}\)\Gamma\(\frac{1}{2}-\frac{\theta}{2i}\)},\;\;\;\;\;
\la{S0S04}
\end{eqnarray}
It has the following large $\theta$ asymptotics:
 \beq\label{ASSS}  i\log S^2_0(\theta)\sim 1/\theta + O(1/\theta^3) \,.  \eeq
The diagonalization of both the L and R factors in the process of
fixing the correct periodicity (\ref{MBAE}) leads  to the following
set of Bethe equations
\begin{eqnarray}
e^{-i\mu \sinh\pi\theta_\a}&=& \prod_{\beta\neq \alpha}\,
S_0^{\,2}\(\th_\a-\th_\b\)
\prod_j\frac{\th_\a-u_j+i/2}{\th_\a-u_j-i/2}\,
\prod_k\frac{\th_\a-v_k+i/2}{\th_\a-v_k-i/2}\,, \label{DBAE1} \\
1&=&\prod_\b\frac{u_j-\th_\b-i/2}{u_j-\th_\b+i/2}
\prod_{i\neq j} \frac{u_j-u_i+i}{u_j-u_i-i}\,,  \label{DBAE2}\\
1&=&\prod_\b\frac{v_k-\th_\b-i/2}{v_k-\th_\b+i/2} \prod_{l\neq k}
\frac{v_k-v_l+i}{v_k-v_l-i}\,, \label{DBAE3}
\end{eqnarray}
where
\begin{eqnarray*}
\a,\b=1,\dots,L,\;\;\;\;\;\;\;\;i,j=1,\dots,J_u,\;\;\;\;\;\;\;\;
k,l=1,\dots,J_v\,.
\end{eqnarray*}
The left and right charges of the Bethe vector, i.e. wave function,
associated with the  two $SU(2)$ spins are given by
\beq  Q_L=L-2J_u\, , \qquad   Q_R=L-2J_v\,,  \la{QQ}\eeq
since a $u$ or $v$ root correspond to a spin flip in the
corresponding $SU(2)$ and since we are normalizing charges so that
$1$ is the unit charge. The absence of $u$'s and $v$'s corresponds
to the ferromagnetic limit with all spins aligned in the up
direction.

Both the construction of the $S$-matrix by means of the bootstrap
program \cite{Zamolodchikov:1977nu} and these Bethe equations (see a
similar example in \cite{Zamolodchikov:1992zr}) describing the
solution to (\ref{MBAE}) are reviewed in Appendix C for the sake of
completeness.

The total momentum and the total energy  of the state are  the sums
of momenta and energies of individual particles. In accordance with
\eq{ACTIONSU2} we take ${\cal L}=2\pi$ (in the units of the string
tension). Then
\begin{eqnarray}
\la{P_SSL} P=\frac{\mu}{2\pi}\sum\sinh(\pi\theta_\alpha) \,,
\end{eqnarray}
\begin{eqnarray}
\la{E_SSL}E=\frac{\mu}{2\pi}\sum\cosh(\pi\theta_\alpha)
\end{eqnarray}

This  model with massive relativistic particles and the
asymptotically free UV behavior cannot look like  a consistent
quantum string theory. Only in the classical limit we can view it as
a string toy model obeying the  classical  conformal symmetry.
   In the classical case it is also easy to impose the Virasoro
   conditions \eq{SKEWV}. In quantum case, we still can try to
   impose the Virasoro conditions, but they will be already some
   conditions selecting  the quantum states. It is not   easy
    on the stage of Bethe equations, but we will later make some
   plausible conjecture concerning this selection.

%%%%%%%%%%%%%%%%%%%%%%%%%%%%%%%%%%%%%%%%%%%%%%%%%
\subsection{The Classical Limit}

In  the classical limit, $\mu={\cal L}m_0\to 0$, the physical mass
of the particle $m_0\sim {\cal L}^{-1} e^{-\sqrt{\lambda}/2}$ is
small in the units of inverse length ${\cal L}^{-1}$ of the circle
on which we put a large number $L\to\infty$ of particles, keeping
$J_u\sim J_v\sim L$ as well. In this limit the Bethe roots
$u_k,v_k,\theta_k$ (and the characteristic distances between them)
are large, of the order of $L$. So we can take $log$'s of Bethe
equations \eq{MBAE} and use the "Coulomb" approximation \eq{ASSS}
for $S_0(\theta)$, as well as for other terms:
$\log\frac{\u+i/2}{u-i/2}= \frac{i}{u}+O(1/u^3).$ \footnote{ This
limit, sometimes called thermodynamical (which is a bit misleading
since the energy is not proportional to the length of the system) in
the context of ferromagnetic spin chains was suggested in
\cite{Sutherland}  and considerably advances in papers
\cite{Shastry,Beisert:2003xu}; see also \cite{Beisert:2005di} where
an important phenomenon of the anomaly cancelation was observed,
vital for the validity of this limit. The antisymmetry  of  the
$\log$'s of all factors in (\ref{DBAE1}-\ref{DBAE3}), including
$\log S(\theta)$, is also important for the validity of the
approximation.} Therefore it is useful to rescale
\begin{eqnarray*}
u=M\, x ,\quad v=M\, y,\quad  \th = M \xi
\end{eqnarray*}
where we chose
\begin{eqnarray*}
M\equiv -\frac{\log\mu}{2\pi}\sim L.
\end{eqnarray*}
Taking $\log$ of each Bethe equation (\ref{DBAE1}-\ref{DBAE3})  we
arrive at the following system of equations
\begin{eqnarray}
\mu \sinh\pi M \xi_\a&=& \frac{1}{M}\sum_{\beta\neq
\alpha}\frac{1}{\xi_{\a}-\xi_{\b}}
-\frac{1}{M}\sum_{j}\frac{1}{\xi_\a-x_j}-\frac{1}{M}\sum_{k}\frac{1}{\xi_\a-y_k}
+ 2\pi m_\alpha \,, \label{LBAE1}\\
0&=&-\frac{1}{M}\sum_\b\frac{1}{x_j-\xi_\b}+\frac{2}{M}\sum_{i\neq
j} \frac{1}{x_j-x_i}
- 2\pi n^u_j \,,\label{LBAE2}\\
0&=&-\frac{1}{M}\sum_\b\frac{1}{y_k-\xi_\b}+\frac{2}{M}\sum_{l\neq
k} \frac{1}{y_k-y_l} - 2\pi n^v_k \,,\label{LBAE3}
\end{eqnarray}
\begin{figure}[t]
    \centering
        \includegraphics[scale=1]{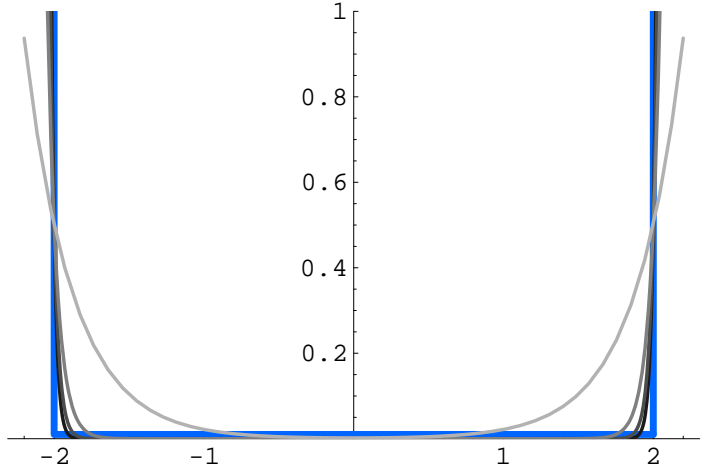}
            \caption{We plot $V(\xi)$ for $M=1,5,9,13$ (lighter to darker gray).
            It is clear that the potential approaches the blue box potential as $M\rightarrow \infty$.}
    \label{fig:box}
\end{figure}
where $m_\alpha,n^u_j,n^v_k$ are integers corresponding to the
choice of a branch of logarithm. We will call them mode numbers
since they are the analogue of the mode numbers of oscillators in
the string theory.

It reminds a system of 2D Coulomb charges of different species in
static equilibrium, each one in its own constant field of  force
$2\pi m_\alpha,2\pi n^u_j,2\pi n^v_k$, respectively, and in a
confining potential for $\theta$'s
\beq  V(\xi)=\mu\cosh(\pi M\xi).  \eeq
In our limit $M\sim\log\mu^{-1} \to\infty$, the potential for
$\xi\sim 1$  basically looks as a box with almost vertical walls
(the ``thickness'' of the wall is of the order $1/M$) - see figure
\ref{fig:box}.  The positions of the walls $\xi=\pm t$ defining the
width of  distribution of $\theta$'s can be estimated with the
logarithmic accuracy from $V(t)\simeq 1$. One has therefore

\begin{eqnarray*}
t =\frac{1}{\pi M}\left(\log{\mu^{-1}}+\mathcal{O}(M^0)\right)\simeq
2\,.
\end{eqnarray*}

All this tells us that we  can write the continuum version of Bethe
equations, similarly to the large $N$ limit of matrix models, in
terms of the resolvents of root distributions:
\begin{eqnarray}
G_\theta(x)\equiv
\frac{1}{M}\sum_{\beta=1}^{L}\frac{1}{z-\xi_\beta}\,\,\, , \,\,\,
\nonumber G_u(z)     \equiv \frac{1}{M}\sum_{i=1}^{J_u}
\frac{1}{z-x_i}     \,\,\, , \,\,\,  \label{BAE_PCF} G_v(z) \equiv
\frac{1}{M}\sum_{l=1}^{J_v}  \frac{1}{z-y_l}. \nonumber
\end{eqnarray}
with the large $z$ asymptotics
\beqa\la{GASYM} G_{u}(z)\simeq
\frac{J_u}{M}\,\frac{1}{z}\,,\;\;\;\;\;G_{v}(z)\simeq
\frac{J_v}{M}\,\frac{1}{z}\,, \;\;\;\;\;G_{\theta}(z)\simeq
\frac{L}{M}\,\frac{1}{z}\,. \eeqa
These equations are defined on the distribution supports
$\mathcal{C}_u,\mathcal{C}_\theta,\mathcal{C}_v$, each characterized
by the corresponding integer mode number $n_u,n_v$ or $m$:
\beqa
\nn2\,\sG_u(z)-G_\theta(z)&=&\;\;\,2\pi n_u,\;\;\;\;\; z\in \mathcal{C}_u\\
\la{BAESU2}\sG_\theta(z)-G_v(z)-G_u(z)&=&-2\pi m,\;\;\;\;\;\, z\in \mathcal{C}_\theta\\
\nn2\,\sG_v(z)-G_\theta(z)&=&\;\;\,2\pi n_v,\;\;\;\;\; z\in
\mathcal{C}_v \eeqa
There can be as many cuts as there are different mode numbers. Note
that we dropped the potential term for $\xi$'s in the second eq.
since, as we explained, it is zero within the infinite walls at
$\xi=\pm 2$. The only trace left by the potential, as shown in
Appendix A, is the  behavior of $G_\theta(z) \sim \frac{1}{\sqrt{z
\pm 2}}$ near the walls of the box which we should impose on the
solutions of \eq{BAESU2}.

Equation (\ref{BAESU2}) can also be written in terms of densities
defined by
\begin{eqnarray*}
\rho_\theta(x)=\lim_{M\rightarrow\infty}\frac{1}{M}\sum_\alpha\delta(z-\xi_\alpha),
\end{eqnarray*}
and similar for $u$'s and $v$'s. Then, for example, second line in
\eq{BAESU2} becomes \beqa \la{rhoBAESU2}\pint_{{\cal
C}_\theta}\frac{\rho_\theta(w)}{z-w}dw-\int_{{\cal C}_u}
\frac{\rho_u(w)}{z-w}dw-\int_{{\cal
C}_v}\frac{\rho_v(w)}{z-w}dw&=&-2\pi m,\;\;\;\;\;\, z\in
\mathcal{C}_\theta\,. \eeqa

We considered only one mode number $m$. Later it will be shown that
in order to reproduce the finite gap solutions for classical strings
on $S^3\times R_t$ we have to assume that we have a single cut
$[-2,2]$ for $\xi$-distribution characterized with all Bethe roots
$\theta_k,\quad k=1,\cdots L$ having the same mode number $m$. As we
will argue in section \ref{Multicuts} the \textit{classical}
multi-cut states with more mode numbers excited correspond to the
longitudinal oscillations of the string. In the string theory these
oscillations are non-physical and forbidden by the Virasoro
constraints, but they are present in the standard relativistic
$\sigma$-model \eq{ACTIONSU2}. We know how to do it in the classical
limit, but it is more difficult to project it in the full quantum
space of states. It is  also plausible to assume that not only in
the  classical, but also in full quantum theory, {\it only the Bethe
states with $\theta_\alpha\,,\,\, \alpha=1,\cdots,L\,,$ having the
same mode number $m$  obey the Virasoro constraints}. We don't have
for the moment any convincing proof of this conjecture, apart from
some examples and the comparison of the classical limit of our Bethe
equations with the classical finite gap solution of section 4.

The total momentum can be calculated exactly, before any classical
limit
\begin{eqnarray}\la{P_SSL}
P=\frac{\mu}{2\pi}\sum_{\a}\sinh(\pi\theta_\alpha)=m L-\sum_p n_p
S_p^u-\sum_p n_p S_p^v
\end{eqnarray}
where $S_p^u,\;S_p^v$ are the filling fractions, or the numbers of
Bethe roots with a given mode numbers $n_{u,p},n_{v,p}$.  To prove
it it suffices to take the sum of logarithms of the \eq{DBAE1} for
all roots $\theta_\alpha$. The contribution of $S_0(\theta)$ terms
cancels due to antisymmetry while the second and third sums in the
r.h.s. of (\ref{LBAE1}) are excluded using  (\ref{DBAE2}) and
(\ref{DBAE3}) respectively.

For the closed string theory we should take $P=0$, which gives the
level matching condition. For the perturbative super YM applications
one should take $S^u_p=0$ \cite{Minahan:2005jq}. Then we have the
well known formula $\sum_p n_p S_p^v=m L$ (see \cite{Kazakov:2004qf}
for details).

Let us now study the system of equations \eq{BAESU2} as a
Riemann-Hilbert problem, first for the highest weight states with
$U(1)$ symmetry, without magnon excitations $u_k$ and $v_k$, and
than for a general classical state. Our goal will be to reproduce
(and to generalize from a single sector to both left and right
sectors) the finite gap results and the corresponding algebraic
curve of the paper \cite{Kazakov:2004qf}.

%%%%%%%%%%%%%%%%%%%%%%%%%%%%%%%%%%%%%
\subsection{Highest Weight States of $U(1)$ Sector in Classical Limit \la{HW}}

If the right and left modes are not excited we have only the states
with  $U(1)$ modes, which means that the currents $j_L=j_R$ are
diagonal matrices.

In the  classical limit, using the Coulomb  approximation,  we have
for this sector the following Bethe equation
\begin{eqnarray}
\nn\mu \sinh\pi M \xi_\a-2\pi\,m&=& -\frac{1}{M}\sum_{\beta\neq
\alpha}^L\frac{1}{\xi_{\a}-\xi_{\b}}\,. \label{BAEnoUV}
\end{eqnarray}
In the continuous limit, the equation for the asymptotic density,
$L\sim M\rightarrow \infty$, is given by \beq\label{THETAG}
\sG_\theta(x)=-2\pi m,\;\;\;\;\;\, x\in \mathcal{C}_\theta\,, \eeq
with inverse square root boundary conditions near $\pm 2$ (see
Appendix A and discussion in previous subsection). The asymptotic
density of rapidities, $\rho_\th(z)$, is obtained through the
resolvent from
\begin{eqnarray*}
\rho_\th(z)=-\frac{1}{2\pi i}\( G_\th(z+i0)-G_\th(z-i0)\)\,.
\end{eqnarray*}
The analytical function $G_\theta(x)$ having a real part on the cut
defined by \eq{THETAG}, with support $[-2,2]$, with inverse square
root boundary conditions and behaving at $z\to\infty$ as
$G_\theta(z)\to \frac{L}{M}\frac{1}{z}$ is completely fixed:
\beq\label{RESTH}   G_\theta(z)=\(\frac{2\pi m\ z
+\frac{L}{M}}{\sqrt{z^2-4}}-2\pi m\)   \eeq
which gives for the density
\beq\label{DENTH}   \rho_\theta(z)=\frac{1}{\pi }\(\frac{2\pi m\ z
+\frac{L}{M}}{\sqrt{4-z^2}}\)  \,. \eeq
Notice that the distribution has a singular behavior at the
endpoints which will be the typical behavior even for the general
multi-cut solution which we study in the next subsection. Notice
also that applying to the \eq{RESTH} the  Zhukovsky map
\beq\label{JUK}   z=x+\frac{1}{x}   \eeq
we obtain
\beq\la{XPOLES}   G_\theta(z(x))=\frac{\frac{L}{2M}+2\pi
m}{x-1}+\frac{\frac{L}{2M}-2\pi m}{x+1} \eeq
 which shows the poles at $x=\pm 1$, typical for the finite gap
solution of the section 1. The Zhukovsky map will be the central
piece of our proof of the identification of the continuous limit of
Bethe ansatz equations with general classical solutions of
$\sigma$-models considered in this paper.

%%%%%%%%%%%%%%%%
\begin{figure}[t]
    \centering
        \includegraphics[scale=0.8]{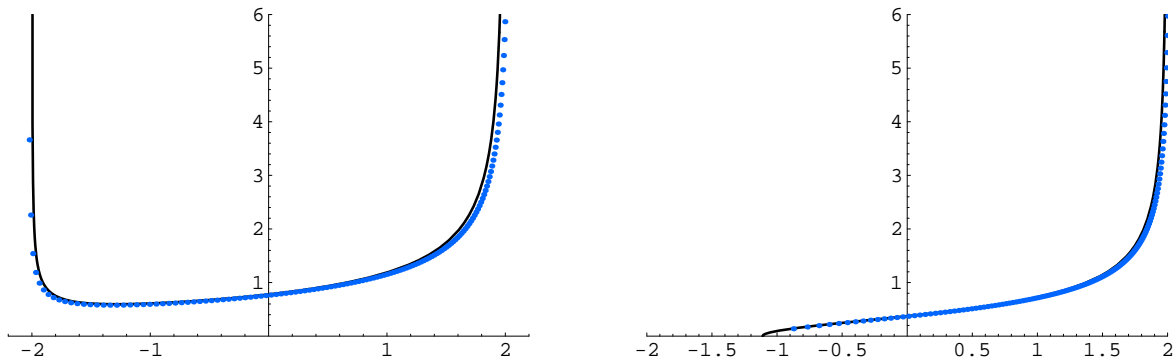}
            \caption{Density of  $\theta$-roots before and after phase transition. Black line - asymptotic
            densities of \eqs{DENTH}{DENTHA}, blue dots - numerical solution for   $L=150$ roots.}
    \label{fig:phases}
\end{figure}
%%%%%%%%%%%%%%%%

 From the general formula \eq{P_SSL} the momentum of such a state is
\beq   P=m L   \eeq

Let us now compute the energy of this state in the continuous limit
(it is derived in greater detail in Section \ref{sec:ener} for a
general state). We have to compute the sum
\beq \nn E\equiv \frac{\mu}{2\pi}\sum_\alpha \cosh(\pi
M\xi_\alpha)\simeq\sum_{\xi_\alpha>0} \frac{\mu}{\pi} \sinh\( \pi
\xi_\alpha M\)- \sum_{\xi_\alpha<0}\frac{\mu}{\pi}\sinh \(\pi
\xi_\alpha M\)\, \la{BIGSUME}\eeq
 where the last equality holds with exponential
precision since the decaying exponents of the $sinh$'es can be
dropped. Then, by the use of Bethe ansatz equations for each of the
$sinh$'s we get
\beq E\simeq \frac{i}{\pi}\!\!\!\!
\sum_{\zeta_\b<0<\zeta_\a}\!\!\!\!\log
S_0^2\(M\[\xi_\alpha-\xi_\beta\]\) +\sum_{\alpha}m\;{\rm
sign}(\xi_\alpha)\nn\,. \la{beforestart} \eeq
 Now we do not have
a problem, as in \eq{BIGSUME}, with badly defined  sums of small
exponentials. In the scaling limit these sums become integrals
\beqa E&\simeq&
-\frac{M}{\pi}\int_{-2}^0 dz\int_{0}^2 dw
\frac{\rho_\theta(z)\rho_\theta(w)}{z-w} +m\,M  \int \rho_\theta(z)
\,{\rm sign}(z)\,dz\label{startingpoint0} \eeqa
 where
$\rho_\theta(z)$ is given by \eq{DENTH}. In fact these integrals can
be easily calculated and the result reads \beq E=\frac{L^2}{4\pi
M}+4\pi M m^2\,. \eeq
 This calculation is generalized to an arbitrary solution of \eq{DBAE1}-(\ref{DBAE3})
  including the left and right excitations
 (with arbitrary $u_k$'s and $v_k$'s) in Section \ref{sec:ener}.

Now notice that if we start \textit{decreasing} the number of
particles $L$, at the point \beq   L^*=4\pi |m| M   \eeq the density
(\ref{DENTH}) will vanish at one of the walls. This point certainly
signifies a phase transition in the effective spin chain and  a
non-analytical change in the classical state. If we continue to
decrease $L$ past this point the new support will now stretch only
on the interval $[-a,2]$ (for $m\ge 0$) and the distribution in the
limit $z\rightarrow -a$ becomes less singular:  $\rho_\theta(z)\sim
\sqrt{a+z}$. The resolvent satisfying the right behavior at infinity
is:
\beq\label{RESTHA} G_\theta(z)=2\pi m \(\frac{\sqrt{z-2+\frac{L}{\pi
m M}}}{\sqrt{z-2}}-1\)     \eeq
which gives for the density
\beq\label{DENTHA}   \rho_\theta(z)=2 m \sqrt{\frac{z+a}{2-z}}
\eeq
where $a=-2+\frac{L}{\pi m M}$.  The phase transition corresponds to
$a=2$. This phase transition fits nicely with the electrostatic
picture point of view since $m$ plays the role of an electric field
which  pushes the particles against one of the walls. It is also interesting to note that $\kappa_+$ and
$\kappa_-$ become of the opposite signs, which,  according to
\ref{PERIODT}, means that the coordinate $Y$ becomes space like on
the world sheet.

Only $L>L^*$ looks compatible with the description of the classical
finite gap solution of section 2 for the $SU(2)_L\times SU(2)_R$
sigma model. Only in that case the poles in $x$ plane will be on the
right place at $x=\pm 1$. For the string theory the total momentum
should be zero $P=2\pi m=0$, hence the classical limit works well,
at least in this highest weight sector.  In the presence of
L,R-excitations with  non-trivial $u_i,v_i$ roots the phase
transition is still possible. It would be very interesting to
understand what are the states beyond this phase transition and what
classical solutions correspond to such states if any.

%%%%%%%%%%%%%%%%%%%%%%%%%%%%%%%%%%%%%%%%%%%%%%%%%%
\subsection{General Classical States and Its Algebraic Curve}

Let us now reproduce the general classical states in terms of an
algebraic curve as a solution of the Bethe ansatz equations
(\ref{BAESU2}) in the scaling limit.

We define the quasi-momenta by the formulas
\begin{eqnarray}
p_1=-p_2&=&G_u-\frac{1}{2}G_\theta \nn \\
p_3=-p_4&=&G_v-\frac{1}{2}G_\theta\,. \la{p1p2p3p4}
\end{eqnarray}
Then the functions $p'_1(z), p'_2(z)$ have the cuts of the type
$C_u,C_\theta$, and $p'_3(z), p'_4(z)$ have the cuts of the type
$C_v,C_\theta$. As in the previous section, we
consider only the situation with a single cut $C_\theta$ and any
number of $C_u,C_v$ cuts. We notice that the quasi-momenta $p'_1(z),
p'_2(z),p'_3(z), p'_4(z)$ form four sheets of the Riemann surface of
an analytical function $p'(z)$ (see fig.\ref{fig:sheets}). Indeed
\begin{eqnarray*}
x\in C_u,&&\;\;\;\;\;{p'_1}^+-{p_2'}^-=2 \sG'_u-G'_\theta\;\,\quad\quad\ \  =0\,,\\
x\in C_\theta,&&\;\;\;\;\;{p'_2}^+-{p_3'}^-=-G'_u-G'_v+ \sG'_\theta\,\,=0\,,\\
x\in C_v,&&\;\;\;\;\;{p'_3}^+-{p_4'}^-=2 \sG'_v-G'_\theta\quad\quad\;\,\,\,\,\,=0\,,\\
x\in C_\theta,&&\;\;\;\;\;{p'_4}^+-{p_1'}^-=-G'_u-G'_v+\sG'_\theta\,\,=0\,,
\end{eqnarray*}
Integrating these equations we restore the mode numbers $n_u$, $n_v$
and $m$, \beqa
x\in C_u,&&\;\;\;\;\;{p_1}^+-{p_2}^-=2\pi  n_u\nn\\
x\in C_\theta,&&\;\;\;\;\;{p_2}^+-{p_3}^-=2\pi m \la{CONTBAE}\\
x\in C_v,&&\;\;\;\;\;{p_3}^+-{p_4}^-=2\pi  n_v \nn\\
x\in C_\theta,&&\;\;\;\;\;{p_4}^+-{p_1}^-=2\pi m  \nn\eeqa These
equations can also be written as holomorphic integrals around the
infinite B-cycles: \beqa \oint_{B^u_{j}}dp &=& 2\pi
n_{u,j}\qquad n_j=1,\ldots,K_u\nn\\
\oint_{B^v_{j}}dp &=& 2\pi
n_{v,j}\qquad n_j=1,\ldots,K_v\la{BCYCLES}\\
\oint_{B^\theta}dp &=& 2\pi m\nn \eeqa
where the the first two conditions correspond to the  equations in
the first and third line of (\ref{CONTBAE}), respectively, while the
last one corresponds to any of the equations of the second and
fourth lines of (\ref{CONTBAE}). The $B$ cycles are defined as in
fig.\ref{fig:sheets}.

 From  (\ref{GASYM},\ref{QQ}) we have the following large $z\to\infty$
asymptotics of quasi-momenta on different sheets
\beqa &p_1(z)\simeq -\frac{Q_L}{2M}\,\frac{1}{z}\,,\qquad
 p_2(z)\simeq \frac{Q_L}{2M}\,\frac{1}{z}\,,\nn\\
&p_3(z)\simeq -\frac{Q_R}{2M}\,\frac{1}{z}\,,\qquad
 p_4(z)\simeq \frac{Q_R}{2M}\,\frac{1}{z}\,,  \la{INFASS}\eeqa

The filling fractions, or the numbers of Bethe $u,v$-roots  forming
each cut, are defined  as follows:
\beqa S_i^v=\frac{M}{2\pi i}\oint_{A^v_i}p(z) dz,\;\;\;\;\;
S_i^u=\frac{M}{2\pi i}\oint_{A^u_i}p(z)dz \la{Filling} \eeqa
These contours are represented in figure \ref{fig:sheets}.

%%%%%%%%%%%%%%%%%%%%%%%%%%%%%%%%%%%%%%%%%%%%%%%%
\subsection{Energy and Momentum\la{sec:ener}}

In this section we will show that the constants $\kappa_\pm$ defined
as the square root singularities of the quasi-momenta \beq
p_{1,3}(z)= \mp\frac{\pi\kappa_\pm}{\sqrt{\pm z-2}},\;\;\;\;\;|z|>2
\label{asympt_p} \eeq are completely fixed by the energy and
momentum. At this point we know that (\ref{P_SSL})
\begin{eqnarray}\la{P_SSL2}
P=\frac{\mu}{2\pi}\sum_{\a}\sinh(\pi\theta_\alpha)=m L-\sum_p n_p
S_p^u-\sum_p n_p S_p^v\,.
\end{eqnarray}
Let us now calculate the energy $E$. As a byproduct we will also
compute the momentum in terms of the singularities at $z=\pm 2$
described by $\kappa_\pm$.

We want to compute the sum
\beq \nn E\equiv \frac{\mu}{2\pi}\sum_\alpha
\cosh(\pi\theta_\alpha)\,, \eeq
but we  \textit{cannot} simply replace this sum by an integral and
use the asymptotic density for $\theta$'s to compute the energy.
This is because the main contribution for the energy comes from
large $\theta$'s, near the walls, where the expression for the
asymptotic density is no longer accurate. It is natural for the
classical limit since the particles become effectively massless, the
contributions of right and left modes is clearly distinguishable and
are located far from $\theta=0$.

The calculation (\ref{P_SSL2}) of momentum was simple because we had
to sum over $\sinh\pi\theta_\a$ which can be taken from the Bethe
equations
\begin{eqnarray}
 \mu \sinh\pi\theta_\a&=&
i\sum_{\beta\neq \alpha}\log S_0^{\,2}\(\th_\a-\th_\b\)
-\sum_j\frac{1}{\theta_\alpha-u_j}-\sum_l\frac{1}{\theta_\alpha-v_l}+2\pi
m\,. \label{BAE_ex_th}
\end{eqnarray}
 Now we have  to sum over $\cosh\pi\theta_\a$. We notice that the energy
is dominated by large $\th$'s where, with exponential precision, we
can replace $\cosh \pi\theta_\a$ by $\pm \sinh \pi \th_\alpha$ for
positive (negative) $\theta_\a$. Furthermore the contribution from
the $\th$'s in the middle of the box is also exponentially
suppressed since $\mu$ is very small. Thus we can pick a point $a$
somewhere in the box not too close to the walls. One can think of
$a$ as being somewhere  in the middle. Then,
\beq \la{ener} E= \sum_{\xi_\alpha>a} \frac{\mu}{\pi} \sinh\( \pi
\xi_\alpha M\)- \sum_{\xi_\alpha<a}\frac{\mu}{\pi}\sinh \(\pi
\xi_\alpha M\)\, , \nn \eeq
 where, let us stress, the result is \textit{correct and
independent of the point $a$ with exponential precision}. Having a
sum of $\sinh \pi \theta_\a$ we can substitute each  of them by the
corresponding Bethe equation (\ref{BAE_ex_th}) obtaining
\beq E\simeq \frac{i}{\pi}\!\!\!\!
\sum_{\zeta_\b<a<\zeta_\a}\!\!\!\!\log
S_0^2\(M\[\xi_\alpha-\xi_\beta\]\) -\sum_{j,\alpha} \frac{{\rm
sign}(\xi_\alpha-a)}{\pi(\xi_\alpha-x_j)M} -\sum_{l,\alpha}
\frac{{\rm sign}(\xi_\alpha-a)}{\pi(\xi_\alpha-y_l)M}
+\sum_{\alpha}m\;{\rm sign}(\xi_\alpha-a)\nn\,. \eeq
 Now we can
safely go to the  continuous limit since in the first term the
distances between $\xi$'s are now mostly of order the $1$. Moreover,
it is very important that the contribution from the $\xi$'s near the
walls $\pm 2$ is now suppressed since
\begin{eqnarray*}
|\log S_0^2(M(2-\xi_\b)|>|\log S_0^2(M(2-a)|\sim 1/M\,.
\end{eqnarray*}
This allows to rewrite the energy, with $1/M$ precision, as follows
\beqa
E&\simeq& -\frac{M}{\pi}\int_{-2}^a dz\int_{a}^2 dw \frac{\rho_\theta(z)\rho_\theta(w)}{z-w}-\frac{M}{2\pi} \int \frac{\rho_\theta(z)\rho_u(w)}{z-w}{\rm sign}(z-a)\,dz\,dw \nn\\
&-&\frac{M}{2\pi} \int \frac{\rho_\theta(z)\rho_v(w)}{z-w}{\rm
sign}(z-a)\,dz\,dw+m\,M  \int \rho_\theta(z) {\rm
sign}(z-a)\,dz\label{startingpoint} \eeqa
where \textit{the densities are the asymptotic densities for the
exact box potential}, given by the integral \eq{rhoBAESU2}.

As a result, by the use of Bethe equations, we managed to transform
the original sum over $\cosh$, highly peaked at the walls, into a
much smoother sum where the main contribution is now softly
distributed along the bulk and where the continuous limit does not
look suspicious.

Let us now take (\ref{startingpoint}) as our starting point. From
the previous discussion we know that this expression does not depend
on $a$ provided $a$ is not too close to the walls. In fact, we can
easily see that it does not depend on $a$ \textit{at all} after
taking the continuous limit leading to the perfect box-like
potential. To see this one can see that, due to Bethe equations
\eq{rhoBAESU2}, the $a$-derivative of this expression is zero for
all $a\in ]-2,2[$. Hence we can even send $a$ close to the wall:
$a=-2+\epsilon$, where $\epsilon$ is very small. Let us calculate
the first term. The main contribution to the integral comes from
$-2\simeq z\sim w$ so that we can use the asymptotics
(\ref{asympt_p}) to get
\beq \nn-\frac{M}{\pi}\int_{-2}^{-2+\epsilon} dz\int_{-2+\epsilon}^2
dw \frac{\rho_\theta(z)\rho_\theta(w)}{z-w}\simeq
-\int_{-2}^{-2+\epsilon} dz\int_{-2+\epsilon}^2 dw \frac{4 M
\kappa_-^2}{\pi(z-w)\sqrt{2+z}\sqrt{2+w}}\simeq  2\pi M \kappa_-^2
\eeq
 The remaining $3$ terms are very simple: since $a\simeq -2$ we
can simply drop the $sign$-functions inside the integrals and obtain
exactly the expression of the momentum in the continuous limit. We
arrive therefore at
\beqa E\simeq 2 M \kappa_-^2 \pi+\(mL-\sum_p n_p^v S^v_p-\sum_p n_p^u S^u_p \) \,.
\la{energy} \eeqa
where the expression in the parentheses is the momentum. If we
compute the $a$-independent integral (\ref{startingpoint}) near the
other wall, i.e. for $a=2-\epsilon$, we find
\begin{eqnarray*}
E\simeq 2 M \kappa_+^2 \pi-\(mL-\sum_p n_p^v S^v_p-\sum_p n_p^u  S^u_p\) \,. %= \nn\\
\end{eqnarray*}
Therefore, equating the results one obtains the desired expressions
for the energy and momentum
\begin{eqnarray}
E\pm P=2\pi\,M\,\kappa_\pm^2 \la{EP}
\end{eqnarray}
through the data $\kappa_\pm$ at the singularities of the curve at
$z=\pm 2$.

\newsection{Matching with the Finite Gap KMMZ Solution \la{sec4}}
\begin{figure}[t]
    \centering
        \includegraphics[scale=0.7]{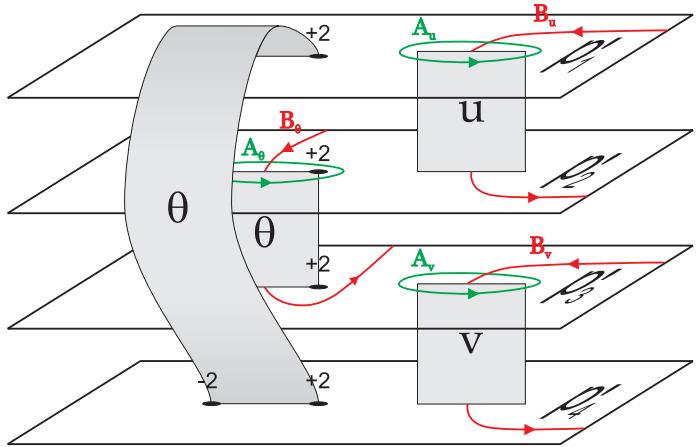}
    \caption{Structure of the curve coming from the Bethe ansatz side. The quasi momenta $p_{1,2,3,4}(z)$ are
    defined in (\ref{p1p2p3p4}). This figure is related with fig.\ref{fig:uvKMMZ} by means of Zhukovsky map.}
    \label{fig:sheets}
\end{figure}
In this section we will demonstrate the main result of our paper:
the equivalence of the large density limit of a system of physical
particles on a circle for the quantum $O(4)$ $\sigma$-model
described in the previous section, to the classical theory of the
same model described in section \ref{sec2}. On the one hand we have
the quantization of the sigma model given by a set of Bethe
equations. For this quantum system Virasoro constraints are somewhat
obscure. We introduce however a natural selection rule for the
quantum states, such that in the scaling limit we obtain the
classical finite gap KMMZ solutions for the $S^3\times R_t$
classical sigma model with classical Virasoro constraints imposed
\cite{Kazakov:2004qf}. Let us also recall that this sector belongs
to the superstring theory on $AdS_5\times S^5$ and is in itself a
consistent classical truncation.

Our result also  means that the considered scaling limit of the
quantum $O(4)$ $\sigma$-model is nothing but the classical limit and
our result gives a decisive demonstration of the correctness of
Zamolodchikovs' physical bootstrap S-matrix approach
\cite{Zamolodchikov:1977nu} to these models, as well as of the
Polyakov--Wiegmann solution of the principle chiral field
\cite{Polyakov:1983tt}\footnote{Although we demonstrate it here only
for the $SU(2)$ case we see no obstacles for the generalization of
our method to any $SU(N)$ principle chiral field.}. According to our
results, both approaches construct indeed the quantization procedure
with the correct classical limit.\footnote{Modulo the phase
transition noticed in the previous section,  to the states which are
hard to identify with the classical solutions of the underlying
classical model.}

More precisely, we will show here that every solution of Bethe
ansatz equations in the scaling limit, described by an algebraic
curve of the quasi-momentum (\ref{DBAE1}-\ref{DBAE3}), is in fact a
finite gap solution for the classical string on $S^3\times R_t$. In
the next section this correspondence will be generalized to the
$O(6)$ sigma model corresponding to the whole $S^5\times R_t$ sector
of the classical superstring. The generalization to all $O(2n)$
sigma models is also straightforward (the corresponding classical
string, unlike the quantum one, is well defined).

To prove this correspondence, we have just to compare the algebraic
curves describing a Bethe state in the scaling limit and the finite
gap curve \eq{FINGAPS}, together with all their moduli and the data
at singularities. The similar goal for the $OSP(m+2n|m)$ model was
achieved in \cite{Mann:2005ab} by the direct solution of integral
Bethe equations.

\begin{figure}[t]
    \centering
        \includegraphics{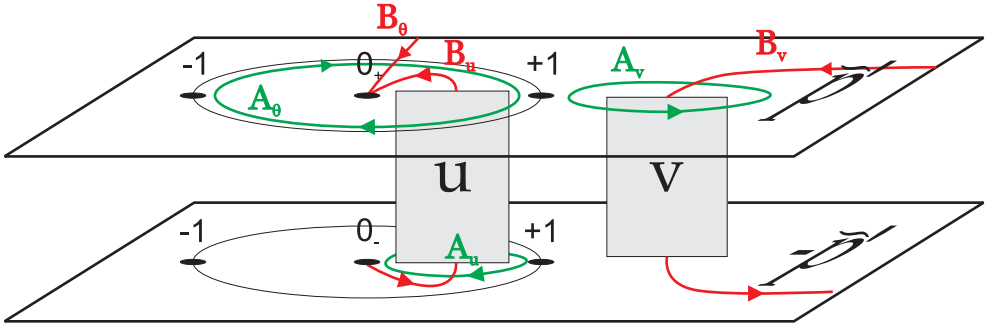}
    \caption{Algebraic curve from the Finite gap method of KMMZ. In this language, $u$ and $v$ cuts correspond to cuts inside and outside the unit circle respectively. This figure is related with fig.\ref{fig:sheets} by means of Zhukovsky map.}
    \label{fig:uvKMMZ}
\end{figure}

\begin{figure}[t]
    \centering
        \includegraphics[scale=0.7]{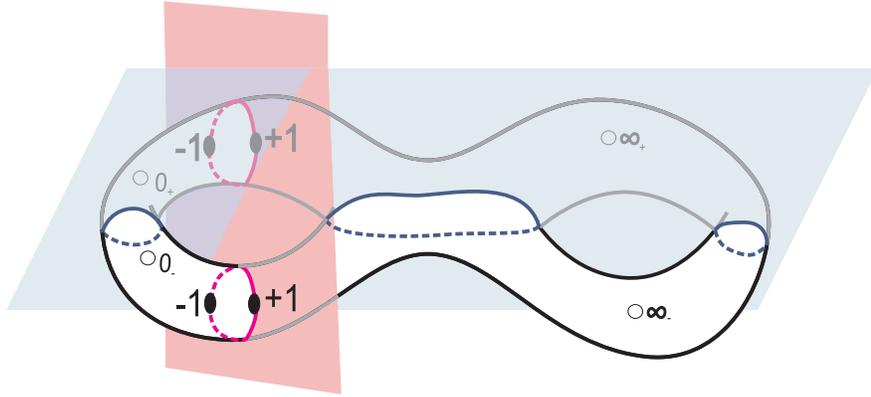}
            \caption{The curves appearing from the finite gap method
            and the Bethe ansatz equations turn out to the different
            projection of the same curve. }
    \label{fig:artistic}
\end{figure}

Central to this comparison will be the Zhukovsky map $z=x+1/x$. We
will show that the Riemann surface in $z$ variable of the scaled
Bethe equations, fig.\ref{fig:sheets}, on the one hand, and the
Riemann surface in $x$ variable of the classical finite gap solution
\eq{FINGAPS}, fig.\ref{fig:uvKMMZ},  on the other hand, are two
different projections of the same algebraic curve related by
Zhukovsky map. We attempted to  present schematically two different
projections by different  colors of projected parts of the curve
fig.\ref{fig:artistic}.

When we apply Zhukovsky map to the Riemann surface of
fig.\ref{fig:sheets} the $\theta$-cuts along $(-2,2)$ disappear on
all 4 sheets and these singular branch points become simple poles,
since under the map
\beq\la{POLESX}   \frac{1}{\sqrt{z\pm 2}} \longleftrightarrow
\frac{1}{x\pm 1}. \eeq
 The Riemann surface in the $x$ projection will consist only of two
sheets, as on fig.\ref{fig:uvKMMZ}. All the $u$-cuts and $v$-cuts
connect now these two sheets.

The inverse map
\beq\la{INVZH}  x_\pm=\hf\(z\pm \sqrt{z^2-4}\)   \eeq
projects the  two upper sheets of fig.\ref{fig:sheets} with $u$-cuts
into the interior of the unite circle in $x$ projection, by means of
$x_-(z)$, where as the two lower cuts of fig.\ref{fig:sheets} with
$v$-sheets are projected into the exterior of this unit circle on
$x$ projection, by means of $x_+(z)$. More precisely, we will show
that \beq \la{stat}p_3(z)=\tilde p\,(x_+(z)),\;\;\;\;\;p_1(z)=2\pi m
-\tilde p\,(x_-(z))\;. \eeq Since, by definition, $p_2=-p_1$ and
$p_4=-p_3$ we see that $p_{\,3,4}$ describe the exterior of the unit
circle in the upper (lower) sheet while $p_{\,2,1}$ describe the
interior of the unit circle in the upper (lower) sheet.

Comparing the singularities  at $x=\pm 1$ after the map we see that
they are again given, due to \eq{asympt_p}, by the energy and
momentum in terms of $\kappa_\pm$ in full correspondence with the
finite gap result \eq{LCH}.

Under this map the $4$ infinities of the $4$ sheets of quasi-momenta
$p_1,p_2,p_3,p_4$, fig.\ref{fig:sheets}, are mapped according to
(see (\ref{INVZH},\ref{stat}))
\begin{eqnarray*}
\(\infty_{2,1},\infty_{3,4}\)  \longleftrightarrow
\(0_{\pm},\infty_{\pm}\)\,.
\end{eqnarray*}
Then, from the first line of (\ref{INFASS}), we read off the
behavior of the quasi-momenta near $x=0_{\pm}$ while from the second
line of (\ref{INFASS}) we infer the behavior of the quasi-momenta at
$x=\infty_{\pm}$. They match the asymptotics  (\ref{PXZERO}) and
(\ref{GLCH}) respectively after imposing
$M=\frac{\sqrt{\lambda}}{4\pi}$ or
\beq -\log\mu=\frac{\sqrt{\lambda}}{2} \, , \eeq
 a  simple and nice
relation exhibiting the dimensional transmutation property of the
asymptotically free theory with the precise coefficient\footnote{
For the $O(N)$ sigma model the beta function for the coupling is
given by $\beta\equiv \frac{\partial}{\partial \log\Lambda}\,
\sqrt{\lambda(\Lambda)}=N-2$ where $\Lambda$ is the cutoff of the
theory. The dynamically generated mass must be of the form
$m=\Lambda\,f(\sqrt{\lambda}\,)$. The functional form of $f$ is
fixed by the $\beta$ function upon imposing independence on the
cutoff of this physical quantity. Thus, for general $N$,
$-\log\mu=\frac{\sqrt{\lambda}}{N-2}+\O(1)$.}.

The filling fractions (\ref{Filling}) coincide with (\ref{ACYCLE})
of the finite gap solution since they are homomorphic integrals,
invariant w.r.t. the change of projection\footnote{The minus sign in
(\ref{ACYCLE}) comes from relation (\ref{stat}) between $p_1(z)$ and
$\tilde p \,(x)$.}. The same can be said about the Bethe equations \eq{BCYCLES})
which we have to compare with the finite gap equations
(\ref{BCYCLE}, with the following definition of the mode numbers
\beq\la{NUREDEF}    n^u_i =  2m-n_i \,\,\, , \,\,\, n^v_i=n_i\,.
\eeq
This redefinition is due to the fact that a $B_u$-cycles of
\eq{INFASS} going from $\infty_1$ on the upper sheet of
fig.\ref{fig:sheets} through the $u$-cut to $\infty_2$ on the next
sheet, are projected by Zhukovsky map into the interior of the unit
circle of the $x$ projection, fig.\ref{fig:uvKMMZ}, with the
$\infty_{1,2}$ projected to $0_\mp$. To complete it to the
$B$-cycles as defined in \eq{BCYCLE} we have to complete them with
paths  $(\infty_-,0_-)$ and $(0_+,\infty_+)$, each of them is equal
to quasi-momentum $-\tilde p(0)=-2\pi m$. This completed cycle will
run in the opposite direction of a usual $B_k$ cycle, hence the
minus sign in \eq{NUREDEF}.

The third equation \eq{BCYCLES} follows immediately since integral
over the $B_\theta$-cycle is mapped into the integral from
$\infty_+$ to $0_+$ yielding $\tilde p(0)$ as result.

We conclude that the two projections, one obtained from the quantum
Bethe ansatz, the other from the classical finite gap solution,
represent the same algebraic curve with the same moduli and thus the
underlying solutions (states) completely coincide.

As a result we can express densities of the Bethe roots thought
classical quasi-momentum $\tilde p(x)$ as\footnote{Last expression
deserves some words. Points $z\pm i0$ below and above the $\theta$
cut are mapped to $x=x_+(z)$ and $x=x_-(z)=1/x_+(z)$ Since
$z\in[-2,2]$ these are conjugate points in the unit circle. Then the
discontinuity of $p(z)$ in the $\theta$ cut is given by the
imaginary part of $\tilde p(x)$.} \beqa
\nn\rho_v(z(x))&=&\rho(x),\;\;\;\;\;x\in C^v\\
\nn\rho_u(z(x))&=&\rho(x),\;\;\;\;\;x\in C^u\\
\nn\rho_\theta(2\cos\phi)&=&-\frac{1}{\pi}\,{\rm Im}\[\tilde
p\(e^{i\phi}\)\],\;\;\;\;\;\phi\in [0,\pi] \eeqa where $2\pi
i\,\rho(x)=\tilde p\,(x-i0)-\tilde p\,(x+i0)$. These densities
automatically satisfy Bethe ansatz equations \eq{rhoBAESU2}.

In Appendix B we provide an alternative, completely algebraic,
derivation of the equivalence between the curves.

%%%%%%%%%%%%%%%%%%%%%%%%%%%%%%%%%%%%%%%%%%%%%%%%%%%%%%%%%%%%%%%%%%%%
\section{ Some Limiting Solutions of the Quantum $O(n)$ Sigma-Model }

Here we will consider three limiting cases: the  BMN solutions from
the quantum Bethe equations, the semi-classical solutions with small
longitudinal amplitudes and the discrete spin chain limit.

%%%%%%%%%%%%%%%%%%%%%%%%%%%%%%%%%%%
\subsection{BMN Limit}

In this section we will consider the BMN limit
\cite{Berenstein:2002jq} which, in our notations, reads $M\sim L\gg
1,\;J_v,\;J_u\sim 1$. We are therefore considering a big $\theta$
cut which will be slightly perturbed by some point like microscopic
$u$ and $v$ cuts, treated as corrections. The position of these
microscopic cuts in the presence of the $\theta$ cut are found from
the first and third equations in (\ref{BAESU2}) with $G_\th$ given
by (\ref{RESTH}) with $m=0$ (the only choice compatible with the
stringy condition $P=0$ for very small filling fractions for $u$'s
and $v$'s.). We find \beq \frac{L}{M}\frac{1}{\sqrt{x_n^2-4}}=2\pi
n,\;\;\;\;\;x_n={\rm sign}\, n\sqrt{4+\frac{L^2}{4\pi^2M^2 n^2}}\,.
\nn \eeq Then, from the second equation in (\ref{BAESU2}), we can
compute the correction $\delta G_\th$ to the resolvent of the main
cut, \beq \delta \sG_\th(x)=\sum_n\frac{N_n}{x-x_n} \nn \eeq where
$N_n=J_n/M$. Since \beq \delta G_\th(x)=\delta
\sG_\th(x)+i\pi\delta\rho_\th(x) \nn \eeq we have \beq
\delta\rho_\th(x)=\sum_n
N_n\(\frac{\sqrt{x_n^2-4}}{\pi\sqrt{4-x^2}}\frac{{\rm
sign}\,x_n}{x-x_n}+\frac{1}{\pi\sqrt{4-x^2}}\)\,. \label{deltarho}
\eeq Let us explain why. The inverse square roots follows from the
known support and behavior of $\rho$ near the walls. Then the first
term inside the brackets follows from cancellation of the poles at
$x_n$ in $\delta G_\th$. Finally the second term in fixed by
requiring $\mathcal{O}(1/x^2)$ decay of $\delta G_\th$ at infinity,
or, equivalently, from requiring that the perturbation of the
density does not change number of particles in the cut, $\int
\delta\rho_\th =0$.

 From (\ref{deltarho}) we can read the change in the behavior near $\pm 2$ of the density of rapidities, i.e.  compute $\delta\kappa_\pm$ and, therefore,
\beqa
\delta P=2\pi M\kappa(\delta\kappa_+-\delta\kappa_-)&=&-\sum n J_n\\
\delta E=2\pi M\kappa(\delta\kappa_++\delta\kappa_-)&=&\frac{L
}{\sqrt{\lambda}}\,\sum J_n\(1-\sqrt{1+\frac{\lambda n^2}{L^2}}\)\,.
\eeqa The same result was obtained in \cite{Kazakov:2004qf}, but in
a different regime, the semi-classical one, where\\ $1\ll
J_v,\,J_u\ll~L\sim~M$. In our case $J_n$ are small {\it integers}
which reflects the quantum nature of the result, as for the original
BMN result \cite{Berenstein:2002jq}.

%%%%%%%%%%%%%%%%%%%%%%%%%%%%%%%%%%%%%%%%%%%%%%%%%%%%%%%%%%%%%%%%
\subsection{Multi Cut Vacuum States \la{Multicuts}}

In this section we will try to understand the meaning of the
multi-cut solutions in $\theta$ and its possible relation with
Virasoro constraint.

We will consider here  the  solutions in the Abelian $U(1)$ sector
described only by $\theta$-variables:
\beq
\la{abb}g=\exp\[-i\tau_3\(T\tau+S\sigma+\sum_{\omega>0}A_\omega\cos(\omega(\tau+\sigma))+\sum_{\omega<0}A_\omega\cos(\omega(\tau-\sigma))\)\]
\eeq
 where $S$ is the winding number, integer to ensure the periodicity
in $\sigma$, and $T=Q_R/\sqrt{\lambda}$ is the  angular momentum.
For small amplitudes $A_m$ we can quantize this solution as a set of
oscillators. The energy in this approximation reads \beq
E=\frac{S^2\sqrt{\lambda}}{2}+\frac{Q_R^2}{2\sqrt{\lambda}}+\sum_{\omega\neq
0} |\omega| N_\omega \eeq
where $N_\omega$ is the mode number of the oscillator with the
frequency $\omega$.
 This formula is a well known expression for
the spectrum of $U(1)$ model. It is very natural since we can use
arguments of BMN to see that in the limit of large angular momentum
the system feels only a small neighborhood of the main circle on
$S^3$ and the longitudinal oscillations are described by a massless
scalar field. This is why our analysis in only valid for small
amplitudes $A_m$.

The worldsheet momentum is
\beq P=\sqrt{\lambda}\,ST+\O(1)\,. \eeq
The terms $\O(1)$ represent quantum corrections which are not
captured by the leading semi-classical approximation.

We will show that the solution \eq{abb} with small amplitudes
corresponds to a multi-cut configuration of $\theta$'s when all cuts
except one are small (with filling fractions of the order $1$). We
will denote by $m_0$ the mode number of the large cut and by
$L_{m_0}$ the number of $\theta$'s in this cut. We assume
$L_0\sim\sqrt{\lambda}$.

We can compute the momentum and the charge in terms of the Bethe
ansatz quantities
\beq P=m_0 L_{m_0}+\sum_{m\neq m_0}m
L_m,\;\;\;\;Q_R=L_{m_0}+\sum_{m\neq m_0} L_m\equiv L \eeq
 and thus
 \beq T=\frac{L}{\sqrt\lambda}
,\;\;\;\;\;S=m_0 \eeq
where we are allowed to drop $1/L_0\sim 1/\sqrt{\lambda}$ terms due
to the winding number quantization, $S\in {\mathbb Z}$. Having
identified $S$ and $T$ in the language of Bethe ansatz we can write
\beq E=\frac{m_0^2\sqrt\lambda }{2}+\frac{L^2}{2\sqrt\lambda}
+\sum_{\omega\neq 0} |\omega| N_\omega \la{Em} \eeq
 where $m_0$ and $L$ are
integers. It is left to  identify the amplitudes with the remaining
filling fraction. Let us denote $N_\omega=L_{\omega+m_0}$, then we
have
\beq E=E_0 +\!\sum_{m\neq m_0} |m-m_0| L_{m}\,. \eeq
Expression (\ref{Em}) can also be established form the Bethe ansatz
side. One has, see section (\ref{HW}), equation (\ref{beforestart})
\begin{eqnarray}
E= \frac{i}{\pi}\!\! \sum_{\zeta_\b<0<\zeta_\a}\!\!\!\!\log
S_0^2\(M\[\xi_\alpha-\xi_\beta\]\) +\sum_{\alpha}m_\alpha\;{\rm
sign}(\xi_\alpha)\,. \label{Emulti}
\end{eqnarray}
The main cut with the mode number $m_0$ occupies most of the box.
Other cuts are squeezed close to the right (left) wall for $m_i$
greater (smaller) than $m_0$. Then the second term in (\ref{Emulti})
reads
 \beq\la{PREVEX}
\sum_{\alpha}m_\alpha\;{\rm sign}(\xi_\alpha)=\sum_{\alpha}
m_0\;{\rm sign}(\xi_\alpha)+\sum_{m} |m-m_0|\,L_m \eeq
 We only have to show that the change in the first term in (\ref{Emulti}) and in the first term in (\ref{PREVEX}) is
small compared to the last term of (\ref{PREVEX}). We will
demonstrate that the change in one of the mode numbers leads to
$1/L$ displacements of $\theta$'s. Consider for example the
situation when we decrease the mode number of the first $\theta$ by
$\delta n\sim 1$. This $\theta$ has the biggest displacement since
the force acting on it increases by $2\pi\delta n$. All other
$\theta$'s are affected only due to the displacement of the first
one. Let us estimate this displacement assuming it to be small. Then
one has the following equilibrium condition
\beq \pi\mu \cosh( \pi\theta_1) \delta\theta_1\sim 1. \eeq
 Since
$\pi\mu \cosh( \pi\theta_1)\gg 1$\footnote{We know that
$E=\frac{1}{2\pi}\mu\sum_\alpha\cosh\pi\theta_\alpha\sim L$. With
exponential precision we can sum only up $\theta_\alpha<-2M+L^{1/2}$
so that number of terms is of order $L^{1/4}$ as one easily sees
from the density computed in \ref{DENTH} . Then we have $E<\mu\cosh(
\pi\theta_1)L^{1/4}$ which means $\mu\cosh( \pi\theta_1) \gg 1$} the
displacement is small $\delta\theta\ll 1$, then $\delta\xi\ll 1/L$.
This means that the change of the density is negligible
$\delta\rho(\zeta)\ll 1/L$ and we can say that the first term in
(\ref{Emulti}) remains intact.

We obtain the precise agreement between semi-classics and Bethe
ansatz calculations for several $\theta$ cuts.  We can conclude that
the longitudinal oscillations, which break the Virasoro conditions
and thus being inadmissible in the string context, manifest
themselves as extra cuts in $\theta$. We observe indeed this
phenomenon, at least  when these oscillations are small. As it was
said already in the introduction, we conjecture on the base of these
arguments and of the comparison of our classical limit with the
direct finite gap approach that the Virasoro conditions impose the
selection rule on the states described by (\ref{DBAE1},\ref{DBAE2})
under which only the solutions having $\theta_k$'s with the same
mode numbers $m_1=\dots=m_L$ are retained.

%%%%%%%%%%%%%%%%%%%%%%%%%%%%%%%%%%%%%%%%%%%%%%%%%%%%%%%%%%%%%%%%
\subsection{XXX Spin Chain Limit}

In this section we will consider the XXX spin chain  limit of the
Bethe equations for the principal chiral field.  Contrary  to the
classical limit considered above it is rather a strong coupling
limit. Namely, we take $\mu\rightarrow \infty$ and arbitrary
$L,J_v,J_u$. In this limit $\theta$'s are squeezed near zero. In the
leading order we can just take them equal to zero so that
(\ref{DBAE2},\ref{DBAE3}) become non-interacting two $SU(2)$ XXX
chains
\begin{eqnarray}\la{XXXBAE}
\(\frac{u_j+i/2}{u_j-i/2}\)^L&=&
\prod_{i\neq j} \frac{u_j-u_i+i}{u_j-u_i-i}\,,\\
\(\frac{v_k+i/2}{v_k-i/2}\)^L&=&\prod_{l\neq k}
\frac{v_k-v_l+i}{v_k-v_l-i}\,.
\end{eqnarray}
These equations  look similar  to the ${\cal N}=4$ SYM XXX spin
chain Bethe equations, the first one  for the sector of  $X,Y$
scalars and the second for  $\bar X,\bar Y$ scalars. To make a more
precise comparison we have to fix the conserved quantities.
Let us firstintroduce the notations
\begin{eqnarray*}
e_s(u)\equiv i \log\frac{u+is/2}{u-is/2}\, , \,\,\,
\end{eqnarray*}
and \beqa \nn
\theta_\a&=&\frac{1}{\mu}\,\theta_\a^0+\frac{1}{\mu^2}\,\theta_\a^1+\cdots
\eeqa
Below we will show that energy of the system given by \eq{E_SSL} is
proportional to that of the XXX spin chain
\beq E_{\rm
xxx}=\sum_j\frac{1}{u_j^2+1/4}+\sum_k\frac{1}{v_k^2+1/4}. \eeq
 To see
this we expand further in powers of $\mu$. The formula for
energy with $\O(\theta^3)$ precision becomes
\beq \la{EEE}E\simeq
\frac{\mu}{2\pi}\sum_{\alpha=1}^L\(1+\frac{\pi^2\theta_\alpha^2}{2}\).
\eeq
 Let us evaluate $\theta^0$ and $\theta^1$. We expand BAE
(\ref{DBAE1}) in powers of $\theta$ up to $\O(\theta^3)$
\begin{eqnarray}
\la{60}\mu \sinh\pi\theta_\a&\simeq&2\pi\, m
+\sum_{\beta\neq\alpha}^L\(\pi\,{\rm
sign}(\theta_\alpha-\theta_\beta)-4\,(\theta_\alpha-\theta_\beta)\log 2\)\\
\nn &-&\sum_k e_1(v_k)-\sum_j e_1(u_j)+\theta_\alpha\, E_{\rm
xxx}+\O(\theta^3).
\end{eqnarray}
The second term in the r.h.s comes from the expansion of $S_0$.
Expanding \eqs{DBAE2}{DBAE3} we have
\begin{eqnarray}
\la{61} 2\pi n_j^u&=&L\,e_1(u_j)-L\,\sum_{i\neq
j}e_2(u_j-u_i)-e'_{1}(u_j)\sum_\alpha\theta_\alpha+\O(\theta^3)\,,\\
\la{62} 2\pi n_k^v&=&L\,e_1(v_k)-L\,\sum_{l\neq
k}e_2(v_k-v_l)-e'_{1}(v_k)\sum_\alpha\theta_\alpha+\O(\theta^3)\,.
\end{eqnarray}
Summing all \eqs{61}{62}, using the level matching condition $P=0$
and \eq{60} one obtains
 \beq \mu \pi\theta_\a \simeq \pi(2\alpha-L-1)
\la{441}+\(\theta_\alpha-\frac{1}{L}\sum_\beta
\theta_\beta\)\(E_{\rm xxx} -4L\log 2 \)+\O(\theta^3), \eeq
so that
\beq
\theta^0_\alpha=2\alpha-L-1,\;\;\;\;\;\theta^1_\alpha=\theta^0_\alpha
\(E_{\rm xxx} -4L\log 2 \) \nn \eeq
 and thus, from \eq{EEE},
 \beqa
E&=&\[\frac{L\mu}{2\pi}+\frac{\pi L(L^2-1)}{12\mu}-\frac{2\pi
L^2(L^2-1)}{3\mu^2}\log2\]+\[\frac{\pi L(L^2-1)}{6\mu^2}\]\,E_{\rm
xxx}\, . \eeqa
This formula, together with \eq{XXXBAE}, reproduces
the energy and the Bethe equations for the
XXX spin chain up to the constant terms inside the square brackets.
Although it is known that the perturbative expansion of the energy $E$
in general does not coincide with that of the eigenvalue of
the dilatation operator in SYM \cite{Serban:2004jf},
it has been confirmed that
they coincide up to two-loop approximation
in the continuous limit\cite{Kazakov:2004qf}.
The above result may serve as a starting point
to elucidate the matching at the discrete level.

%%%%%%%%%%%%%%%%%%%%%%%%%%%%%%%%%%%%%%%%%%%%%%%%%%%%%%%%%%%%%%%%%%%%%%%%%%%%%%
\newsection{$O(6)$ Sigma-Model}

Let us now move on to a larger subsector of the superstring theory,
namely let us consider the $O(6)$ non-linear sigma model. As before
we describe a general quantum state by the system of vector
particles which live on a circle of length ${\cal L}=2\pi$. The wave
function is then specified by a set of rapidities
$\{\theta_\alpha\}$ which determine the coordinate part of the wave
function, plus  a set of Bethe roots
$\{u_j^{(1)},u_j^{(2)},u_j^{(3)}\}$ encoding its color
structure\footnote{For a general simple group one would have $r$
different kinds of roots where $r$ is the rank of the group, i.e.
the number of simple roots.}. When imposing periodicity of the wave
function one obtains the quantization of these parameters. They are
are constrained by Bethe ansatz equations (see Appendix D for the
derivation and for the generalization to $O(2n)$)
\begin{eqnarray}
\nn
e^{-i\mu\sinh\frac{\pi\theta_\a}{2}}&=&\prod_{\beta\neq\alpha}^{L}S_0(\theta_\a-\theta_\b)
\prod_{j=1}^{K_2}\frac{\theta_\a-u^{(2)}_j+i/2}{\theta_\a-u^{(2)}_j-i/2} \\
\nn 1 &=& \prod_{j\neq
i}^{K_1}\frac{u^{(1)}_i-u^{(1)}_j+i}{u^{(1)}_i-u^{(1)}_j-i}
\prod_{j=1}^{K_2}\frac{u^{(1)}_i-u^{(2)}_j-i/2}{u^{(1)}_i-u^{(2)}_j+i/2}\\
\la{qBAE_SO6}\prod_{\alpha=1}^{L}\frac{u^{(2)}_i-\theta_\alpha+i/2}{u^{(2)}_i-\theta_\alpha-i/2}
&=& \prod_{j\neq
i}^{K_2}\frac{u^{(2)}_i-u^{(2)}_j+i}{u^{(2)}_i-u^{(2)}_j-i}
\prod_{j=1}^{K_3}\frac{u^{(2)}_i-u^{(3)}_j-i/2}{u^{(2)}_i-u^{(3)}_j+i/2}
\prod_{j=1}^{K_1}\frac{u^{(2)}_i-u^{(1)}_j-i/2}{u^{(2)}_i-u^{(1)}_j+i/2} \\
\nn 1 &=& \prod_{j\neq
i}^{K_3}\frac{u^{(3)}_i-u^{(3)}_j+i}{u^{(3)}_i-u^{(3)}_j-i}
\prod_{j=1}^{K_2}\frac{u^{(3)}_i-u^{(2)}_j-i/2}{u^{(3)}_i-u^{(2)}_j+i/2}
\end{eqnarray}
where \beqa S_0(\theta)
=-\frac{\Gamma\left(\frac{1}{4}-i\frac{\theta}{4}\right)\Gamma\left(\frac{1}{2}-i\frac{\theta}{4}\right)\Gamma\left(\frac{3}{4}+i\frac{\theta}{4}\right)\Gamma\left(1+i\frac{\theta}{4}\right)}
{\Gamma\left(\frac{1}{4}+i\frac{\theta}{4}\right)\Gamma\left(\frac{1}{2}+i\frac{\theta}{4}\right)\Gamma\left(\frac{3}{4}-i\frac{\theta}{4}\right)\Gamma\left(1-i\frac{\theta}{4}\right)}\,.
\eeqa

In the next subsection we will show that, in the classical scaling
limit $L\sim \log\frac{1}{\mu}\rightarrow \infty$, the solutions of
this system of Bethe ansatz equations are in one to one
correspondence with classical solutions of $O(6)$ sigma model,
classified by means of an algebraic curve \cite{Beisert:2004ag}. The
derivation here will be similar to the $SU(2)$ principal chiral
field considered in previous section. In the classical limit when
$\theta's$ will be large we can use Coulomb approximation and
substitute $S_0(\theta)$ by its large $\theta$ asymptotics
\begin{eqnarray*}
i\,\log\,S_0(\theta)=\frac{1}{\theta}+\mathcal{O}(1/\theta^3)\,.
\end{eqnarray*}
Furthermore, the potential becomes a square box and
$\theta_\alpha/M$ will therefore be distributed, at leading order,
between $-2$ and $2$ provided we define the variable $z=\theta/M$
with $M=-\frac{\log\mu}{\pi}$. Thus, in this limit, we can recast
(\ref{qBAE_SO6}) as \beqa
\nn2\sG_1-G_2&=&2\pi n^{(1)}\\
\la{BAE_SO6}2\sG_2-G_1-G_3-G_\theta&=&2\pi n^{(2)},\;\;\;\;\;\;\;\;\;\;\sG_\theta-G_2=2\pi m\\
\nn2\sG_3-G_2&=&2\pi n^{(3)} \eeqa where the resolvents are given by
\beq
G_\theta(z)=\frac{1}{M}\sum_\alpha\frac{1}{z-\theta_\alpha/M},\;\;\;\;\;G_{l}(z)=\frac{1}{M}\sum_{i}^{K_{l}}\frac{1}{z-u^{(i)}/M},\;\;\;\;\;l=1,\dots,3\,,
\la{defGG} \eeq and each equation holds for $z$ belonging to a
$\theta$ or $u^{(i)}$ cut if $G_\theta$ or $G_i$ is slashed,
respectively. These equations, as we will show in the next
subsection, can be used to define some algebraic curve which maps
onto the classical curve of \cite{Beisert:2004ag} after
Z-transformation $z=x+1/x$.

The quantum state (Bethe vector) corresponding to the solutions of
(\ref{qBAE_SO6}) carries the following $\alg{so}(6)$ spins
\begin{eqnarray}
\nn J_1&=&L-K_2,\\
\la{J2K}J_2&=&K_2-K_1-K_3,\\
\nn J_3&=&K_1-K_3 \,,
\end{eqnarray}
where $(J_1,J_2,J_3)$ are measured by the orthonormal basis of the
Cartan subalgebra.

\subsection{Algebraic Curve}
To relate solutions of BAE (\ref{qBAE_SO6}) with classical solutions
classified by algebraic curves \cite{Beisert:2004ag} we will
construct possible curves corresponding to (\ref{BAE_SO6}). This
means that  the resolvents $G_{1,2,3}(z)$ and $G_\theta(z)$, when
taken in appropriate linear combinations, are in fact different
branches of some unique analytical function. In fact there are
several possible combinations, each of them corresponding to
different representation of $SO(6)$.

\subsubsection{Curve for the Vector Representation}

\begin{figure}[t]
    \centering
        \includegraphics[scale=0.85]{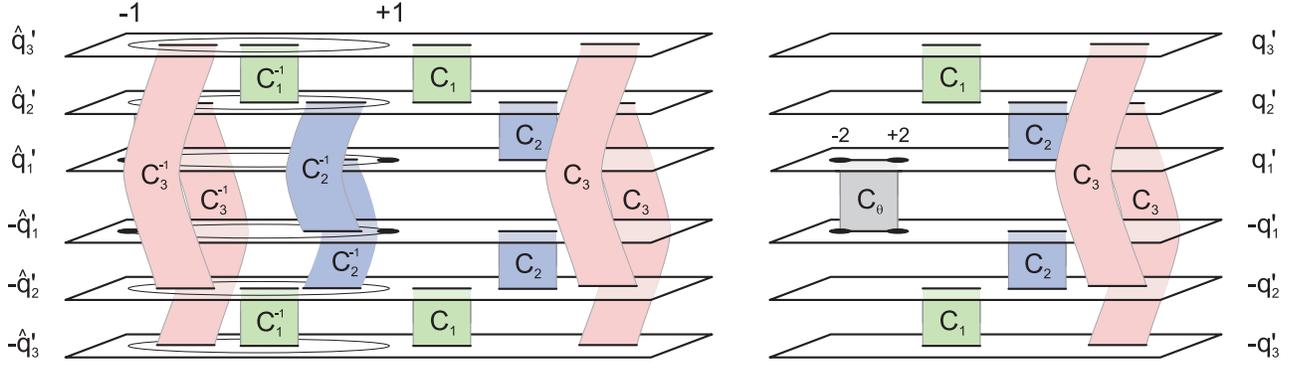}
            \caption{Riemann surfaces resulting from classics and from the
             Bethe ansatz match after Zhukovsky map. On the left  picture
             we plot some possible cuts coming from the classical curves analysis,
             (\ref{qqh}). For each cut outside the unit circle there is a mirror
cut inside the unit circle, (\ref{propq3}). In this picture we did
not plot some possible cuts, for sake of simplicity. There is no
cut, for instance connecting $q_1$ and $q_3$. On the right picture
we present the curve obtained from the Bethe ansatz point of view,
(\ref{BAESO}). Here it might seem that we have plotted every
possible type of cuts. It is not the case -- due to the existence of
stacks \cite{Beisert:2005di} roots of different kinds can attract
each other forming some extra cuts. This extra cuts would be the
image, under Zhukovsky map, of the cuts which we did not plot on the
left figure.}
    \label{fig:BKS}
\end{figure}

The curve in vector representation is the most simple one.
Introducing quasi-momenta
\begin{eqnarray}
\nn q_1&=&G_\theta-G_2 \\
\la{CvPedro}q_2&=&G_2-G_3-G_1  \\
\nn q_3&=&G_1-G_3
\end{eqnarray}
one can easily see from \eq{BAE_SO6} that $(q'_{1,2,3},-q'_{1,2,3})$
are branches of the same function. Riemann surface corresponding to
this function is depicted in fig.\ref{fig:BKS}, to the right. In
terms of these quasi-momenta \eq{BAE_SO6} takes the form \beqa
\nn \sq_1-\sq_2&=&\;\;\,2\pi n^{(2)}\\
\la{BAESO}\sq_3-\sq_2&=&\;\;\,2\pi n^{(1)},\;\;\;\;\;\;\;\;\;\;\sq_1=2\pi m\\
\nn \sq_2+\sq_3&=&-2\pi n^{(3)} \eeqa

Let us see how this curve appears from the results of
\cite{Beisert:2004ag}, reviewed in Section \ref{AnPrSO6}. There one
has the curve
$(\hat{q}'_1,\hat{q}'_2,\hat{q}'_3,-\hat{q}'_1,-\hat{q}'_2,-\hat{q}'_3)$
with the properties (\ref{singinf},\ref{singpm1},\ref{x1/x}),
\begin{eqnarray}
\hat{q}_k(x)&=&\delta_{k=1}\frac{2\pi\,\kappa_{\pm}}{x\mp 1}+\mathcal{O}(1)\,,  \label{propq1}\\
\hat{q}_k(x)&=&\frac{4\pi\,J_k}{\sqrt{\lambda}}\frac{1}{x}+\mathcal{O}(1/x^2)\,, \label{propq2}\\
\hat{q}_k(1/x)&=&\delta_{k,1}\(4\pi m-\hat{q}_1(x)\) +
(1-\delta_{k,1}) \hat{q}_k(x)\,. \label{propq3}
\end{eqnarray}
Charges $J_i$ are properly normalized to be integers.

Let us check that Z-transformation maps this classical curve to the
curve (\ref{CvPedro}) corresponding to a solution of BAE
(\ref{qBAE_SO6}). Namely\footnote{We notice that this is the only
possible normalization -- (\ref{BAESO}) and (\ref{qqh}) forbid us to
multiply $q_k$ by any constant coefficient.}
\begin{eqnarray}
\la{defqq}q_k(z)&=&\hat{q}_k(x_+)\;\;\;\;\;x_+=\frac{1}{2}\(z+\sqrt{z^2-4}\)\,.
\end{eqnarray}
Where the square root is defined in such a way that $|x_+|>1$ for
$z\in \mathbb C$. Since the values of $\hat q_i$ inside the unit
circle are related with the values of $\hat q_i$ outside unit circle
by \eq{propq3} we can always reconstruct $\hat q(x)$ in the whole
$x$ plane from $q(z)$. From (\ref{propq1},\ref{propq3}), one has,
for small $\epsilon$,
\begin{eqnarray}
\hat{q}_1(1+\epsilon)&=&2\pi\,m+\frac{2\pi\,\kappa_+}{\epsilon}+\sum_{k=1}^{\infty} C^{(1)}_k \epsilon^{2k-1} \la{qepsilon1}\\
\hat{q}_{2,3}(1+\epsilon)&=&\sum_{k=0}^{\infty} C^{(2,3)}_k
\epsilon^{2k} \la{qepsilon2}
\end{eqnarray}
Since, after Zhukovsky map,
$q_k(2+\epsilon)=\hat{q}_k(1+\sqrt{\epsilon})$, one sees that a new
cut appears connecting $\pm\,{q}'_1$ while no singularity appears
for $q'_{2,3}$ (see fig.\ref{fig:BKS}). This is consistent with
\eq{CvPedro} where only $q_1$ has $\theta$ cut.

Let us now understand this Z-map in greater detail. We start with six sheets $(\pm
\hat q'_{1,2,3})$. Then we apply Zhukovsky map by means of which
each of these sheets is mapped to a new pair of sheets, one coming
from points located originally inside the unit circle while the
other is the map of the points in the exterior. We designate this
set of sheets by $(\pm q'^{\,in}_{1,2,3},\pm q'^{\,out}_{1,2,3})$.
Now, from (\ref{propq3}), we know that,
\begin{eqnarray*}
(\mp q'^{\,out}_1, \mp q'^{\,out}_{2,3})= (\pm q'^{\,in}_1 , \mp
q'^{\,in}_{2,3} )\,.
\end{eqnarray*}
Thus we can keep only the sheets in the left hand side. Their cut
structure, figure \ref{fig:BKS}, is inherited from the original ones
with an additional cut between $\pm q'_1$ as explained above. It is
exactly the structure we find from the BAE point of view
(\ref{CvPedro}).

A small remark is in order here. From (\ref{qqh}) we see that
classical curve can have more cuts than are listed in \eq{BAESO} and
depicted  in fig. \ref{fig:BKS}. For example $\hat q'_1$ and $\hat
q'_3$ can be connected by a cut. This apparent discrepancy is solved
by the introduction of stacks \cite{Beisert:2005di}. In the
thermodynamical limit Bethe roots of different types can group near
the same extremum and  form cuts. To describe them one can also
define resolvents. As a result, quasi-momenta \eq{CvPedro} can have
cuts connecting for example $q_1$ and $q_3$. For more details see
\cite{Beisert:2005di}, Section 4.

It is now clear, that the equations in the first row in
(\ref{BAESO}) follow from (\ref{qqh}) using the map (\ref{defqq}).
It is also the case for the equation to the right row. Indeed, from
\eq{propq3}, one has
\beq 2\sq_1(z)=q_1(z+i0)+q_1(z-i0)=\hat
q_1(x_+[z+i0])+\hat q_1(x_+[z-i0])=4\pi m \eeq
 where for the last
equality we use $x_+(z+i0)=1/x_+(z-i0)$. Other equalities are just
restatements of our definitions.

From (\ref{qepsilon1},\ref{qepsilon2}) and (\ref{CvPedro}) we see
that \beq \la{rhoth2}\rho_\theta\equiv -\frac{1}{2\pi
i}\(G_\theta(z+i0)-G_\theta(z-i0)\)\simeq
\frac{2\kappa_\pm}{\sqrt{2\mp z}},\;\;\;\;z\rightarrow \pm 2 \eeq

Now, having related $q_{1,2,3}(z)$ and $\hat q_{1,2,3}(x)$ we can
relate $\sqrt{\lambda}$ and $M$. Namely from the definitions
(\ref{CvPedro},\ref{defGG}) we immediately see that, for large $z$,
\beqa
\nn q_1(z)&\simeq& \frac{1}{z}\,\frac{L-K_2}{M}\,,\\
q_2(z)&\simeq& \frac{1}{z}\,\frac{K_2-K_3-K_1}{M}\,,\\
\nn q_3(z)&\simeq& \frac{1}{z}\,\frac{K_1-K_3}{M}\,. \eeqa Comparing
with (\ref{propq2},\ref{J2K}) we obtain
$M=\frac{\sqrt\lambda}{4\pi}$ or, since $M=-\frac{\log\mu}{\pi}$,
\beq \la{SO6bta} -\log\mu=\frac{\sqrt{\lambda}}{4} \,. \eeq This
result matches perfectly the 1-loop $\beta$-function of $O(6)$
sigma model (see footnote in Section \ref{sec4}). It is a strong
confirmation of our correspondence between the quantum Bethe ansatz
\eq{qBAE_SO6} in the scaling limit and the classical sigma model.

\subsubsection{Curve for the Spinor Representation}
In this subsection we will construct the curve associated with the
spinor representation. Having already seen the mechanism at work in
the previous section we will now proceed rather briefly. An 8-sheet
surface can be build by introducing quasi-momenta
\beqa
\nn p_1&=&\;\;\,\frac{1}{2}\,G_\theta-G_1\\
\la{SO6crv}p_2&=&\;\;\,\frac{1}{2}\,G_\theta+G_1-G_2\\
\nn p_3&=&-\frac{1}{2}\,G_\theta- G_3+G_2\\
\nn p_4&=&-\frac{1}{2}\,G_\theta+G_3  \,. \eeqa
It is then easy to see that $(p'_{1,2,3,4},-p'_{1,2,3,4})$ are
branches of the same analytical function (see fig.\ref{fig:so6}).
Indeed \beqa
\nn\left. p_1'-{p'_2}\right|_{C_1}&=&-2\sG'_1+G'_2=0\\
\nn\left. {p'_1}+{p'_2}\right|_{C_\theta}&=&\;\;\;\sG'_\theta-G'_1=0\\
\la{KoCurve}\left. {p'_2}-{p'_3}\right|_{C_2}&=&-2\sG'_2+G'_\theta+G'_1+G'_3=0\\
\nn\left. {p'_3}+{p'_4}\right|_{C_\theta}&=&-\sG'_\theta+ G'_3=0\\
\nn\left. {p'_3}-{p'_4}\right|_{C_3}&=&-2\sG'_3+G'_2=0 \eeqa

\begin{figure}[t]
    \centering
        \includegraphics[scale=1]{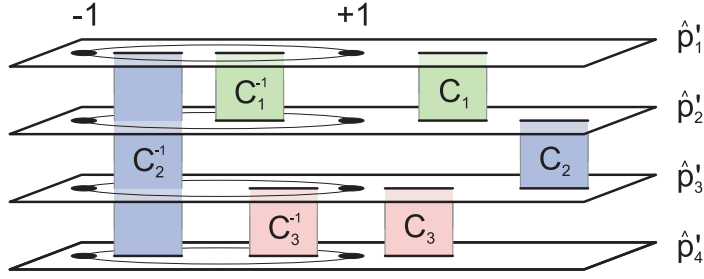}
            \caption{Classical curve from \cite{Beisert:2004ag}. For each cut outside the unit circle there is a mirror cut inside the unit circle -- eq.(\ref{vec_sym}). In each sheet there are poles at $\pm 1$ which will lead to $\theta$ cuts after Z-transformation -- see fig.\ref{fig:so6}.}
    \label{fig:S06bks}
\end{figure}

\begin{figure}[t]
    \centering
        \includegraphics[scale=1]{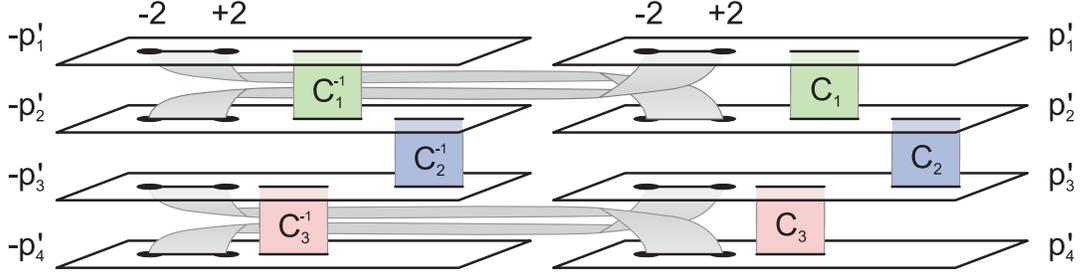}
            \caption{Curve corresponding to the classical limit of the quantum Bethe ansatz. It is related by Z-transformation with the curve on fig.\ref{fig:S06bks}. }
    \label{fig:so6}
\end{figure}

This curve corresponds to the classical curve in spinor
representation \cite{Beisert:2004ag}. As in the previous subsection
they are related by Z-transformation (see fig.\ref{fig:S06bks}).
After Zhukovsky transformation 4 sheets split into 8. Since there
are poles at $\pm 1$ in each of the original 4 sheets the resulting
8 sheets will be pairwise connected by 4 $\theta$ cuts as shown in
fig.\ref{fig:so6}.

Let us use the spinor curves to relate $M$ and $\sqrt\lambda$ as in
previous subsection. The vector representation (of the previous
subsection) and the spinor representation are related by
\cite{Beisert:2004ag}
\begin{eqnarray*}
p_1&=&\frac{q_1+q_2-q_3}{2}  \,,\,\,\,\,\,\,\,\,\,\,\,\, p_2\,\,=\,\,\frac{q_1-q_2+q_3}{2}\\
p_3&=&\frac{-q_1+q_2+q_3}{2} \,,\,\,\,\,\,\,\,
p_4\,\,=\,\,\frac{-q_1-q_2-q_3}{2}\,,
\end{eqnarray*}
with the same expression for hatted quantities. Then the properties
for the $\hat{q}$'s lead to \beqa \hat p_1(1/\epsilon)&\simeq
\la{vec_asy}&\frac{2\pi\epsilon}{\sqrt{\lambda}}(J_1+J_2-J_3),\;\;\;\;\;\;\;\;\;\;
\hat p_2(1/\epsilon)\;\;\simeq \;\;\,\frac{2\pi\epsilon}{\sqrt{\lambda}}(J_1+J_3-J_2)\\
\hat p_3(1/\epsilon)&\simeq
&\frac{2\pi\epsilon}{\sqrt{\lambda}}(J_2+J_3-J_1),\;\;\;\;\;\;\;\;\;\;
\hat p_4(1/\epsilon)\;\;\,\simeq
\nn-\frac{2\pi\epsilon}{\sqrt{\lambda}}(J_1+J_2+J_3) \eeqa
and
\begin{eqnarray}
\shp_k\mp \shp_l=2\pi\,n_a\,\,\, , \,\,\, x\in \mathcal{C}_a\,.
\la{BAE_clas_SO6}
\end{eqnarray}
Furthermore $x\leftrightarrow 1/x$ relates different sheets of the
surface \beq \hat p_{1,2}(1/x)=2\pi m - \la{vec_sym}\hat
p_{2,1}(x),\;\;\;\;\;\hat p_{3,4}(1/x)=-2\pi m - \hat p_{4,3}(x)
\eeq

Again we state the following relation
\beq p_{1,2,3,4}(z)=\hat
p_{1,2,3,4}\(\frac{1}{2}\[z+\sqrt{z^2-4}\]\) \nn \eeq
Now we compute $p_i$ on $C_\theta$. From \eq{vec_sym} we have
equations for $\theta$-cut \beq \la{mSO6}\sp_1+\sp_2=2\pi
m,\;\;\;\;\;\sp_3+\sp_4=-2\pi m \,.\eeq Among the equations
(\ref{BAE_clas_SO6}) one has
\begin{eqnarray*}
\shp_2-\shp_1=2\pi n^{(1)},\;\;\;\;\;x\in C_1\\
\shp_3-\shp_2=2\pi n^{(2)},\;\;\;\;\;x\in C_2\\
\shp_4-\shp_3=2\pi n^{(3)},\;\;\;\;\;x\in C_3
\end{eqnarray*}
which give us \eq{BAE_SO6}. Remaining equations are seen in the
Bethe ansatz side through the introduction of stacks  (see
discussion in previous subsection.). Finally from (\ref{SO6crv}) and
(\ref{vec_asy}) we again obtain (\ref{SO6bta}).

\subsection{Energy and Momentum}
In this subsection let us observe the perfect matching of both
energy and momentum from classical and Bethe ansatz calculations.
Energy is again defined by
\beq
E=\frac{\mu}{2\pi}\sum_\alpha\cosh\(\frac{\pi\theta_\alpha}{2}\) \,.
\eeq
Repeating the same calculations as for $SO(4)$ case we see that once
again the result depends on the behaviour of $\rho_\theta(z)$ near
points $z=\pm 2$, \eq{rhoth2}. In fact everything goes exactly as
before so that we will arrive at the equivalent of (\ref{energy}),
namely \beq E=2\pi M\kappa_-^2+\(-\sum_{a=1,2,3}\sum_p
n_p^{(a)}S_p^{(a)}-m L \)\,. \nn \eeq Let us recall that this was
obtained by calculating the $a$ independent integral for
$a=-2+\epsilon$. Had we calculated it for $a=2-\epsilon$ and we
would have obtained \beq E=2\pi M\kappa_+^2-\(-\sum_{a=1,2,3}\sum_p
n_p^{(a)}S_p^{(a)}-m L \)\,. \nn \eeq Both results should be equal.
This gives us not only the value of the energy
\beq E=\pi
M\(\kappa_-^2+\kappa_+^2\)\,, \la{enso6} \eeq
 but also the value of
the expression in the parentheses which is nothing but the total
momentum $P$ obtained as before by summing Bethe equations
\beq
P=\frac{\mu}{2\pi}\sum_\alpha\sinh(\pi\theta_\alpha)=-\sum_{a=1,2,3}\sum_p
n_p^{(a)}S_p^{(a)}-m L=\pi M\(\kappa_+^2-\kappa_-^2\)\,. \la{momso6}
\eeq
 Since we have already identified
$M=\frac{\sqrt{\lambda}}{4\pi}$ we observe perfect matching between
(\ref{enso6},\ref{momso6}) and the classics (\ref{ENMOMCL}).

%%%%%%%%%%%%%%%%%%%%%%%%%%%%%%%%%%%%%%%

\section{Conclusions and Prospects}

The  $O(2n)$  $\sigma$-models considered in this paper cannot be
viewed of course as  string sigma models in its full quantum
version, although the $O(6)$ $\sigma$-model in the classical scaling
limit perfectly describes the compact bosonic sector of classical
superstring on $AdS_5\times S^5$, when the Virasoro conditions are
imposed. We rather consider  these quantum theories as  toy models
bearing many realistic features of a quantum string sigma model and
shedding some light on the quantization of the full
Green-Schwarz-Metsaev-Tseytlin superstring. These models show some
important features of, yet unknown in full detail, string Bethe
equations: the nested nature of these equations, following from the
symmetry algebra of the target space, the underlying discrete
quantum degrees of freedom whose role is played in our case by
physical particles and by the magnon type excitations on a dynamical
inhomogeneous lattice created by particles etc.

Let us point out that the starting point of our construction, the
physical S-matrix, can be viewed as describing the
anti-ferromagnetic states of a specific inhomogeneous lattice spin
chain, as pointed out in \cite{Faddeev:1985qu}\footnote{ We thank
D.~Serban for the discussion on this issue.}, what might be related
to the proposals of \cite{Polyakov:2005ss}. The resulting
inhomogeneous dynamical spin chain of our paper is however treated
in the ferromagnetic way in order to restore the classical limit.

This dynamical lattice becomes regular and rigid in the strong
coupling limit of $\sigma$-model giving the XXX-type  spin chain,
similar (up to some normalization factors) to the spin chain
describing the 1-loop ${\cal N}=1$ SYM dilatation operator. This
observation might  lead to the right mechanism reproducing the
all-loop SYM Bethe ansatz of the papers
\cite{Beisert:2005fw,Beisert:2004hm,Rej:2005qt}, including the
periodicity with respect to  momentum due to the regular lattice.
The Hubbard model interpretation of the paper \cite{Rej:2005qt}
should be somehow incorporated in the full quantum string.

The asymptotic freedom of the quantum $O(2n)$ models  cannot be a
part of this full string $\sigma$-model on $AdS_5\times S^5$. It
should be rather a finite theory, with zero beta function and
without any logarithmic divergencies in the world sheet perturbation
theory.

What could this string sigma model look like? We find the
construction of quantum string states out of the physical particles
based on the Yang periodicity equation of \eq{MBAE} very promising.
Of course there can be no massive particles there, and the
$\theta$-filling of the vacuum described here and in
\cite{Mann:2005ab} should be based not on the confining box
potential  but on a subtle equilibrium between all  roots of the
Bethe ansatz equations.

It is plausible to think that the  system of Bethe equations for the
full theory is based on the $\alg{psu}(2,2|4)$ algebra and may look
like an affine extension of it, as we saw in the bosonic cases. Even
though the scattering of particles in the critical string theory
could be qualitatively different from that of massive particles, we
still  expect that such Bethe ansatz equations may serve as a
quasi-particle description \cite{Kedem:1992jv,Kuniba:1992sp} and
provide us with a basis for the Fock-space like construction of the
quantum Hilbert space.

Apart from the string theory applications, we think that our method
of restoring the classical limit of integrable sigma models is
interesting in itself and can be applied to many problems of
two-dimensional field theories, especially for the construction of
the quasi-classical approximation. It is also useful for proving the
validity of more heuristic approaches to such field theories, like
of the Zamolodchikovs' bootstrap S-matrices. In particular, we
consider the coincidence  of our classical limit in the Bethe ansatz
with the finite gap solution of the $O(2n)$ sigma models as a
decisive proof of correctness of the bootstrap approach.

Another interesting possibility based on our observations is the
construction of  quantization schemes of integrable models starting
from the algebraic curves of their finite gap solutions
(quantization of KdV system in \cite{Smirnov}  provides an
interesting example). To quantize the algebraic curve, we have to
choose its particular projection (Riemann surface) and introduce the
Bethe roots which form the cuts of the Riemann surface in the
classical limit. The quantization might be possible in different
projections, but some of them should be more convenient than the
others. We saw for example from our construction that  quantization
of the finite gap algebraic curve of  $O(2n)$ sigma model looks more
natural after a change of projection by Zhukovsky map.

An interesting application of our methods might be the calculation
of quantum $1/L$ corrections in the compact sector of  superstring.
It was pointed out in \cite{Beisert:2005mq} that the direct string
calculations of \cite{Frolov:2002av}  of this effects can be
reproduced from the compact sector only by the use of a natural
$\zeta$-functional regularization, without a direct computation of
fermionic contributions. Our Bethe ansatz is a good starting point
for such calculation.

%%%%%%%%%%%%%%%%%%%%%%%%%%%%%%%%%%%%%%%%%%%%%%%%%%%%%%%%%%
\subsection*{Acknowledgements}

We would like to thank   I.~Kostov, J.~Penedones, D.~Serban,
J.~Troost, Al.~Zamolodchikov and especially F.~Smirnov and K.~Zarembo for
discussions. The work of V.K. was partially supported by European
Union under the RTN contracts MRTN-CT-2004-512194 and by
INTAS-03-51-5460 grant. The work of N.G. was partially supported by
French Government PhD fellowship and by RSGSS-1124.2003.2. P.~V. is
supported by the Funda\c{c}\~ao para a Ci\^encia e Tecnologia
fellowship SFRH/BD/17959/2004/0WA9.

%%%%%%%%%%%%%%%%%%%%%%%%%%%%%%%%%%%%%%%%%%%%%%%%%%%%%%%%%%
\subsection*{Note Added:}

We thank  T.~Klose and K.~Zarembo for  informing us on their
forthcoming work \cite{KloseZarembo} where they also developed an
approach to the quantization of stringy sigma models in various
subsectors of the superstring theory on $AdS_5\times S^5$. Their
approach seems to be very different from ours, although both give
the right classical string limit. Hopefully, both quantizations
describe the same system and provide an important complementary
information on it.

\setcounter{section}{0}

%%%%%%%%%%%%%%%%%%%%%%%%%%%%%%%%%%%%%%%%

\appendix{Charged Particles in a Box}

Let us analyse Bethe equations \eq{BAEnoUV} with no $u$'s nor $v$'s
present. Furthermore we will considering
 a single mode number for $\theta$'s (which in the continuous
 limit means only one $\theta$ cut).

Has we explained in section 3, the $\mu\cosh M\pi x$ potential is
essentially a box when $\mu=e^{-\frac{M}{2\pi}}$ is small, see
 figure \ref{fig:box}. Furthermore, provided we are considering
 $\theta$ roots not to close to the edges of the distribution,
 i.e. the walls of the box, coulomb interaction in the
 thermodynamical limit, where $L\sim M \rightarrow\infty$, becomes exact.

To gain some intuition about this system we will find exact
distribution of the system of Coulomb particles inside square box.
For the sake of generality we consider a box will charged walls with
charge
 $q\,M$. Then the equilibrium
condition is
\begin{eqnarray*}
\frac{1}{M}\sum_{j\neq i}\frac{1}{x_i-x_j}=2q\frac{x_i}{4-x_i^2}\,.
\end{eqnarray*}
%For actual problem there is a factor of $1/L$ is the lhs.
%Furthermore $q$ is just a regulator, i.e. the amount of charge on
%what we call \textit{wall} one should kill at the end. Hence the the
%$1/L$ is implicit in the definition of $q$.

The solutions of this discrete equation, i.e. the positions of the
particles are the zeros of the Jacobi polynomials \cite{Shastry}.
Indeed, defining $Q(x)=\prod_j(x-x_i)$, one has
\begin{eqnarray*}
\frac{Q''(x_i)}{Q'(x_i)}=2\sum_{j\neq i}\frac{1}{x_i-x_j}=4q
M\frac{x_i}{4-x_i^2}\,,
\end{eqnarray*}
which implies that the $L$ degree polynomial
$R(x)=Q''(x)(4-x^2)-4q\,M\,x\, Q'(x)$ has the same zeros as $Q(x)$
so they must be the same up to a multiplicative constant. Comparing
the $x^L$ coefficient one finds $R(x)=(-L\,(L-1)-4\,q\,M\,L)\,Q(x)$
so that
\begin{eqnarray*}
Q''(x)(4-x^2)-4q\,M\,x\, Q'(x)+(L(L-1)+4q\,M\,L)Q(x)=0\,.
\end{eqnarray*}
Thus
\begin{eqnarray*}
Q(x)\propto\,P_L^{2q-1,2q-1}\(\frac{x}{2}\)
\end{eqnarray*}
where $P_{n}^{a,b}(z)$ are the Jacobi polynomials. Particles will be
located at the zeros of this polynomials.

Let us analyse the large $M,L$ limit. From the differential equation
for $Q$ one has, for the resolvent $G=\frac{1}{M}\frac{Q'}{Q}$,
\begin{eqnarray*}
\frac{1}{M}\frac{dG(x)}{dx}=-G^2(x)+4\,q\,\frac{x}{4-x^2}G(x)-\left(\frac{L^2}{M^2}+\frac{4q\,L}{M}-\frac{L}{M^2}\right)\frac{1}{4-x^2}\,.
\end{eqnarray*}
Then in the large $L\sim M$ limit
\begin{eqnarray}
G(x)=\frac{2\,q\,x-2\sqrt{\(2q+\frac{L}{M}\)^2x^2-4\(\frac{L^2}{M^2}+\frac{4\,q\,L}{M}\)}}{4-x^2}\,.
\label{Gq}
\end{eqnarray}
Thus, for small $q$, the density has $2$ peaked maxima near the walls at $\pm 2$. It vanishes before reaching the wall as a square root. In the limiting case, when $q\rightarrow 0$, the support goes to [-2,2] as expected and the density becomes
\begin{eqnarray*}
\rho(x)\rightarrow \frac{1}{\pi\sqrt{4-x^2}}\,.
\end{eqnarray*}
which reproduces \eq{DENTH} for $m=0$ in proper normalization.

We considered this exactly solvable example to demonstrate that
influence of the walls results in the inverse square root behavior
of the density near the walls.

\appendix{Relating resolvents}

Equivalence between (\ref{BAESU2}) and (\ref{BAESU2cl}) will be estabilished in this
section. We define resolvents in the $x$ and $z$ planes as \beq
H_\pm(x)\equiv \int_{C_\pm}\frac{\rho[y]}{x-y}dy,\;\;\;\;\;
G_\pm(z)\equiv \int_{C_\pm}\frac{\rho[y_\pm(w)]}{z-w}dw,\;\;\;\;\;
y_\pm(w)=\frac{1}{2}\(w\pm\sqrt{w^2-4}\) \eeq where $C_{\pm}$ are cuts
outside (inside) the unit circle in the $x$ plane. Since
\beq
G_\pm(z)=\int_{C_\pm}\frac{\rho[y]}{z-w(y)}w'(y)dy=
\int_{C_\pm}\rho(y)\(\frac{1}{x(z)-y}+\frac{1}{y}+\frac{1}{\frac{1}{x(z)}-y}\)dy
\eeq
we have
\beq
G_\pm(z)=H_\pm[x_\pm(z)]+H_\pm\[x_\mp(z)\]-H_\pm(0) \,. \la{G2G}
\eeq

\subsubsection*{Derivation of BAE from Classical Equations}
It is natural, see (\ref{G2G}), to define
\beqa
\nn G_v(z)&\equiv&H_+(x_+)+H_+(x_-)-H_+(0)\,,\\
\la{otherlabel} -G_u(z)&\equiv&H_-(x_+)+H_-(x_-)-H_-(0)\,,\\
\nn G_\theta(z)&\equiv&\frac{4\pi\kappa}{\sqrt{z^2-4}}+2H_+(x_-)-2H_-(x_+)-2H_+(0)\,.
\eeqa
These functions are indeed resolvents, i.e.
they have only cuts and behave as $1/z$ at infinity. Indeed from
$x_+(\infty)=\infty,\;x_-(\infty)=0$ we see that they vanish
for large $z$. Now, $G_v$ ($G_u$) have only $v$ ($u$) cuts since at the cut $[-2,2]$ one has $x_+\leftrightarrow x_-$. On the other hand $G_\th(z)$ has only got $\th$ cut.

Then, it is easy to see that (\ref{BAESU2}) is satisfied if
\beq
\la{BAESU2cl}\sH(x)=\frac{\pi\kappa}{x-1}+\frac{\pi\kappa}{x+1}+\pi
n,\;\;\;\;\;H(0)=2\pi m
\eeq

\subsubsection*{\la{KMMZ2BAE}Derivation of Classical Equations from BAE}
Now let us assume that \eq{BAESU2} is satisfied and let us show that
$H(x)$ defined by \beqa
H(x_+)&\equiv& \;\;\,G_v(z)-\frac{G_\theta(z)}{2}+\frac{2\pi\kappa}{\sqrt{z^2-4}}\\
H(x_-)&\equiv&
-G_u(z)+\frac{G_\theta(z)}{2}-\frac{2\pi\kappa}{\sqrt{z^2-4}}+2\pi m
\eeqa satisfies \eq{BAESU2cl}. To see this in suffices to notice that
$\pm\frac{2}{\sqrt{z^2-4}}=\frac{1}{x_\pm-1}+\frac{1}{x_\pm+1}$.
Moreover since $x_-=0$ means $z=\infty$ we immediately have
$H(0)=2\pi m$. Since $\sG_\theta-G_u-G_v=-2\pi m$, $H(x)$
is continuous on the unit circle and is thus a well defined function of $x$. Furthermore it is obviously zero at infinity.

%%%%%%%%%%%%%%%%%%%%%%%%%%%%%%%%%%%%%%%%%%%%%%%%%%%%%%%%%%%%%%%%%%%%%%%%%%%%
\appendix{Bootstrap and Bethe Ansatz}

To make the account of this paper self-contained, we decided to
review in this appendix  the original bootstrap program of
\cite{Zamolodchikov:1977nu} for the $SO(4)=SU(2)\times SU(2)$
non-linear sigma model. We follow closely, apart from some minor
changes, the original notation. Furthermore we will use the obtained
$S$--matrix to carry the algebraic Bethe ansatz procedure and
quantize the sigma model \cite{Zamolodchikov:1992zr}.

Particles are associated to the noncommutative symbols $A_i(\theta)$
where $i=1,\dots,4$ stands for the isovector index and the rapidity
$\theta$ parametrizes the momentum as $p=m_0\,\cosh(\pi\theta)$.
``In" states are identified with products arranged in the order or
decreasing rapidity while ``out" states should be arranged in the
opposite order. The transition from a $2$ particle ``in"--state to an
``out"--state (assuming no particle creation, which is indeed a
condition for integrability) is written as
\begin{eqnarray}
A_i(\theta)A_j(\theta')=S_2(\Delta\theta)\!\(\!g^{-1}(\Delta\theta)
\delta_{ij}\sum_{k=1}^{4}A_k(\theta')A_k(\theta)+A_j(\theta')A_i(\theta)+h^{-1}(\Delta
\theta)A_i(\theta')A_j(\theta)\!\) \la{AA}
\end{eqnarray}
where $\Delta\th=\theta'-\theta$. In this expression the first term
represents annihilation, the second one is the transmission
contribution and the third one the reflection term.
\begin{figure}[t]
    \centering
        \includegraphics[scale=0.8]{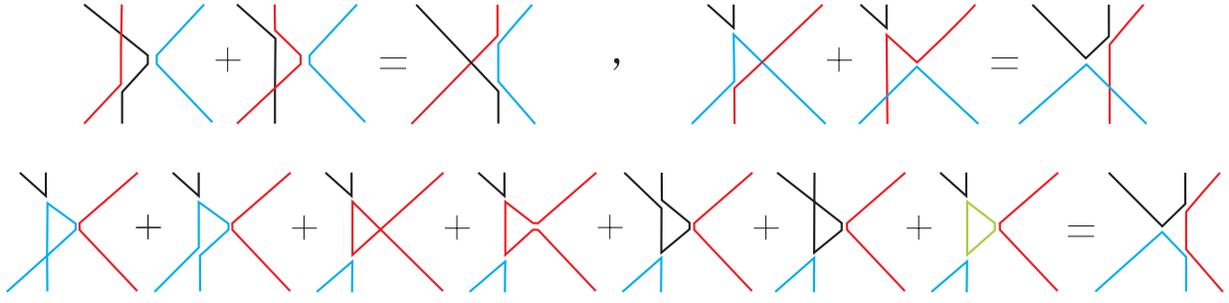}
            \caption{Pictorial representation of Yang Baxter equation.
            In each diagram time flows in the up direction and initial
            particles are aligned, from left to right, according to its
            rapidity (the faster the more to the left).
Consider for instance the last diagram in the lhs of the last
equation. It represents the annihilation of two blue particles to
form any possible pair of green particles. One of the particles of
this pair will be reflected back by a red particle to meet again its
green twin and annihilate to give rise to a pair of black particles.
The fastest ``in"--particles are the blue ones while the slowest
``out"--particles are the black ones.}
    \label{fig:YBaxter}
\end{figure}
In general, the transition of the ``in"-state
\begin{eqnarray*}
A_i(\theta_1)\,A_j(\theta_2)\,A_k(\theta_3)
\end{eqnarray*}
to the ``out"-state is not so simple. However, for integrable
theories where one has a large quantity of conserved charges it
happens that the existence of these charges imposes that, in every
region where the particles are far away from each other, the wave
function is a sum of plane waves where in each region the set of momenta $\{p_i\}$ is
the same. Then for these
integrable theories we can obtain the ``out"--state by applying
(\ref{AA}) consecutively to the pairs in this triplet until the
$A$'s are ordered by decreasing rapidities. This is called
factorizability. Now, consistency requires that the final result
should not depend on whether we arrive at region
$\theta_3<\theta_2<\theta_1$ by scattering first $A_j$ at first with
either $A_k$ or $A_i$. By doing it in both ways and equating both
results one obtains for each choice of isospin indexes some
constraint on the functions in (\ref{AA}). These equations can be
obtained in a straightforward fashion and have a nice physical
interpretation -- see caption in figure \ref{fig:YBaxter} -- all
other possible choices of isovector indexes (i.e. colors) give rise
either to the same constraints or to some trivial identities where
the lhs and the rhs are already identically equal. One obtains:
\begin{eqnarray}
h(\theta)+h(\theta')&=&h(\theta+\theta') \la{hs} \\
h(\theta)+g(\theta+\theta')&=&g(\theta') \nn
\end{eqnarray}
plus another similar, yet bigger, equation corresponding to the
third identity if figure \ref{fig:YBaxter}. The solution of the 2
equations yield $h(\theta)$ and $g(\theta)$ in terms of two
constants $\lambda$ and $\kappa$. When plugging these solutions into
the third equation one obtains $\kappa(\lambda)$. At the end of the
day one has
\begin{eqnarray}
h(\theta)=g(i\lambda-\theta)=\frac{i\theta}{\lambda}\,. \la{handg}
\end{eqnarray}
Thus, out of 3 functions we are now left with one function and one
constant. From (\ref{AA}) one reads off the S--matrix
\begin{eqnarray}
S_{ik}^{j\,l}(\theta)=
S_2(\theta)\left[\,\delta_{ik}\delta_{jl}\,g^{-1}(\theta)+\delta_{ij}\delta_{lk}+\delta_{il}\delta_{jk}\,
  h^{-1}(\theta)\right] \la{SMatrix} % + \(i\leftrightarrow k,p_1 \leftrightarrow p_2\)
\end{eqnarray}
Changing $i\leftrightarrow j$ and the channel
$s=4m_0^2\cosh^2(\pi\theta/2) \leftrightarrow u$ (i.e. $\theta
\rightarrow i-\theta$) should leave the S-matrix invariant. This is
crossing--symmetry. It implies
\begin{eqnarray}
S_2(\theta)=S_2(i-\theta) \la{cross}
\end{eqnarray}
and $h(\theta)=g(i-\theta)$, or $\lambda=1$. Finally we must impose
the most natural requirement, namely, unitarity. Setting
\begin{eqnarray*}
S_{mn}^{j\,l}(-\theta)
S_{ik}^{mn}(\theta)=\delta_{ij}\delta_{kl}\,S_2(-\theta)S_2(\theta)\(1+h^{-1}(-\theta)h^{-1}(\theta)\)+
  \delta_{ik}\delta_{jl}\( \dots \)+\delta_{il}\delta_{kj}\( \dots \)
\end{eqnarray*}
to be equal to $\delta_{ij}\delta_{kl}$ one obtains 3 equations. The
exact expressions inside the parentheses are not relevant for our
discussion. It suffices to say that, for the $g$ and $h$ we found,
(\ref{handg}), they vanish identically. Thus one is left with
\begin{eqnarray}
S_2(-\theta)S_2(\theta)=\frac{\theta^2}{\theta^2+1}\,. \la{unit}
\end{eqnarray}
From (\ref{cross}) and (\ref{unit}) it follows that $S_2(\th)$ is
given by
\begin{eqnarray}
\frac{\theta}{\theta-i}\,S_0^2(\theta) \,\,\, , \,\,\,
S_0(\theta)=i\,\frac{\Gamma\(-\frac{\theta}{2i}\)\Gamma\(\frac{1}{2}
+\frac{\theta}{2i}\)}{\Gamma\(\frac{\theta}{2i}\)\Gamma\(\frac{1}{2}-\frac{\theta}{2i}\)}\,
\la{minimal}
\end{eqnarray}
times a CDD factor
\begin{eqnarray}
f(\theta)=\prod_{k=1}^{L}\frac{\sinh{\pi\theta}+i\sin\alpha_k}{\sinh{\pi\theta}-i\sin\alpha_k}
\la{CDD}
\end{eqnarray}
where $\alpha_k$ are arbitrary real numbers. The form
(\ref{minimal}) is needed to have the right pole and zero structure
according to (\ref{cross}) and (\ref{unit}) while the ambiguity
(\ref{CDD}) is unfixed since
$f(\theta)/f(i-\theta)=f(\theta)\,f(-\theta)=1$\,. Absence of
additional bound states forces one not only to exclude this factor
but also to introduce the $i$ in $S_0(\theta)$
\cite{Karowski:1978ps,Ogievetsky:1984pv}. For the non-linear sigma
model the correct S-matrix is indeed given by the minimal choice
(\ref{minimal}). To verify this claim some convincing cross checks
are done \cite{Zamolodchikov:1977nu,Berg:1978dz}. Nevertheless
notice that for what concerns the classical analysis we did not need
to know anything about this CDD factor since for large $\theta$'s it
is given by $1+\O(e^{-2\pi\th})$ being therefore irrelevant in the
scaling limit. If this CDD factor was present it could, however, in
principle contribute into quasi-classical corrections.

Finally since $SO(4)=SU(2)\times SU(2)$ we can replace $i,k,j,l$ by
$(\a,\dot\a),(\b,\dot\b),(\a',\dot\a'),(\b',\dot\b')$ and write
(\ref{SMatrix}) as
\begin{eqnarray*}
&&\frac{S_0^2(\theta)}{(\theta-i)^2}\left(i\th\,\epsilon_{\a\b}\epsilon_{\dot\a\dot\b}\epsilon^{\a'\b'}\epsilon^{\dot\a'\dot\b'}
+\th(\th-i)\,\delta_{\a}^{\a'}\delta_{\dot\a}^{\dot\a'}\delta_{\b}^{\b'}\delta_{\dot\b}^{\dot\b'}
-i(\th-i)\,\delta_{\a}^{\b'}\delta_{\dot\a}^{\dot\b'}\delta_{\b}^{\a'}\delta_{\dot\b}^{\dot\a'}\right)\\
&=&\frac{S_0^2(\theta)}{(\theta-i)^2} \left( \th\,
\delta_{\a}^{\a'}\delta_{\b}^{\b'}-i
\delta_{\a}^{\b'}\delta_{\b}^{\a'} \right) \left(
\th\,\delta_{\dot\a}^{\dot\a'}\delta_{\dot\b}^{\dot\b'}-i\delta_{\dot\a}^{\dot\b'}\delta_{\dot\b}^{\dot\a'}
\right)\,.
\end{eqnarray*}
or
\begin{eqnarray}
\hat{S}(\theta)=\hat{S}_R(\theta)\times \hat{S}_L(\theta) \,\,\,,
\,\,\,  \hat{S}_{L,R}(\theta)=S_0(\theta)\,\hat{\mathcal{R}}(\theta)
\,\,\, , \,\,\,  \hat{\mathcal{R}}=\frac{\th \hat I- i\hat
P}{\theta-i} \label{SinSU(2)}
\end{eqnarray}
where $\hat P$ is the permutation operator in ${\mathbb C}^2\times
{\mathbb C}^2$. Notice that, since $SO(4)=SU(2)\times SU(2)$, we
could have started with this ansatz where
\begin{eqnarray*}
\hat{S}_{L,R}(\theta)=\hat{S}_0(\theta)\,\frac{\th}{\theta-i}\(\hat{I}-
h^{-1}(\th) \,\hat{P}\)\,.
\end{eqnarray*}
instead of starting with (\ref{AA}). The absence of the channel
where two particles annihilate is natural since the particles are
charged with left/right charge. Consistency relation would be
represented by the first equation in figure \ref{fig:YBaxter} only,
or equation (\ref{hs}), which states that $h(\th)$ is linear in
$\theta$. Then we would continue by imposing unitarity and
crossing-symmetry to arrive at the same results we did for the form
of $S_0$ and the proportionality constant in $h(\th)$. The results
of this appendix could be easily generalized to both $O(N)$ and
$SU(N)_L\times SU(N)_R$ non-linear sigma models or to many other
groups.

Let us now start from the obtained S-matrix (\ref{SinSU(2)}) and
carry out the algebraic Bethe ansatz program
\cite{Zamolodchikov:1992zr}. We introduce a ghost particle in the
auxiliary space $0$ and scatter it through the other particles. We
want to diagonalize
\begin{eqnarray*}
\Tr \,\hat{T}(\theta) \,\,\, , \,\,\,
\hat{T}(\theta)=\prod_{k=1}^{L}\hat{S}_{\,0k}(\theta-\theta_k)
\end{eqnarray*}
where we trace over the auxiliary space. This is a relevant problem
because if we solve it for any $\theta$ and then set
$\theta=\theta_k$ then this means that the ghost particle will
change its quantum numbers with particle $k$ since
$\hat{S}_{\,0k}(0)=-P_{\,0k}\times P_{\,0k}$. In other words,
\begin{eqnarray*}
-\,\Tr
\,\hat{T}(\theta_k)=\hat{S}_{\,k,k-1}(\theta_k-\theta_{k-1})\dots
\hat{S}_{\,k,1}(\theta_k-\theta_{1})\hat{S}_{\,k,N}(\theta_k-\theta_{N})\dots
\hat{S}_{\,k,k+1}(\theta_k-\theta_{k+1})\,
\end{eqnarray*}
so that the problem of imposing periodicity of the wave function
reads
\begin{eqnarray}
-\,e^{im_0\mathcal{L}\sinh(\pi\theta_k)}\,\Tr
\,\hat{T}(\theta_k)\,|\Psi\rangle=|\Psi\rangle\,. \la{periodicity}
\end{eqnarray}
Let us now perform the diagonalization of \ref{periodicity}. We consider
\begin{eqnarray}
|\Psi\rangle =\prod_{i=1}^{J_u} \hat{B}_L(u_i)\prod_{j=1}^{J_v}
\hat{B}_R(v_j) |\,\Omega(\theta_1,\dots,\theta_L)\rangle \la{Psi}
\end{eqnarray}
where $\Omega$ is the state with $L$ particles, where the right and
left spin of every particle is pointing in the up direction, and
\begin{eqnarray}
\hat{T}_R(\theta)=\prod_{k=1}^{L}\hat{S}_{R,\,0k}(\theta-\theta_k)=\(
\begin{array}{cc}
\hat{A}_{R}(\th) & \hat{B}_{R}(\th) \\
\hat{C}_{R}(\th) & \hat{D}_{R}(\th) \end{array}\)
 \la{Tmatrix} %=\prod_{\a=1}^{L}S_0(\th-\theta_\a)\mathcal{R}_{0\a}(\theta-\theta_a)
\end{eqnarray}
with a similar definition for the left sector. Acting on $\Omega$
one has
\begin{eqnarray*}
\mathcal{R}_{0k}(\theta)|\,\Omega\rangle=\frac{1}{\th-i}\(
\begin{array}{cc}
\th-\frac{i}{2}(\tau^3+1) & -\frac{i}{2}\tau^-  \\
-\frac{i}{2}\tau^+ & \th+\frac{i}{2}(\tau^3-1)
\end{array}\)|\,\Omega\rangle=\frac{1}{\th-i}\( \begin{array}{cc}
\(\th-i\)|\,\Omega\rangle & *  \\
0 & \th|\,\Omega\rangle \end{array}\)
\end{eqnarray*}
The upmost right element is not important for our discussion.
However the vanishing of the element in the left down corner is. It
implies that $|\,\Omega\rangle$ is eigenvalue of both $A$ and $D$
with eigenvalues
\begin{eqnarray*}
A(\theta)|\,\Omega\rangle&=&\prod_{\a=1}^{L}S_0(\th-\th_\a) \,|\,\Omega\rangle \\
D(\theta)|\,\Omega\rangle&=&\prod_{\a=1}^{L}S_0(\th-\th_\a)\frac{\th-\th_a}{\th-\th_a-i}
\,|\,\Omega\rangle
\end{eqnarray*}
So now we have to understand how $A$ and $D$ pass through the $B$'s.
Consistency relations can be cast as
$S_{12}(\theta)S_{13}(\theta+\theta')S_{23}(\theta')=S_{23}(\theta')S_{13}(\theta+\theta')S_{12}(\theta)$
which imply (cf. \cite{Faddeev:1996iy})
\begin{eqnarray*}
T^{a}_{R,L}(\theta)\,T^{a'}_{R,L}(\theta')\,\mathcal{S}_{aa'}(\theta'-\theta)=\mathcal{S}_{aa'}(\th'-\th)\,T^{a'}_{R,L}(\theta')\,T^{a}_{R,L}(\theta)
\,.
\end{eqnarray*}
where $a$ and $a'$ are two ${\mathbb C}^2$ auxiliary spaces. This
gives us the commutation relations between the elements of the
transfer matrix (\ref{Tmatrix}). In particular
\begin{eqnarray}
[B(\theta),B(\theta')]&=&0  \nn\\
A(\theta)\,B(\theta')&=&\frac{\theta'-\theta-i}{\theta'-\theta}\,B(\theta')\,A(\theta)+\frac{i}{\theta'-\theta}\,B(\theta)\,A(\theta')\, \label{AB}\\
D(\theta)\,B(\theta')&=&\frac{\theta'-\theta+i}{\theta'-\theta}\,B(\theta')\,D(\theta)-\frac{i}{\theta'-\theta}\,B(\theta)\,D(\theta')\,.
\label{DB}
\end{eqnarray}
for symbols in the same right or left sector. Symbols in different
sectors commute to zero of course. Then, acting on (\ref{Psi}), one
has
\begin{eqnarray*}
-\,\Tr\,\hat{T}(\th)\,|\Psi\rangle&=&(\hat{A}_R(\th)+\hat{D}_R(\th)) \times (\hat{A}_L(\theta)+\hat{D}_L(\th))\prod_{i=1}^{J_u}  \hat{B}_L(u_i) \times\prod_{j=1}^{J_v} \hat{B}_R(v_j) |\,\Omega(\theta_1,\dots,\theta_L)\rangle \\
&=&\prod_{\a=1}^{L}S_0^2(\th-\th_a)\(
\prod_{i=1}^{J_u}\frac{u_i-\th-i}{u_i-\th}+\prod_{i=1}^{J_u}\frac{u_i-\th+i}{u_i-\th}\prod_{\a=1}^{L}\frac{\th-\th_a}{\th-\th_\a-i}\)
\\&\times& \(
\prod_{i=1}^{J_v}\frac{v_i-\th-i}{v_i-\th}+\prod_{i=1}^{J_v}\frac{v_i-\th+i}{v_i-\th}\prod_{\a=1}^{L}\frac{\th-\th_a}{\th-\th_\a-i}\)|\,\Psi\rangle+\dots
\end{eqnarray*}
where dots stand for \textit{undesirable} terms which would make
$|\Psi\rangle$ not to be an eigenvector of $\Tr\,\hat{T}$ while the
displaced term is the one we obtain ignoring the second term in the
rhs of both (\ref{AB}) and (\ref{DB}). The condition that these
\textit{undesirable} terms vanish gives us a set of equations for
$u_i$ and $v_j$. There is however a shortcut to arrive at these
equations provided we know that these terms can indeed be killed.
The argument is the following -- each of the two last terms inside
the big parentheses came from the diagonalization of a product of
$\mathcal{Q}=\th-i P$. The diagonalization of such a product of
operators must yield a polynomial in $\theta$ therefore the residues
of the apparent poles which seem to be part of the eigenvalue for
$\th=u_i$ (or $v_j$) must vanish. This implies
\begin{eqnarray}
1&=&\prod_{i\neq j}^{J_u}\frac{u_j-u_i+i}{u_j-u_i-i}\prod_{\a=1}^{L}\frac{u_j-\th_a}{u_j-\th_a+i} \la{baeu}\\
1&=&\prod_{i\neq
j}^{J_v}\frac{v_j-v_i+i}{v_j-v_i-i}\prod_{\a=1}^{L}\frac{v_j-\th_a}{u_j-\th_a+i}
\la{baev}
\end{eqnarray}
Furthermore (\ref{periodicity}) reads
\begin{eqnarray}
e^{im_0\mathcal{L}\sinh\pi\theta_\b}\prod_{\a\neq\b}^{L}S_0^2(\th_\b-\th_a)
\prod_{i=1}^{J_u}\frac{\th_\b-u_i+i}{\th_\b-u_i}
\prod_{i=1}^{J_v}\frac{\th_\b-v_i+i}{\th_\b-v_i}=1 \la{baeth}\,.
\end{eqnarray}
Equations (\ref{baeu}-\ref{baeth}) coincide precisely with
(\ref{DBAE1}-\ref{DBAE3}) after the trivial shift $(u,v)\rightarrow
(u-i/2,v-i/2)$ in the former.

%%%%%%%%%%%%%%%%%%%%%%%%%%%%%%%%%%%%%%%%%%%%%%%%%%%%%%%%%%%%%%%%%%%%%%%%%%%%%
\appendix{Bethe Ansatz Equations for $O(2n)$ Sigma-Model\label{App:geneO2n}}

Zamolodchikovs' S-matrix for the $O(2n)$ sigma-model
takes the form\cite{Zamolodchikov:1977nu,Zamolodchikov:1978xm}
\begin{eqnarray}
\hS_{a\phantom{'}b}^{a'b'}(\theta)
\Eqn{=}\sigma_2(\theta)
\left[
\hI_{a\phantom{'}b}^{a'b'}
-\frac{i}{\theta}\hP_{a\phantom{'}b}^{a'b'}
-\frac{i}{i(n-1)-\theta}\hK_{a\phantom{'}b}^{a'b'}
\right]
\end{eqnarray}
where the overall factor is given by
\begin{equation}
\sigma_2(\theta)=
\frac{\Gamma(\Delta+\varphi)\Gamma(1-\varphi)
      \Gamma(\tfrac{1}{2}+\varphi)
      \Gamma(\tfrac{1}{2}+\Delta-\varphi)}
     {\Gamma(1+\Delta-\varphi)\Gamma(\varphi)
      \Gamma(\tfrac{1}{2}-\varphi)
      \Gamma(\tfrac{1}{2}+\Delta+\varphi)},
\end{equation}
\begin{equation}
\varphi=\frac{-i\theta}{2n-2},
\quad \Delta=\frac{1}{2n-2},
\end{equation}
and
\begin{equation}
\hI_{a\phantom{'}b}^{a'b'}
  =\delta_{a}^{a'}\delta_{b}^{b'},\quad
\hP_{a\phantom{'}b}^{a'b'}
  =\delta_{a}^{b'}\delta_{b}^{a'},\quad
\hK_{a\phantom{'}b}^{a'b'}
  =\delta_{ab}\delta^{a'b'}
\end{equation}
are $O(2n)$-invariant tensor bases.
Let us restrict ourselves to $n\ge 2$.
The above form is determined as a minimal solution
satisfying the Yang--Baxter equation
\begin{equation}
\hS_{c_1c_2}^{b_1b_2}(\theta)
\hS_{a_1c_3}^{c_1b_3}(\theta+\theta')
\hS_{a_2a_3}^{c_2c_3}(\theta')
=
\hS_{a_1a_2}^{c_1c_2}(\theta)
\hS_{c_1a_3}^{b_1c_3}(\theta+\theta')
\hS_{c_2c_3}^{b_2b_3}(\theta'),
\end{equation}
the unitarity condition
\begin{equation}
\hS_{b_1b_2}^{c_1c_2}(\theta)
\hS_{a_1a_2}^{b_1b_2}(-\theta)=\hI_{a_1a_2}^{c_1c_2},
\end{equation}
and the crossing symmetry
\begin{equation}
\hS_{a\phantom{'}b}^{a'b'}\bigl(i(n-1)-\theta\bigr)
=\hS_{a\phantom{'}b'}^{a'b}(\theta).
\end{equation}
By non-trivial comparison with perturbation theory
\cite{Zamolodchikov:1977nu,Zamolodchikov:1978xm,Hasenfratz:1990zz,Hasenfratz:1990ab},
this S-matrix is believed to describe
the scattering of vector particles
in the $O(2n)$ sigma-model.
Let us consider the system of $L$ particles
put into a one-dimensional periodic box.
The system is described by the Bethe wave function
$\psi^{a_1\cdots a_L}(\theta_1,\ldots,\theta_L)$
where $\{\theta_\alpha\}$ are
rapidities associated with each of the particles.
The wave function reflects
the elastic, factorisable property of the scattering as
\begin{equation}
\psi^{b_2b_1a_3\cdots a_L}(\theta_2,\theta_1,\theta_3,\ldots,\theta_L)
=\hS_{a_1a_2}^{b_1b_2}(\theta_1-\theta_2)
\psi^{a_1a_2a_3\cdots a_L}(\theta_1,\theta_2,\theta_3,\ldots,\theta_L).
\end{equation}
It also satisfies the periodic boundary condition
\begin{equation}
\psi^{a_2\cdots a_La_1}(\theta_2,\ldots,\theta_L,\theta_1)
=e^{-ip(\theta_1)}
\psi^{a_1\cdots a_L}(\theta_1,\ldots,\theta_L).
\end{equation}
The momentum of the $i$th particle is given by
$p(\theta_\alpha)=\mu\sinh(\frac{\pi}{n-1}\theta_\alpha\bigr)$.
Following from these properties, the wave function
obeys the Yang equations
\begin{equation}
\label{YangEqs}
\left[
e^{-ip(\theta_\alpha)}\bigl(\delta_{a_1}^{b_1}\cdots\delta_{a_L}^{b_L}\bigr)
+\hT(\theta_\alpha)_{a_1\cdots a_L}^{b_1\cdots b_L}
\right]
\psi^{a_1\cdots a_L}(\theta_1,\ldots,\theta_L)=0,
\end{equation}
for $\alpha=1,\ldots,L$. We introduced the transfer matrix
$\hT(u)_{a_1\cdots a_L}^{b_1\cdots b_L}$,
given by the trace
\begin{equation}
\hT(u)_{a_1\cdots a_L}^{b_1\cdots b_L}
=\hOmega_a^a(u)_{a_1\cdots a_L}^{b_1\cdots b_L}
\end{equation}
where
\begin{equation}
\hOmega_a^b(u)_{a_1\cdots a_L}^{b_1\cdots b_L}=
\hS_{a\phantom{_1}a_1}^{c_1b_1}(u-\theta_1)
\hS_{c_1a_2}^{c_2b_2}(u-\theta_2)\cdots
\hS_{c_La_L}^{b\phantom{_L}b_L}(u-\theta_L)
\end{equation}
is the monodromy matrix.

The matrix-valued equations
(\ref{YangEqs}) can be diagonalized
with the help of the nested Bethe ansatz\cite{Zamolodchikov:1992zr}.
Supplied with an appropriately defined Bethe vector (wave function)
\cite{deVega:1986xj},
the transfer matrix $\hT(u)$
is diagonalized independently of the value of $u$.
The lowest eigenvalue is given by
\begin{equation}
T(u)=W(u)
\sum_{k=1}^n\bigl(t_k(u)
           +\bar{t}_k(u)\bigr),
\end{equation}
where
\begin{eqnarray}
t_k(u)\Eqn{=}
  Y_{k-1}\bigl(u-\tfrac{ki}{2}\bigr)^{-1}
  Y_k\bigl(u-\tfrac{(k-1)i}{2}\bigr),
  \quad\mbox{for}\quad k=1,\ldots,n-2,\nonumber\\[.8ex]
t_{n-1}(u)\Eqn{=}
  Y_{n-2}\bigl(u-\tfrac{(n-1)i}{2}\bigr)^{-1}
  Y_{n-1}\bigl(u-\tfrac{(n-2)i}{2}\bigr)
  Y_{n}\bigl(u-\tfrac{(n-2)i}{2}\bigr),\nonumber\\[.8ex]
t_n(u)\Eqn{=}
  Y_{n-1}(u-\tfrac{ni}{2})^{-1}
  Y_{n}(u-\tfrac{(n-2)i}{2}),\nonumber\\[.8ex]
\bar{t}_k(u)\Eqn{=}
  Y_{k-1}\bigl(u-\tfrac{(2n-2-k)i}{2}\bigr)
  Y_k\bigl(u-\tfrac{(2n-1-k)i}{2}\bigr)^{-1},
  \quad\mbox{for}\quad k=1,\ldots,n-2,\nonumber\\[.8ex]
\bar{t}_{n-1}(u)\Eqn{=}
  Y_{n-2}\bigl(u-\tfrac{(n-1)i}{2}\bigr)
  Y_{n-1}\bigl(u-\tfrac{ni}{2}\bigr)^{-1}
  Y_{n}\bigl(u-\tfrac{ni}{2}\bigr)^{-1},\nonumber\\[.8ex]
\bar{t}_n(u)\Eqn{=}
  Y_{n-1}(u-\tfrac{(n-2)i}{2})
  Y_{n}(u-\tfrac{ni}{2})^{-1}.
\end{eqnarray}
Scattering terms $W(u),Y_k(u)$ are defined as
\begin{eqnarray}
W(u)\Eqn{=}\prod_{\alpha=1}^{K_0}\sigma_2\bigl(u-u^{(0)}_\alpha\bigr),\\
Y_k(u)\Eqn{=}\prod_{j=1}^{K_k}
  \frac{u-u_j^{(k)}+\frac{i}{2}}{u-u_j^{(k)}-\frac{i}{2}},
\quad\mbox{for}\quad k=0,\ldots,n.
\end{eqnarray}
We sometimes use the notations $u^{(0)}_\alpha:=\theta_\alpha$
and $K_0:=L$ for the sake of simplicity.
$\{u^{(k)}_j\}$ are rapidities of magnons
introduced in the construction of the Bethe vector\cite{deVega:1986xj}.

\begin{figure}[t]
%
%%% D_n Dynkin diagram %%%
%
\begin{center}
\unitlength=2.0pt
\begin{picture}(160,50)
\thicklines
\put(24,19){\line(2,-1){5}}
\put(24,21){\line(2,1){5}}
\put(24,19){\line(1,0){12}}
\put(24,21){\line(1,0){12}}
\put(44,20){\line(1,0){12}}
\dashline[50]{4.8}(64,20)(96,20)
\put(104,20){\line(1,0){12}}
\put(122.82,22.82){\line(1,1){8.48}}
\put(122.82,17.18){\line(1,-1){8.48}}
\put(20,20){\circle{8}}
\put(40,20){\circle{8}}
\put(60,20){\circle{8}}
\put(100,20){\circle{8}}
\put(120,20){\circle{8}}
\put(134.14,5.86){\circle{8}}
\put(134.14,34.14){\circle{8}}
\put(7,27){$\theta=u^{(0)}$}
\put(36,27){$u^{(1)}$}
\put(56,27){$u^{(2)}$}
\put(92,27){$u^{(n-3)}$}
\put(113,27){$u^{(n-2)}$}
\put(140,33){$u^{(n-1)}$}
\put(140,4){$u^{(n)}$}
\end{picture}
\end{center}
\caption{Dynkin diagram for
the $B^{(1)}_n$ root system\label{fig:Dynkin}}
\end{figure}
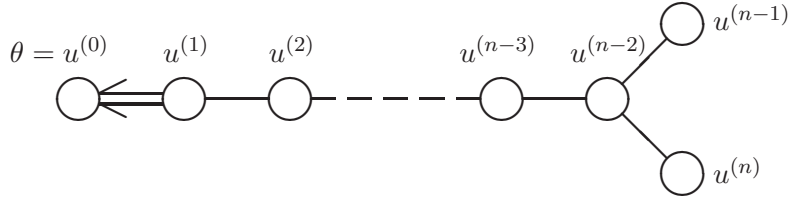

The diagonalized Yang equations are then given by
\begin{equation}
\label{diagYangEq}
e^{-i\mu\sinh(\frac{\pi}{n-1}\theta_\alpha)}
=\prod_{\beta\ne\alpha}^{L}S_0\bigl(\theta_\alpha-\theta_\beta\bigr)
\prod_{j=1}^{K_1}
  \frac{\theta_\alpha-u_j^{(1)}+\frac{i}{2}}
       {\theta_\alpha-u_j^{(1)}-\frac{i}{2}},
\end{equation}
\begin{equation}
\label{geneS0}
S_0(\theta)
=\frac{\theta-i}{\theta}\sigma_2(\theta)
=-\frac{\Gamma(\Delta+\varphi)\Gamma(1-\varphi)
        \Gamma(\tfrac{1}{2}+\varphi)
        \Gamma(\tfrac{1}{2}+\Delta-\varphi)}
       {\Gamma(\Delta-\varphi)\Gamma(1+\varphi)
        \Gamma(\tfrac{1}{2}-\varphi)
        \Gamma(\tfrac{1}{2}+\Delta+\varphi)}.
\end{equation}
To make the diagonalization complete,
the rapidities $\{u^{(k)}_j\}$ have to satisfy
the Bethe ansatz equations.
Ignoring the common scalar factor $W(u)$,
$T(u)$ is essentially a polynomial in $u$
by construction and thus has no pole
at any finite value of $u$.
The requirement of cancellation of poles
between two consecutive terms in $T(u)$
gives rise to the Bethe ansatz equations
\begin{equation}
\label{conciseBAEs}
-1=\prod_{l=0}^n\prod_{j=1}^{K_l}
\frac{u^{(k)}_i-u^{(l)}_j+\frac{i}{2}(\alpha_k|\alpha_l)}
     {u^{(k)}_i-u^{(l)}_j-\frac{i}{2}(\alpha_k|\alpha_l)}.
\end{equation}
$\{\alpha_k\}\ (k=1,\ldots,n)$ denote the simple roots
of the $D_n=\alg{so}(2n)$ Lie algebra
and $(\alpha_k|\alpha_l)$ expresses inner product
defined in the root space.
The non-zero elements read explicitly
$(\alpha_k|\alpha_k)=2$ $(k=1,\ldots,n)$,
$(\alpha_k|\alpha_{k+1})=-1$ $(k=1,\ldots,n-2)$ and
$(\alpha_{n-2}|\alpha_{n})=-1$,
which are encoded in the Dynkin diagram (see Fig.\ref{fig:Dynkin}).
In addition we formally introduced an extra root $\alpha_0$
giving $(\alpha_0|\alpha_k)=-\delta_{1k}$,
so that the interaction between $\theta_\alpha$'s
and $u^{(k)}_j$'s are included
in the common notation (\ref{conciseBAEs}).
Note that the norm of $\alpha_0$ can be fixed
as $(\alpha_0|\alpha_0)=1$,
if we take account of the asymptotics
of the self interaction between $\theta_\alpha$'s:
Observing the asymptotics
\begin{equation}
S_0(\theta)
=1-\frac{i}{\theta}+{\cal O}(\theta^{-2}),
\end{equation}
we see that the self interaction between $\theta_\alpha$'s is
half the strength of that of $u_i$'s.
In the Sutherland limit,
the Yang equation (\ref{diagYangEq}) and
the Bethe ansatz equations (\ref{conciseBAEs})
are unified into the form
\begin{equation}
\sum_{l=0}^n(\alpha_k|\alpha_l)
\,\sG_l\!\bigl(u^{(k)}_i/M\bigr)=2\pi n^{(k)}_i,
\quad\mbox{for}\quad k=0,\ldots,n,
\end{equation}
with mode numbers $n^{(k)}_i\in\bbZ$ and resolvents
\begin{equation}
\label{geneReso}
G_k(z)=\frac{1}{M}\sum_{j=1}^{K_k}\frac{1}{z-u^{(k)}_j/M}.
\end{equation}
Note that $\alpha_0$ precisely corresponds
to the affine root which extends the root system of $D_n$
to that of the $B^{(1)}_n$ affine Lie algebra.
This enhanced symmetry may be viewed as a part of the symmetry
of the conformal field theory arising at the renormalization fixed point.

%%%%%%%%%%%%%%%%%%%%%%%%%%%%%%%%%%%%%%%%%%%%%%%%%%%%%%%%%%%%%%%%%%%%%%%%%%%%


\begin{thebibliography}{100}
\renewcommand{\itemsep}{0ex}
\small


\bibitem{Maldacena:1997re}
%\cite{Maldacena:1997re}
J.~M.~Maldacena,
%``The large N limit of superconformal field theories and supergravity,''
Adv.\ Theor.\ Math.\ Phys.\  {\bf 2} (1998) 231 [Int.\ J.\ Theor.\
Phys.\  {\bf 38} (1999) 1113] [arXiv:hep-th/9711200].
%%CITATION = HEP-TH 9711200;%%

%_std
\bibitem{Witten:1998qj}
%\cite{Witten:1998qj}
  E.~Witten,
  %``Anti-de Sitter space and holography,''
  Adv.\ Theor.\ Math.\ Phys.\  {\bf 2} (1998) 253
  [arXiv:hep-th/9802150].
  %%CITATION = HEP-TH 9802150;%%


%_std
\bibitem{Gubser:1998bc}
%\cite{Gubser:1998bc}
S.~S.~Gubser, I.~R.~Klebanov and A.~M.~Polyakov,
%``Gauge theory correlators from non-critical string theory,''
Phys.\ Lett.\ B {\bf 428} (1998) 105 [arXiv:hep-th/9802109].
%%CITATION = HEP-TH 9802109;%%

%_std
\bibitem{Aharony:1999ti}
%\cite{Aharony:1999ti}
O.~Aharony, S.~S.~Gubser, J.~M.~Maldacena, H.~Ooguri and Y.~Oz,
%``Large N field theories, string theory and gravity,''
Phys.\ Rept.\  {\bf 323} (2000) 183 [arXiv:hep-th/9905111].
%%CITATION = HEP-TH 9905111;%%

%_std
\bibitem{Metsaev:1998it}
%\bibitem{MetsaevT}
%\cite{Metsaev:1998it}
R.~R.~Metsaev and A.~A.~Tseytlin,
%``Type IIB superstring action in AdS(5) x S(5) background,''
Nucl.\ Phys.\ B {\bf 533} (1998) 109 [arXiv:hep-th/9805028].
%%CITATION = HEP-TH 9805028;%%

%_std
\bibitem{Minahan:2002ve}
%\cite{Minahan:2002ve}
J.~A.~Minahan and K.~Zarembo,
%``The Bethe-ansatz for N = 4 super Yang--Mills,''
JHEP {\bf 0303} (2003) 013 [arXiv:hep-th/0212208].
%%CITATION = HEP-TH 0212208;%%

%_std
\bibitem{Lipatov:1994yb}
L.~N.~Lipatov,
%\textit{``High-energy asymptotics of multicolor QCD and exactly solvable
%  lattice models''},
\textsf{JETP~Lett.~59,~596~(1994)}, {\texttt{hep-th/9311037}}.
%%CITATION = HEP-TH 9311037;%%

%_std
\bibitem{Faddeev:1995zg}
L.~D.~Faddeev and G.~P.~Korchemsky,
%\textit{``High-energy QCD as a completely integrable model''},
\textsf{Phys.~Lett.~B342,~311~(1995)}, {\texttt{hep-th/9404173}}.
%%CITATION = HEP-TH 9404173;%%


%_std
\bibitem{Bena:2003wd}
%\cite{Bena:2003wd}
I.~Bena, J.~Polchinski and R.~Roiban,
%``Hidden symmetries of the AdS(5) x S**5 superstring,''
Phys.\ Rev.\ D {\bf 69} (2004) 046002 [arXiv:hep-th/0305116].
%%CITATION = HEP-TH 0305116;%%

%_std
\bibitem{Gubser:2002tv}
%\cite{Gubser:2002tv}
  S.~S.~Gubser, I.~R.~Klebanov and A.~M.~Polyakov,
  %``A semi-classical limit of the gauge/string correspondence,''
  Nucl.\ Phys.\ B {\bf 636} (2002) 99
  [arXiv:hep-th/0204051].
  %%CITATION = HEP-TH 0204051;%%

%_std
\bibitem{Frolov:2002av}
%\cite{Frolov:2002av}
S.~Frolov and A.~A.~Tseytlin,
%``Semiclassical quantization of rotating superstring in AdS(5) x S(5),''
JHEP {\bf 0206} (2002) 007 [arXiv:hep-th/0204226].
%%CITATION = HEP-TH 0204226;%%

%_std
\bibitem{Frolov:2003qc}
%\cite{Frolov:2003qc}
S.~Frolov and A.~A.~Tseytlin,
%``Multi-spin string solutions in AdS(5) x S**5,''
Nucl.\ Phys.\ B {\bf 668} (2003) 77 [arXiv:hep-th/0304255].
%%CITATION = HEP-TH 0304255;%%

%_std
\bibitem{Beisert:2003tq}
%\cite{Beisert:2003tq}
N.~Beisert, C.~Kristjansen and M.~Staudacher,
%``The dilatation operator of N = 4 super Yang--Mills theory,''
Nucl.\ Phys.\ B {\bf 664} (2003) 131 [arXiv:hep-th/0303060].
%%CITATION = HEP-TH 0303060;%%

%_std
\bibitem{Beisert:2003ys}
%\cite{Beisert:2003ys}
N.~Beisert,
%``The su(2$|$3) dynamic spin chain,''
Nucl.\ Phys.\ B {\bf 682} (2004) 487 [arXiv:hep-th/0310252].
%%CITATION = HEP-TH 0310252;%%

%_std
\bibitem{Beisert:2003yb}
%\cite{Beisert:2003yb}
N.~Beisert and M.~Staudacher,
%``The N = 4 SYM integrable super spin chain,''
Nucl.\ Phys.\ B {\bf 670} (2003) 439 [arXiv:hep-th/0307042].
%%CITATION = HEP-TH 0307042;%%

%_std
\bibitem{Sutherland}
B.~Sutherland,
%\textit{``Low-Lying Eigenstates of the
%One-Dimensional Heisenberg Ferromagnet
%  for any Magnetization and Momentum''},
\textsf{Phys.~Rev.~Lett.~74,~816~(1995)}.


%_std
\bibitem{Shastry}
  B.~S.~Shastry and A.~Dhar
  %``Solution of a generalized Stieltjes problem ''
    J.\ Phys.\ A:\ Math. \ Gen.\ {\bf 34} (2001) 6197.
    [arxiv:cond-mat/0101464]

%_std
\bibitem{Beisert:2003xu}
%\cite{Beisert:2003xu}
N.~Beisert, J.~A.~Minahan, M.~Staudacher and K.~Zarembo,
%``Stringing spins and spinning strings,''
JHEP {\bf 0309} (2003) 010 [arXiv:hep-th/0306139].
%%CITATION = HEP-TH 0306139;%%

%_std
\bibitem{Kazakov:2004qf}
%\cite{Kazakov:2004qf}
%\bibitem{KMMZ}
V.~A.~Kazakov, A.~Marshakov, J.~A.~Minahan and K.~Zarembo,
%``Classical / quantum integrability in AdS/CFT,''
JHEP {\bf 0405} (2004) 024 [arXiv:hep-th/0402207].
%%CITATION = HEP-TH 0402207;%%

%_std
\bibitem{Kazakov:2004nh}
%\cite{Kazakov:2004nh}
V.~A.~Kazakov and K.~Zarembo,
%``Classical / quantum integrability in non-compact sector of AdS/CFT,''
JHEP {\bf 0410} (2004) 060 [arXiv:hep-th/0410105].
%%CITATION = HEP-TH 0410105;%%

%_std
\bibitem{Beisert:2004ag}
%\cite{Beisert:2004ag}
%\bibitem{BKS}
N.~Beisert, V.~A.~Kazakov and K.~Sakai,
%``Algebraic curve for the SO(6) sector of AdS/CFT,''
arXiv:hep-th/0410253.
%%CITATION = HEP-TH 0410253;%%

%_std
\bibitem{Beisert:2005bm}
%\cite{Beisert:2005bm}
N.~Beisert, V.~A.~Kazakov, K.~Sakai and K.~Zarembo,
%``The algebraic curve of classical superstrings on AdS(5) x S**5,''
arXiv:hep-th/0502226.
%%CITATION = HEP-TH 0502226;%%

%_std
\bibitem{Schafer-Nameki:2004ik}
%\cite{Schafer-Nameki:2004ik}
S.~Schafer-Nameki,
%``The algebraic curve of 1-loop planar N = 4 SYM,''
Nucl.\ Phys.\ B {\bf 714} (2005) 3 [arXiv:hep-th/0412254].
%%CITATION = HEP-TH 0412254;%%

%_std
\bibitem{Beisert:2005di}
%\cite{Beisert:2005di}
%\bibitem{BKSZ2}
N.~Beisert, V.~A.~Kazakov, K.~Sakai and K.~Zarembo,
%``Complete spectrum of long operators in N = 4 SYM at one loop,''
JHEP {\bf 0507} (2005) 030 [arXiv:hep-th/0503200].
%%CITATION = HEP-TH 0503200;%%

%_std
\bibitem{Staudacher:2004tk}
%\cite{Staudacher:2004tk}
M.~Staudacher, ``The factorized S-matrix of CFT/AdS,'' JHEP {\bf
0505} (2005) 054 [arXiv:hep-th/0412188].
%%CITATION = HEP-TH 0412188;%%

%_std
\bibitem{Beisert:2004hm}
%\cite{Beisert:2004hm}
N.~Beisert, V.~Dippel and M.~Staudacher,
%``A novel long range spin chain and planar N = 4 super Yang--Mills,''
JHEP {\bf 0407} (2004) 075 [arXiv:hep-th/0405001].
%%CITATION = HEP-TH 0405001;%%

%_std
\bibitem{Beisert:2005fw}
%\cite{Beisert:2005fw}
N.~Beisert and M.~Staudacher,
%``Long-range PSU(2,2$|$4) Bethe ansaetze for gauge theory and strings,''
Nucl.\ Phys.\ B {\bf 727} (2005) 1 [arXiv:hep-th/0504190].
%%CITATION = HEP-TH 0504190;%%

%_std
\bibitem{Serban:2004jf}
%\cite{Serban:2004jf}
D.~Serban and M.~Staudacher,
%``Planar N = 4 gauge theory and the Inozemtsev long range spin chain,''
JHEP {\bf 0406} (2004) 001 [arXiv:hep-th/0401057].
%%CITATION = HEP-TH 0401057;%%

%_std
\bibitem{Frolov:2003tu}
%\cite{Frolov:2003tu}
S.~Frolov and A.~A.~Tseytlin,
%``Quantizing three-spin string solution in AdS(5) x S**5,''
JHEP {\bf 0307} (2003) 016 [arXiv:hep-th/0306130].
%%CITATION = HEP-TH 0306130;%%

%_std
\bibitem{Parnachev:2002kk}
%\cite{Parnachev:2002kk}
A.~Parnachev and A.~V.~Ryzhov,
%``Strings in the near plane wave background and AdS/CFT,''
JHEP {\bf 0210} (2002) 066 [arXiv:hep-th/0208010].
%%CITATION = HEP-TH 0208010;%%

%_std
\bibitem{Callan:2003xr}
%\cite{Callan:2003xr}
  C.~G.~.~Callan, H.~K.~Lee, T.~McLoughlin, J.~H.~Schwarz, I.~J.~Swanson and X.~Wu,
%``Quantizing string theory in AdS(5) x S**5: Beyond the pp-wave,''
  Nucl.\ Phys.\ B {\bf 673}, 3 (2003)
  [arXiv:hep-th/0307032].
  %%CITATION = HEP-TH 0307032;%%

%_std
\bibitem{Fuji:2005ry}
%\cite{Fuji:2005ry}
H.~Fuji and Y.~Satoh,
%``Quantum fluctuations of rotating strings in AdS(5) x S**5,''
arXiv:hep-th/0504123.
%%CITATION = HEP-TH 0504123;%%

%_std
\bibitem{Beisert:2005mq}
%\cite{Beisert:2005mq}
N.~Beisert, A.~A.~Tseytlin and K.~Zarembo,
%``Matching quantum strings to quantum spins: One-loop vs. finite-size
%corrections,''
Nucl.\ Phys.\ B {\bf 715} (2005) 190 [arXiv:hep-th/0502173].
%%CITATION = HEP-TH 0502173;%%

%_std
\bibitem{Hernandez:2005nf}
  R.~Hernandez, E.~Lopez, A.~Perianez and G.~Sierra,
 % \textit{``Finite size effects in ferromagnetic spin chains and quantum  corrections
 % to classical strings''},
  \textsf{JHEP {\bf 0506}, 011 (2005)}
  [arXiv:hep-th/0502188].
  %%CITATION = HEP-TH 0502188;%%



%_std
\bibitem{Beisert:2005bv}
%\cite{Beisert:2005bv}
N.~Beisert and L.~Freyhult,
%``Fluctuations and energy shifts in the Bethe ansatz,''
Phys.\ Lett.\ B {\bf 622} (2005) 343 [arXiv:hep-th/0506243].
%%CITATION = HEP-TH 0506243;%%

%_std
\bibitem{Gromov:2005gp}
%\cite{Gromov:2005gp}
N.~Gromov and V.~Kazakov,
%``Double scaling and finite size corrections in sl(2) spin chain,''
Nucl.\ Phys.\ B {\bf 736} (2006) 199 [arXiv:hep-th/0510194].
%%CITATION = HEP-TH 0510194;%%

%_std
\bibitem{Polyakov:2005ss}
%\cite{Polyakov:2005ss}
  A.~M.~Polyakov,
 % ``Supermagnets and sigma models,''
  arXiv:hep-th/0512310.
  %%CITATION = HEP-TH 0512310;%%

%_std
\bibitem{Zamolodchikov:1977nu}
%\cite{Zamolodchikov:1977nu}
A.~B.~Zamolodchikov and A.~B.~Zamolodchikov,
%``Relativistic Factorized S Matrix In Two-Dimensions Having O(N) Isotopic
%Symmetry,''
Nucl.\ Phys.\ B {\bf 133} (1978) 525 [JETP Lett.\  {\bf 26} (1977)
457].
%%CITATION = NUPHA,B133,525;%%

%_std
\bibitem{Polyakov:1983tt}
  A.~M.~Polyakov and P.~B.~Wiegmann,
 % ``Theory Of Nonabelian Goldstone Bosons In Two Dimensions,''
  Phys.\ Lett.\ B {\bf 131}, 121 (1983).
  %%CITATION = PHLTA,B131,121;%%

%_std
\bibitem{Wiegmann:1984ec}
%\cite{Wiegmann:1984ec}
  P.~Wiegmann,
 % ``Exact Factorized S Matrix Of The Chiral Field In Two-Dimensions,''
  Phys.\ Lett.\ B {\bf 142}, 173 (1984).
  %%CITATION = PHLTA,B142,173;%%

%_std
\bibitem{Wiegmann:1984pw}
  P.~B.~Wiegmann,
  %``On The Theory Of Nonabelian Goldstone Bosons In Two-Dimensions: Exact
  %Solution Of The O(3) Nonlinear Sigma Model,''
  Phys.\ Lett.\ B {\bf 141}, 217 (1984).
  %%CITATION = PHLTA,B141,217;%%

%_std
\bibitem{Faddeev:1985qu}
%\cite{Faddeev:1985qu}
  L.~D.~Faddeev and N.~Y.~Reshetikhin,
  %``Integrability Of The Principal Chiral Field Model In (1+1)-Dimension,''
  Annals Phys.\  {\bf 167}, 227 (1986).
  %%CITATION = APNYA,167,227;%%

%_std
\bibitem{Ogievetsky:1984pv}
%\cite{Ogievetsky:1984pv}
  E.~Ogievetsky, N.~Reshetikhin and P.~Wiegmann,
  %``The Principal Chiral Field In Two-Dimension And Classical Lie Algebra,''
NORDITA-84/38

%_std
\bibitem{Wiegmann:1984mk}
%\cite{Wiegmann:1984mk}
P.~B.~Wiegmann,
%``Exact Solution For The SU(N) Main Chiral Field In Two-Dimensions,''
JETP Lett.\  {\bf 39} (1984) 214 [Pisma Zh.\ Eksp.\ Teor.\ Fiz.\
{\bf 39} (1984) 180].
%%CITATION = JTPLA,39,214;%%

%_std
\bibitem{Fateev:1994ai}
%\cite{Fateev:1994ai}
  V.~A.~Fateev, V.~A.~Kazakov and P.~B.~Wiegmann,
 % ``Principal chiral field at large N,''
  Nucl.\ Phys.\ B {\bf 424}, 505 (1994)
  [arXiv:hep-th/9403099].
  %%CITATION = HEP-TH 9403099;%%

%_std
\bibitem{Shankar:1977cm}
%\cite{Shankar:1977cm}
R.~Shankar and E.~Witten,
%``The S Matrix Of The Supersymmetric Nonlinear Sigma Model,''
Phys.\ Rev.\ D {\bf 17} (1978) 2134.
%%CITATION = PHRVA,D17,2134;%%

%_std
\bibitem{Saleur:2001cw}
%\cite{Saleur:2001cw}
H.~Saleur and B.~Wehefritz-Kaufmann,
%``Integrable quantum field theories with OSP(m/2n) symmetries,''
Nucl.\ Phys.\ B {\bf 628} (2002) 407 [arXiv:hep-th/0112095].
%%CITATION = HEP-TH 0112095;%%

%_std
\bibitem{Mann:2004jr}
%\cite{Mann:2004jr}
  N.~Mann and J.~Polchinski,
  %``Finite density states in integrable conformal field theories,''
  arXiv:hep-th/0408162.
  %%CITATION = HEP-TH 0408162;%%

%_std
\bibitem{Dorn:1994xn}
%\cite{Dorn:1994xn}
  H.~Dorn and H.~J.~Otto,
  %``Two and three point functions in Liouville theory,''
  Nucl.\ Phys.\ B {\bf 429}, 375 (1994)
  [arXiv:hep-th/9403141].
  %%CITATION = HEP-TH 9403141;%%


%_std
\bibitem{Zamolodchikov:1995aa}
%\cite{Zamolodchikov:1995aa}
A.~B.~Zamolodchikov and A.~B.~Zamolodchikov,
%``Structure constants and conformal bootstrap in Liouville field theory,''
Nucl.\ Phys.\ B {\bf 477} (1996) 577 [arXiv:hep-th/9506136].
%%CITATION = HEP-TH 9506136;%%


%_std
\bibitem{Teschner:1997ft}
  J.~Teschner,
  %`On structure constants and fusion rules in the SL(2,C)/SU(2) WZNW  model,''
  Nucl.\ Phys.\ B {\bf 546}, 390 (1999)
  [arXiv:hep-th/9712256].
  %%CITATION = HEP-TH 9712256;%%

%_std
\bibitem{FZZ} V.~Fateev, Al.~Zamolodchikov and A.~Zamolodchikov,
unpublished (see in P.~Baseilhac and V.~A.~Fateev,
  %``Expectation values of local fields for a two-parameter family of
  %integrable models and related perturbed conformal field theories,''
  Nucl.\ Phys.\ B {\bf 532}, 567 (1998)
  [arXiv:hep-th/9906010]).

%_std
\bibitem{Polyakov:1998ju}
%\cite{Polyakov:1998ju}
  A.~M.~Polyakov,
  %``The wall of the cave,''
  Int.\ J.\ Mod.\ Phys.\ A {\bf 14} (1999) 645
  [arXiv:hep-th/9809057].
  %%CITATION = HEP-TH 9809057;%%

%_std
\bibitem{Kazakov:1988ch}
%\cite{Kazakov:1988ch}
  V.~A.~Kazakov and A.~A.~Migdal,
  %``Recent Progress In The Theory Of Noncritical Strings,''
  Nucl.\ Phys.\ B {\bf 311}, 171 (1988).
  %%CITATION = NUPHA,B311,171;%%

%_std
\bibitem{Kazakov:2000pm}
%\cite{Kazakov:2000pm}
  V.~Kazakov, I.~K.~Kostov and D.~Kutasov,
 % ``A matrix model for the two-dimensional black hole,''
  Nucl.\ Phys.\ B {\bf 622}, 141 (2002)
  [arXiv:hep-th/0101011].
  %%CITATION = HEP-TH 0101011;%%

%_std
\bibitem{Maldacena:2002eg}
%\cite{Maldacena:2002eg}
  J.~M.~Maldacena,
  %``Lectures on AdS/CFT,''
%\href{http://www.slac.stanford.edu/spires/find/hep/www?irn=6239110}{SPIRES entry}
{\it Prepared for Theoretical Advanced Study Institute in Elementary
Particle Physics (TASI 2002): Particle Physics and Cosmology: The
Quest for Physics Beyond the Standard Model(s), Boulder, Colorado,
2-28 Jun 2002}

%_std
\bibitem{Arutyunov:2004vx}
%\cite{Arutyunov:2004vx}
  G.~Arutyunov, S.~Frolov and M.~Staudacher,
  %``Bethe ansatz for quantum strings,''
  JHEP {\bf 0410} (2004) 016
  [arXiv:hep-th/0406256].
  %%CITATION = HEP-TH 0406256;%%

%_std
\bibitem{Arutyunov:2004vx}
%\cite{Arutyunov:2004vx}
  G.~Arutyunov, S.~Frolov and M.~Staudacher,
 % ``Bethe ansatz for quantum strings,''
  JHEP {\bf 0410}, 016 (2004)
  [arXiv:hep-th/0406256].
  %%CITATION = HEP-TH 0406256;%%



%_std
\bibitem{Schafer-Nameki:2005tn}
  S.~Schafer-Nameki, M.~Zamaklar and K.~Zarembo,
 % \textit{``Quantum corrections to spinning strings in AdS(5) x S**5 and Bethe ansatz:
 % A comparative study,''}
  arXiv:hep-th/0507189.
  %%CITATION = HEP-TH 0507189;%%


%_std
\bibitem{Frolov:2006cc}
%\cite{Frolov:2006cc}
  S.~Frolov, J.~Plefka and M.~Zamaklar,
 % ``The AdS(5) x S**5 superstring in light-cone gauge and its Bethe
 % equations,''
  arXiv:hep-th/0603008.
  %%CITATION = HEP-TH 0603008;%%


%_std
\bibitem{Faddeev:1979gh}
%\cite{Faddeev:1979gh}
  L.~D.~Faddeev, E.~K.~Sklyanin and L.~A.~Takhtajan,
 % ``The Quantum Inverse Problem Method. 1,''
  Theor.\ Math.\ Phys.\  {\bf 40}, 688 (1980)
  [Teor.\ Mat.\ Fiz.\  {\bf 40}, 194 (1979)].
  %%CITATION = TMPHA,40,688;%%



%_std
\bibitem{Kulish:1981gi}
  P.~P.~Kulish, N.~Y.~Reshetikhin and E.~K.~Sklyanin,
 % \textit{``Yang-Baxter Equation And Representation Theory: I,''}
  \textsf{Lett.\ Math.\ Phys.\  {\bf 5}, 393 (1981).}
  %%CITATION = LMPHD,5,393;%%

%_std
\bibitem{Beisert:2005tm}
%\cite{Beisert:2005tm}
  N.~Beisert,
 % ``The su(2$|$2) dynamic S-matrix,''
  arXiv:hep-th/0511082.
  %%CITATION = HEP-TH 0511082;%%

%_std
\bibitem{Mann:2005ab}
%\cite{Mann:2005ab}
  N.~Mann and J.~Polchinski,
  %``Bethe ansatz for a quantum supercoset sigma model,''
  Phys.\ Rev.\ D {\bf 72} (2005) 086002
  [arXiv:hep-th/0508232].
  %%CITATION = HEP-TH 0508232;%%

%_std
\bibitem{Rej:2005qt}
%\cite{Rej:2005qt}
  A.~Rej, D.~Serban and M.~Staudacher,
  %``Planar N = 4 gauge theory and the Hubbard model,''
  arXiv:hep-th/0512077.
  %%CITATION = HEP-TH 0512077;%%

%_std
\bibitem{Dorey:2006zj}
%\cite{Dorey:2006zj}
  N.~Dorey and B.~Vicedo,
  %``On the dynamics of finite-gap solutions in classical string theory,''
  arXiv:hep-th/0601194.
  %%CITATION = HEP-TH 0601194;%%

%_std
\bibitem{Novikov:1984id}
  S.~Novikov, S.~V.~Manakov, L.~P.~Pitaevsky and V.~E.~Zakharov,
  \textit{``Theory Of Solitons. The Inverse Scattering Method,''}
%\href{http://www.slac.stanford.edu/spires/find/hep/www?irn=1274929}{SPIRES entry}

%_std
\bibitem{ZHUK} N.~E.~Zhukovsky,
Trudi Otdelenia fizicheskih nauk O.L.E., t.XIII, vip.2 (1906) (In
Russian).

%_std
\bibitem{Minahan:2005jq}
%\cite{Minahan:2005jq}
%\bibitem{Minahan}
  J.~A.~Minahan,
 % \textit{``The SU(2) sector in AdS/CFT,''}
  \textsf{Fortsch.\ Phys.\  {\bf 53}, 828 (2005)}
  [arXiv:hep-th/0503143].
%%CITATION = HEP-TH 0405243;%%

%_std
\bibitem{Green:1981yb}
%\cite{Green:1981yb}
  M.~B.~Green and J.~H.~Schwarz,
  %\textit{``Supersymmetrical String Theories,''}
  \textsf{Phys.\ Lett.\ B {\bf 109} (1982) 444.}
%%CITATION = PHLTA,B109,444;%%

%_std
\bibitem{Zamolodchikov:1992zr}
%\cite{Zamolodchikov:1992zr}
A.~B.~Zamolodchikov and A.~B.~Zamolodchikov,
%``Massless factorized scattering and sigma models with topological terms,''
Nucl.\ Phys.\ B {\bf 379} (1992) 602.
%%CITATION = NUPHA,B379,602;%%

%_std
\bibitem{Berenstein:2002jq}
%\cite{Berenstein:2002jq}
D.~Berenstein, J.~M.~Maldacena and H.~Nastase,
%``Strings in flat space and pp waves from N = 4 super Yang Mills,''
JHEP {\bf 0204} (2002) 013 [arXiv:hep-th/0202021].
%%CITATION = HEP-TH 0202021;%%

%_std
\bibitem{Kedem:1992jv}
%\cite{Kedem:1992jv}
R.~Kedem, T.~R.~Klassen, B.~M.~McCoy and E.~Melzer,
%``Fermionic quasiparticle representations for characters of G(1)1 x G(1)1 /
%G(1)2,''
Phys.\ Lett.\ B {\bf 304} (1993) 263 [arXiv:hep-th/9211102].
%%CITATION = HEP-TH 9211102;%%

%_std
\bibitem{Kuniba:1992sp}
%\cite{Kuniba:1992sp}
A.~Kuniba, T.~Nakanishi and J.~Suzuki,
%``Characters in conformal field theories from thermodynamic Bethe ansatz,''
Mod.\ Phys.\ Lett.\ A {\bf 8} (1993) 1649 [arXiv:hep-th/9301018].
%%CITATION = HEP-TH 9301018;%%

%_std
\bibitem{Smirnov}
  F.~A.~Smirnov,
 % \textit{``Quasi-classical study of form factors in finite volume''},
  arXiv:hep-th/9802132.
  %%CITATION = HEP-TH 9802132;%%

%_std
\bibitem{KloseZarembo}
%\cite{KloseZarembo}
T.~Klose and K.~Zarembo,
%``Bethe Ansatz in Stringy Sigma Models,''
to appear.
%arXiv:hep-th/0603***.
%%CITATION = HEP-TH 0603***;%%



%_std
\bibitem{Karowski:1978ps}
%\cite{Karowski:1978ps}
  M.~Karowski,
 % ``On The Bound State Problem In (1+1)-Dimensional Field Theories,''
  Nucl.\ Phys.\ B {\bf 153} (1979) 244.
  %%CITATION = NUPHA,B153,244;%%

%_std
\bibitem{Berg:1978dz}
%\cite{Berg:1978dz}
  B.~Berg, M.~Karowski, V.~Kurak and P.~Weisz,
  %``Scattering Amplitudes Of The Gross-Neveu And Nonlinear Sigma Models In
  %Higher Orders Of The 1/N Expansion,''
  Phys.\ Lett.\ B {\bf 76} (1978) 502.
%%CITATION = PHLTA,B76,502;%%

%_std
\bibitem{Faddeev:1996iy}
%\cite{Faddeev:1996iy}
  L.~D.~Faddeev,
% ``How Algebraic Bethe Ansatz works for integrable model,''
  arXiv:hep-th/9605187.
  %%CITATION = HEP-TH 9605187;%%

%_std
\bibitem{Zamolodchikov:1978xm}
%\cite{Zamolodchikov:1978xm}
  A.~B.~Zamolodchikov and A.~B.~Zamolodchikov,
  %``Factorized S-Matrices In Two Dimensions As The Exact Solutions Of  Certain
  %Relativistic Quantum Field Models,''
  Annals Phys.\  {\bf 120} (1979) 253.
  %%CITATION = APNYA,120,253;%%

%_std
\bibitem{Hasenfratz:1990zz}
%\cite{Hasenfratz:1990zz}
  P.~Hasenfratz, M.~Maggiore and F.~Niedermayer,
  %``The Exact Mass Gap Of The O(3) And O(4) Nonlinear Sigma Models In D = 2,''
  Phys.\ Lett.\ B {\bf 245} (1990) 522.
  %%CITATION = PHLTA,B245,522;%%

%_std
\bibitem{Hasenfratz:1990ab}
%\cite{Hasenfratz:1990ab}
  P.~Hasenfratz and F.~Niedermayer,
  %``The Exact Mass Gap Of The O(N) Sigma Model For Arbitrary N Is >= 3 In D =
  %2,''
  Phys.\ Lett.\ B {\bf 245} (1990) 529.
  %%CITATION = PHLTA,B245,529;%%

%_std
\bibitem{deVega:1986xj}
%\cite{deVega:1986xj}
H.~J.~de Vega and M.~Karowski,
%``Exact Bethe Ansatz Solution Of 0(2n) Symmetric Theories,''
Nucl.\ Phys.\ B {\bf 280} (1987) 225.
%%CITATION = NUPHA,B280,225;%%

%_std
\end{thebibliography}
\end{document}